\documentclass{aa}  
\pdfoutput=1
\usepackage{graphicx}
\usepackage{txfonts}
\usepackage{xcolor}
\usepackage{appendix}
\usepackage{indentfirst}
\usepackage{parskip}
\usepackage{multirow}
\usepackage{soul}
\usepackage{bm}

\begin{document}

   \title{ The GUAPOS project: G31.41+0.31 Unbiased ALMA sPectral Observational Survey   }

   \subtitle{ I. Isomers of $\mathrm{C_{2}H_{4}O_{2}}$ }

   \author{C. Mininni \inst{1,}\inst{2,}\thanks{Contact: mininni@arcetri.astro.it}, M. T. Beltr\'an\inst{2},  V. M. Rivilla\inst{3,2}, A. S\'anchez-Monge\inst{4}, F. Fontani\inst{2}, T. M\"oller\inst{4},\\ 
   R. Cesaroni\inst{2}, P. Schilke\inst{4}, S. Viti\inst{5}, I. Jim\'enez-Serra\inst{3}, L. Colzi\inst{3,2}, A. Lorenzani\inst{2}, L. Testi\inst{6}
          }

   \institute{Dipartimento di Fisica e Astronomia, Universit\`a degli Studi di Firenze, Via Sansone 1, 50019 Sesto Fiorentino, Italy 
               \and INAF Osservatorio Astrofisico di Arcetri, Largo E. Fermi 5, 50125 Firenze, Italy 
            \and Centro de Astrobiolog\'ia (CSIC, INTA), Ctra. de Ajalvir, km. 4, Torrej\'on de Ardoz, E-28850 Madrid, Spain
           \and I. Physikalisches Institut, Universit\"at zu K\"oln, Z\"ulpicher Str. 77, 50937 K\"oln, Germany
           \and Dep. of Physics and Astronomy, UCL, Gower Place, London WC1E 6BT, UK
           \and European Southern Observatory (ESO), Karl-Schwarzschild-Str. 2, D-85748 Garching, Germany
             }


 
  \abstract
   {Understanding the degree of chemical complexity that can be reached in star-forming regions, together with the identification of precursors of the building blocks of life in the interstellar medium, is one of the goals of astrochemistry. To answer these questions, unbiased spectral surveys with large bandwidth and high spectral resolution are needed, to resolve line blending in chemically rich sources and identify each molecule (especially for complex organic molecules). This kind of observations has been successfully carried out,  mainly towards the Galactic Center, a region that shows peculiar environmental conditions.  } 
   {We present an unbiased spectral survey of one of the most chemically rich hot molecular cores located  outside the Galactic Center, in the high-mass star-forming region G31.41+0.31. The aim of this 3mm spectral survey is to identify and characterize the physical parameters of the gas emitting in different molecular species, focusing on complex organic molecules. In this first paper, we present the survey and discuss the detection and relative abundances of the 3 isomers of $\mathrm{C_{2}H_{4}O_{2}}$: methyl formate, glycolaldehyde and acetic acid.  }
   {Observations were carried out with the interferometer ALMA and cover the entire Band 3 from 84 to 116 GHz ($\sim 32$ GHz bandwidth) with an angular resolution of $1.2\arcsec\times1.2\arcsec$ ($\sim4400\,\mathrm{au}\times4400\,\mathrm{au}$) and a spectral resolution of $\sim0.488\,$ MHz ($\sim1.3-1.7\,\mathrm{km\,s^{-1}}$). The transitions of the 3 molecules have been analyzed with the software XCLASS to determine the physical parameters of the emitting gas. }
   {All three isomers were detected with abundances of $(2\pm0.6)\times10^{-7}$, $(4.3-8)\times10^{-8}$ and $(5.0\pm1.4)\times10^{-9}$ for methyl formate, acetic acid and  glycolaldehyde, respectively. Methyl formate and acetic acid abundances are the highest detected up to now, if compared to sources in literature. The size of the emission varies among the three isomers with acetic acid showing the most compact emission while methyl formate exhibits the most extended emission. Different chemical pathways, involving both grain-surface chemistry and cold or hot gas-phase reactions, have been proposed for the formation of these molecules, but the small number of detections, especially of acetic acid and glycolaldehyde, made it very difficult to confirm or discard the predictions of the models. The comparison with chemical models in literature suggests the necessity of grain-surface routes for the formation of methyl formate in G31, while for glycolaldehyde both scenarios could be feasible. Proposed grain-surface reaction for acetic acid is not able to reproduce the observed abundance in this work, while gas-phase scenario should be further tested due to large uncertainties. 
   }
   {}


  \titlerunning{The GUAPOS project - I. Isomers of $\mathrm{C_{2}H_{4}O_{2}}$ }
  \authorrunning{Mininni et al. }
   \maketitle
%

\section{Introduction}
\begin{figure*}[t]
\begin{centering}
\includegraphics[width=16cm]{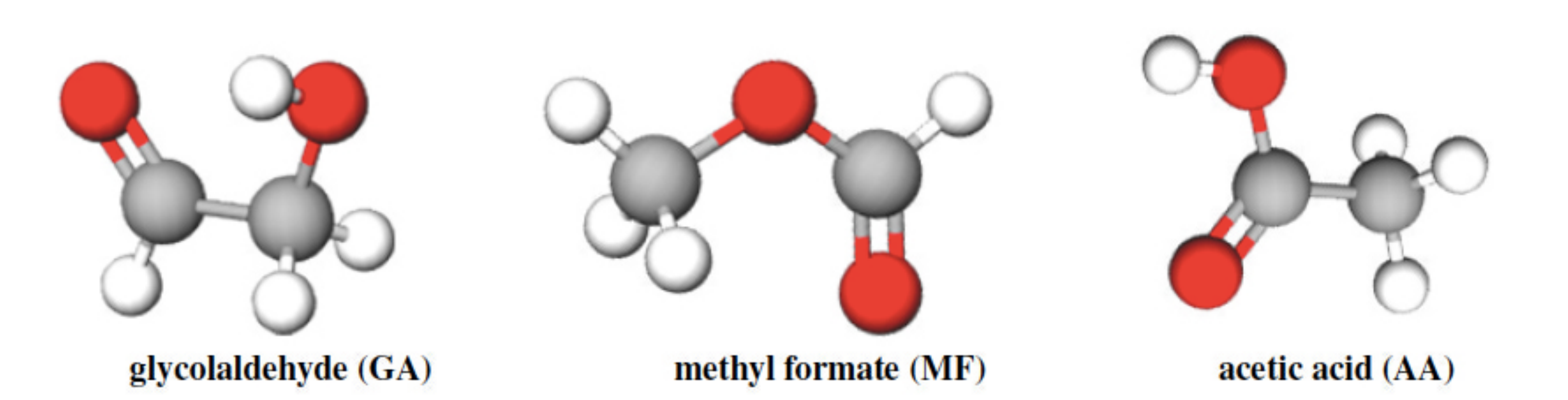}
\vspace{3mm} 
\caption{Chemical structure of the 3 isomers of $\mathrm{C_2H_4O_2}$.}
\label{figisomeri}
\end{centering}
\end{figure*}
\setlength{\parindent}{3ex}
\par Hot Molecular Cores (HMCs), the birthplace of massive stars, are the sources that show the highest chemical complexity in the Galaxy \citep{ComRevHerbst}. Their spectra include emission from a large variety of molecules, starting from the simplest diatomic species, to reach complex organic molecules (COMs), i.e. molecules containing carbon with 6 or more atoms. In recent years, our view and comprehension of the chemistry of the interstellar medium (ISM) has substantially improved, and species with up to 13 atoms
have been unambiguously detected (c-C$_6$H$_5$CN, \citealt{McGuire2018}).
The advent of more sensitive instruments with high spectral and spatial resolution such as ALMA (Atacama Large Millimeter Array), has allowed to detect emission from faint molecular species, in particular of heavy COMs, since they typically emit a large number of transitions that could be faint and/or blended with those of other molecules.\newline\indent
This chemically rich environment and the presence of several COMs is thought to be the result of the evaporation of the products of grain-surface reactions, thanks to the presence of an already formed proto-stellar object(s) which starts to heat up the surrounding medium, and of the subsequent hot gas-phase chemistry. 
The detection of COMs also in cold environments \citep{Ob2010,bac2012,vas2014,jimenez2016} opened new scenarios, and other possible chemical routes in cold gas-phase were proposed to explain these observations (i.e. \citealt{vasy13,balucani2015, vas2017}). It is thus possible that also in HMCs  some of the observed COM emission is inherited from the cold chemistry at early stages of the star-formation process. Hence, these new routes need to be considered in chemical models. \newline\indent
To investigate the maximum degree of chemical complexity in star-forming regions, spectral surveys have been carried out towards a few sources, both high-mass and low-mass cores \citep{kal2010,bell2016,Jorge2016}. One of the most studied sources is SgrB2 \citep{bell13,bell2016,bell19,alvaro2017}, although the environmental conditions of this source, catalogued as a mini-starbust region, cannot be considered as typical of HMCs because such a peculiar environment has likely an impact on the chemistry.\newline\indent
With this in mind we decided to perform a spectral survey  covering the entire ALMA band 3 towards one of the most chemically rich HMCs outside the Galactic Center (GC), G31.41+0.31 (hereafter G31), where glycolaldehyde ($\mathrm{CH_2OHCHO}$, GA) has been detected for the first time outside the GC \citep{beltran2009}.\newline\indent
The target of the  G31.41+0.31 Unbiased ALMA sPectral Observational Survey (GUAPOS) is a well-known and studied HMC located at a distance of 3.75 kpc \citep{immer2019} with a luminosity of $4.4\times 10^{4} \, \rm{L}_{\odot}$ (from \citealt{oso2009}) and a mass $M \sim 70\,\rm{M}_{\odot}$ (\citealt{cesa2019}), after rescaling to the new distance estimate  (the previous distance estimate was of 7.9 kpc). The source was first detected  in  $\mathrm{NH_{3}(4,4)}$ \citep{cesa1994a}, whose emission was co-spatial to the distribution of water masers  \citep{hofner93}, and separated by $\sim 5 \arcsec$ from an already known ultra compact (UC) HII region. \citet{cesa1994b} discovered a velocity gradient of $\sim 400 \,\mathrm{km\, s^{-1}\, pc^{-1}}$ across the core in the emission of  methyl cyanide ($\mathrm{CH_{3}CN}$) at 3mm, along the SW-NE direction.
The nature of this velocity gradient was further investigated in more recent studies, and is most likely associated with a rotating toroid with a spin up toward the center \citep{beltran2004,beltran2005,beltran2018,cesa2010,cesa2011,cesa2017}. Although another interpretation in terms of expansion was proposed \citep{Gibb2004,araya2008}, this is inconsistent with recent polarization observations \citep{beltran2019}. These studies revealed also the presence of several outflows, traced by CO and SiO emission \citep{cesa2011,beltran2018}, of infalling material (firstly found by \citealt{gir2009}; see also \citealt{may2014} and \citealt{beltran2018}) and of 2 free-free sources at 0.7 and 1.3 cm separated by $0\farcs19$ embedded in the core \citep{cesa2010}.\newline\indent
G31 has also been studied in polarized emission at mm wavelengths by \cite{gir2009} and, recently, by \cite{beltran2019}. The reconstructed shape of the magnetic field follows an hourglass morphology, with the best model of \cite{beltran2019} suggesting a predominant poloidal field, oriented perpendicular to the SW-NE velocity gradient previously detected in G31, lending strong support to the rotating toroid scenario, as already mentioned. \newline\indent
G31 has also been studied from a chemical point of view, revealing one of the most striking features of this HMC: its chemical richness. As previously mentioned, the first detection of glycolaldehyde outside the GC was obtained towards G31 \citep{beltran2009}, and confirmed by \cite{Calcutt2014} and \citet{rivilla2017a}. Other heavy COMs such as ethyl cyanide $\mathrm{C_{2}H_{5}CN}$, dimethyl ether $\mathrm{CH_{3}OCH_{3}}$, ethanol $\mathrm{C_{2}H_{5}OH}$ and ethylene glycol $\mathrm{ (CH_{2}OH)_{2} }$ (9, 9, 9 and 10 atoms respectively) were also detected here (\citealt{beltran2005,fontani2007,iso2013,rivilla2017a,coletta}). The chemical richness of this HMC has been quantified by Cesaroni et al. (2017), who reported that  the fraction of channels with line emission was 0.74 (the largest in the sample of high-mass young stellar objects studied by these authors) in the 217 - 237 GHz range. G31 is thus one of the best candidates to investigate the degree of chemical complexity that can be reached in typical high-mass star-forming regions. \newline\indent
Among COMs, a special class of molecules is that of isomers, i.e. molecules with the same chemical composition, but with a different molecular structure. These molecules can be used to study which chemical formation pathway is more efficient, leading to a predominance of one isomer with respect to the other(s), helping to constrain the chain of chemical reactions involved in their formation. 
Among isomers, those of $\mathrm{C_{2}H_{4}O_{2}}$ (see Fig. \ref{figisomeri}), namely glycolaldehyde ($\mathrm{CH_{2}OHCHO}$, hereafter GA), methyl formate ($\mathrm{CH_{3}OCHO}$, hereafter MF) and acetic acid ($\mathrm{CH_{3}COOH}$, hereafter AA), are especially interesting because of their relevance for the formation of prebiotic molecules. In fact, GA is the simplest sugar-related molecule and can react with propenal to form ribose, an essential constituent of RNA. The 3 molecules were all firstly detected towards SgrB2 \citep{brown75,mehr1997,hollis2000}.
 MF has been detected in a large number of objects including high-mass star-forming regions \citep{beuther2007,fontani2007,favre14,sakai15,bell2016}, low-mass star-forming regions \citep{cazaux2003, bottinelli2007,Jorge2012,jacob19}, prestellar cores \citep{bac2012,jimenez2016}, cold envelopes around protostars \citep{Ob2010,cerni2012}, and outflow and shock regions \citep{arce2008, Csengeri19shock}. On the other hand, the number of detections of GA and AA is limited compared to methyl formate, despite the presence of dedicated surveys  (e.g. \citealt{remijan2003}).
In literature there are few star-forming regions in which all 3 isomers have been detected: SgrB2(N)-LMH \citep{hollis2001,bell13,xue2019}, the two HMCs NGC 6334I MM1 \& MM2 \citep{el2019} and the low-mass sources IRAS16293-2422B and IRAS16293-2422A \citep{Jorge2012, Jorge2016, manigand2020}. To better constrain and compare the predictions of the chemical models with observations more sources in which all of the 3 isomers have been detected are needed.\newline\indent  GA was firstly detected outside the GC towards G31 \citep{beltran2009}, where \cite{iso2013}, \cite{Calcutt2014} and \citet{rivilla2017a} detected MF.\newline\indent 
The aim of this work is to present the GUAPOS project{\footnote{\ webpage of the project: http://www.arcetri.astro.it/$\sim$guapos/}} and to focus on the he simulaneous analysis of the 3 C$_2$H$_4$O$_2$ isomers: GA, MF and AA.
In Sect. 2 we describe the observations and the data reduction process to obtain the final spectra. In Sect. 3  we analyze the continuum emission and describe the methodology for the spectral analysis of the 3 isomers of $\mathrm{C_{2}H_{4}O_{2}}$. In Sect. 4 we show the result for MF, AA and GA. In Sect. 5 we discuss the abundances and the  column density ratios among the three isomers and compare them with previous values in the literature and with the predictions of different chemical models to better understand how this COMs are formed in the ISM of star-forming regions. Finally in Sect. 6 we summarize the conclusions.
\begin{table}[t]
     \caption[]{Coordinates used for the observations and main parameters of the source G31.}
         \label{SourcePar}
     
     \scalebox{0.85}{
         \begin{tabular}{cccccc}
            \hline
        
                   R.A.(J2000) & Dec.(J2000)  &  $\mathrm{v}_{LSR}$  & $d^{a}$ & $L^{b}$ & $M^{c}$  \\
                    \small{[h m s]} & \small{[$^{\circ}$ \arcmin \hspace{0.6mm} \arcsec]} & \small{[$\rm{km\,s^{-1}}$]} &  \small{[kpc] } &  \small{[$\rm{L}_{\odot}$]} &  \small{[$\rm{M}_{\odot}$]}\\
            
            \hline
       
              18 47 34& -01 12 45 &  96.5     &  $3.75$ & $4.4\times10^{4}$ & $70$\\
         \hline
         \end{tabular}}
     
  \tablefoot{a) \citet{immer2019}; b) from \cite{oso2009}, rescaled to a distance of 3.75 kpc; c) from \cite{cesa2019}, after rescaling to a distance of 3.75 kpc.}
\end{table}

\section{Observations and data reduction}

\subsection{ Observations}
\par The observations were carried out with ALMA during Cycle 5, between the $18^{th}$ of January  and the $7^{th}$ of September 2018 (project 2017.1.00501.S, P.I.: M. T. Beltr\'an), using 43 antennas. The coordinates of phase center, together with other properties of G31, are given in Table 1. More details about the spectral setup, the baselines, and the flux and phase calibrators are given in Table \ref{SpecSet}. The survey covered the complete spectral range of ALMA band 3, between 84.05 GHz and 115.91 GHz ( $\sim 32\,\mathrm{GHz}$ bandwidth), with a spectral resolution of $\sim0.488\,\mathrm{MHz}$ ($\sim1.3\,-\, 1.7\,\mathrm{km\,s^{-1}}$). We used 9 correlator configurations (\textit{spec}) and for each of them, 4 contiguous basebands of $\sim937\,\mathrm{MHz}$ were observed simultaneously. In order to create a single spectrum starting from the 36 spectra of the respective basebands, an overlap in frequency ranging from $\sim7.3$ to   $\sim 29\,\mathrm{MHz}$ was chosen for each pair of adjacent basebands. The original angular resolution requested in the proposal (1\arcsec) was not achieved for two out of nine specs  (\textit{spec} 3 and 4) during the observations between January and March 2018. Although we were granted additional observing time for these two specs, the new observations did not reach the angular resolution requested either, but only $\sim1\farcs2$. We decided to degrade the angular resolution of all the \textit{specs} to the lowest one ($\sim1\farcs2$) to have the same spatial resolution for all of them. For all the \textit{specs} the source used as flux and bandpass calibrator is J1751+0939, while J1851+0035 is the phase calibrator. The uncertainties in the flux calibration are  $\sim5\%$ (from Quality Assesment 2 reports), in good agreement with flux uncertainties of other ALMA band 3 calibrators reported in \citet{bonatoALMAcalibrator}.\newline\indent 
The data were calibrated and imaged with  CASA\footnote{https://casa.nrao.edu} (the \textit{Common Astronomy Software Applications} package, \citealt{CASAmcmullin}). The maps were created using a robust parameter of \citet{briggs} set equal to 0 and a restoring sinthesized beam of $1\farcs2\times1\farcs2$. $rms$ of the maps varies between 0.5 mJy/beam and 1.9 mJy/beam.

\subsection{Continuum determination}
\label{excontinuum}
\begin{figure}
   \centering
   \includegraphics[width=9cm]{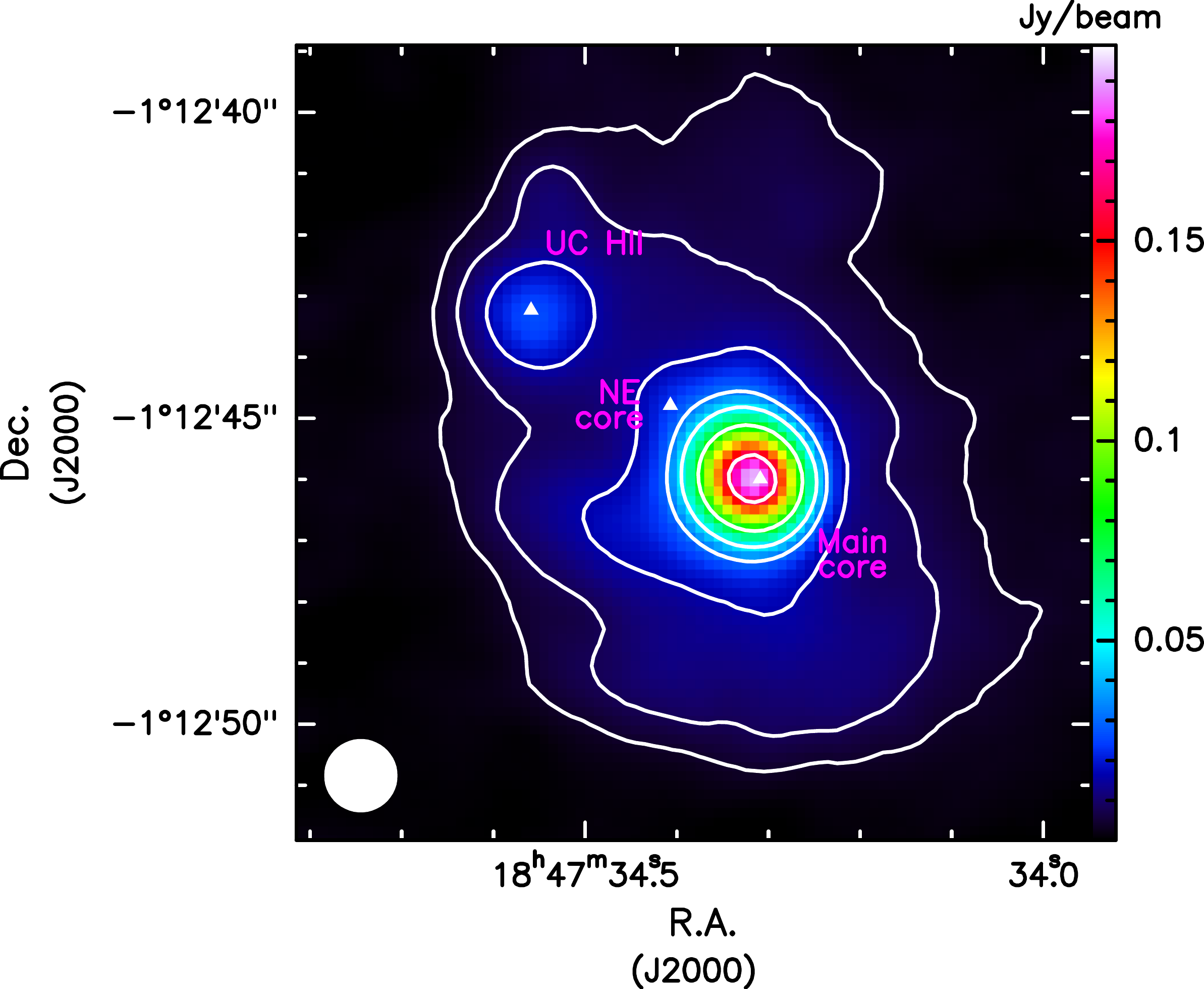}
      \caption{Continuum map of the HMC G31.41+0.31 and the nearby UC HII region. Contour levels are at 5, 10, 20, 40, 60, 100 and 200 times the value of the $rms = 0.8\, \mathrm{mJy/beam}\,$. The 3 white triangles indicate the coordinates of the Main Core and NE core by \citet{beltran2018}, and of the UC HII region by \citet{cesa98}. The beam is shown in white in the lower left corner.
              }
         \label{FigContinuum}
\end{figure}

Due to the large number of lines in the spectra, it was not possible to find a sufficient number of channels without line emission that could be used to obtain a map of the continuum emission. To overcome this problem we used STATCONT \citep{alvaro2018}, a python-based tool designed to determine the continuum emission level in spectral data, in particular for sources with a very rich spectrum. This tool determines the continuum level by using a statistical approach on the intensity distribution of the spectrum, and produces a continuum map\newline\indent We obtained the continuum map (Fig. \ref{FigContinuum}) from the line+continuum map of the spec5 BB1 (see Table 2, $\nu : 98499.749 - 99436.626$ MHz), in order to have the continuum extracted from a frequency close to the center of the frequency range covered by the GUAPOS observations ($\sim 84-116\,\mathrm{GHz}$).
\begin{table*}
\centering
\caption{Spectral setup of observations: number of spec, number of baseband, central observed frequency of each baseband $\nu_{0}$, spectral resolution in observed frequency $\Delta\nu$, $rms$ of the maps for channel and baselines range in meters.}
\label{SpecSet}
     
         \begin{tabular}{llccccccc}
            \hline
        
                  & baseband& $\nu_{0}$ &$\Delta\nu$ & $ rms $ & baselines   \\
                  & & \small{[MHz]} &  \small{[MHz]} &  \small{[mJy/beam]}& [m]  \\
            
            \hline
             spec 1  & BB 1 & 84520.801        & 0.48824 & 1.1 & 15 - 1397   \\
                       & BB 2 &  85378.837       & 0.48824 & 0.8 & \\
                       & BB 3 &  86284.458       & 0.48824 & 0.8 &\\
                       & BB 4 &  87190.496       & 0.48824 & 0.8 &\\
            \hline
            spec 2 & BB 1 &  88096.667        &     0.48823 & 0.7 & 15 - 1397  \\
                       & BB 2 &  89002.644          & 0.48823 & 0.6  & \\
                       & BB 3 &  89908.312          & 0.48823 & 0.6  &\\
                       & BB 4 &  90814.305          & 0.48823 & 0.5  &\\
             \hline
            spec 3 & BB 1 &   91720.492        & 0.48821 & 1.9 & 15 - 783 &  \\
                       & BB 2 &    92626.501   & 0.48821 & 1.8  & \\
                       & BB 3 &    93532.124   & 0.48821 & 1.9  & \\
                        & BB 4 &   94438.102  &  0.48821 & 1.9  & \\
                        
             \hline
            spec 4      & BB 1 & 95344.296       & 0.48829 & 1.2 & 15 - 783 & \\
                        & BB 2 &    96250.320    & 0.48829 & 1.6  &\\
                        & BB 3 &    97155.974    & 0.48829 & 1.2  &\\
                        & BB 4 &   98061.998     & 0.48829 & 1.0  &\\
             \hline
            spec 5      & BB 1 &  98968.187        & 0.48821 & 1.7 & 15 - 783 \\
                       & BB 2 &   99874.167        & 0.48821 & 1.7  &\\
                       & BB 3 &   100780.279       & 0.48821 & 1.5  &\\
                        & BB 4 &   101686.288      & 0.48821 & 1.4  &\\
             \hline
            spec 6 & BB 1 &   102592.110           & 0.48821 & 1.5 & 15 - 783 \\
                       & BB 2 &    103498.090      & 0.48821 & 1.4  & \\
                       & BB 3 &    104404.324      & 0.48821 & 1.5  & \\
                        & BB 4 &   105310.303      & 0.48821 & 1.5  & \\
            \hline
            spec 7 & BB 1 &   106216.143          & 0.48821 & 1.4 & 15 - 783  & \\
                       & BB 2 &    107122.124     & 0.48821 & 1.6  & \\
                       & BB 3 &   108028.359      & 0.48821 & 1.5  & \\
                        & BB 4 &   108934.339     & 0.48821 & 1.4  & \\
           \hline
            spec 8 & BB 1 &   109840.176         & 0.48821 & 1.5 & 15 - 783  \\
                       & BB 2 &   110746.157     & 0.48821 & 1.1  & \\
                       & BB 3 &   111652.391     & 0.48821 & 1.5  & \\
                        & BB 4 &   112558.371    & 0.48821 & 1.1  & \\
            \hline
            spec 9 & BB 1 &   113363.505         & 0.48821 & 0.9 & 15 - 783 \\
                       & BB 2 &    114170.246    & 0.48821 & 1.0 & \\
                       & BB 3 &    114943.241    & 0.48821 & 1.3 & \\
                        & BB 4 &   115443.023    & 0.48821 & 1.3 & \\
            
            \hline
         \end{tabular}
\end{table*}
\subsection{Combination of the spectral windows}
\label{onefinalspectrum}
\begin{figure*}
\centering
\includegraphics[trim = 12 0 0 0 , clip,height=10cm]{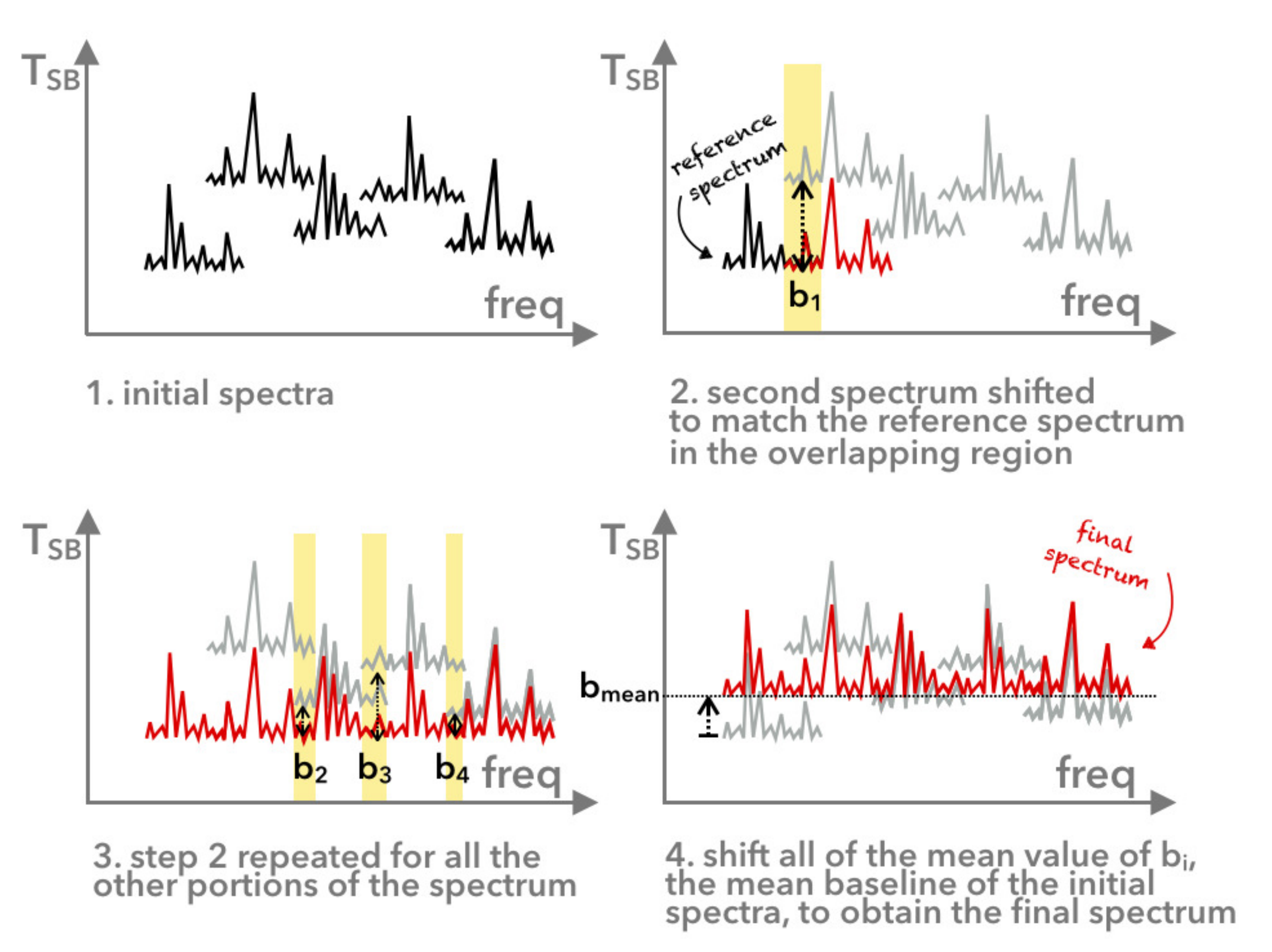}
\caption{Graphic scheme of the steps followed to obtain the final spectrum}
\label{sketch}
\end{figure*}

From each baseband we extracted the mean spectrum inside an area equal to the beam, centered at the position of the continuum peak of the HMC (see coordinates in Sect. \ref{ancontinuum}). This area is smaller than the size of the continuum core (see Sect.\ref{ancontinuum} and Fig. \ref{FigContinuum}). The conversion from flux  $I_{\rm{\nu}}$ [mJy/beam] to synthesized beam brightness temperature $T_{\rm{SB}}$ [K] is given in the Rayleigh-Jeans approximation by the equation:
\begin{equation}
\label{eqTK}
T_{\rm{SB}}=1.22\times10^{3} \frac{I_{\rm{\nu}}}{\nu^{2}\theta_{\rm{a}}\theta_{\rm{b}}}\:,
\end{equation}
where $\nu$ is the frequency in units GHz and $\theta_{\rm{a}}$ and $\theta_{\rm{b}}$ are the major and minor axes of the synthesized beam in arcsec. The discrepancy between the Rayleigh-Jeans approximation and the true value is of the order of $\sim15\%$ for the channels with only continuum emission, and only of the $\sim3\%$ for the channels with bright-line emission. Adjacent spectra have been merged to produce a single final spectrum, thanks to the overlapping regions between each pair of adjacent basebands. Since jumps up to $3\,\mathrm{K}$ are present between partially overlapping adjacent spectra, we adopt the following procedure to create the merged spectrum. 
We align the spectra from the different basebands using the spectrum at longer $\lambda$ (spec1 BB1, see Table \ref{SpecSet}) as reference. In practice, we forced the channels of spec1 BB2 in the overlapping region to match those of spec1 BB1, by subtracting from spec1 BB2 the mean value of the difference between spec1 BB2 and spec1 BB1. We define this mean value as $b_{1}$.
Then the procedure was repeated for all the other pairs (spec1 BB2 and spec1 BB3, spec1BB3 and spec1 BB4,spec1 BB4 and spec2 BB1 etc.) thus obtaining a single spectrum without artificial jumps and the values $b_{i}$ of the baselines of the spectra of all the basebands with respect to the reference one. As a last step we shifted the final spectrum without jumps by a value equal to the mean of $b_{i}$, to obtain a final spectrum without jumps with baseline equal to the mean value of the baselines of the original spectra. Figure \ref{sketch} summarizes and graphically illustrates the main steps described above.
\begin{sidewaysfigure*}

\centering
\includegraphics[width=24cm]{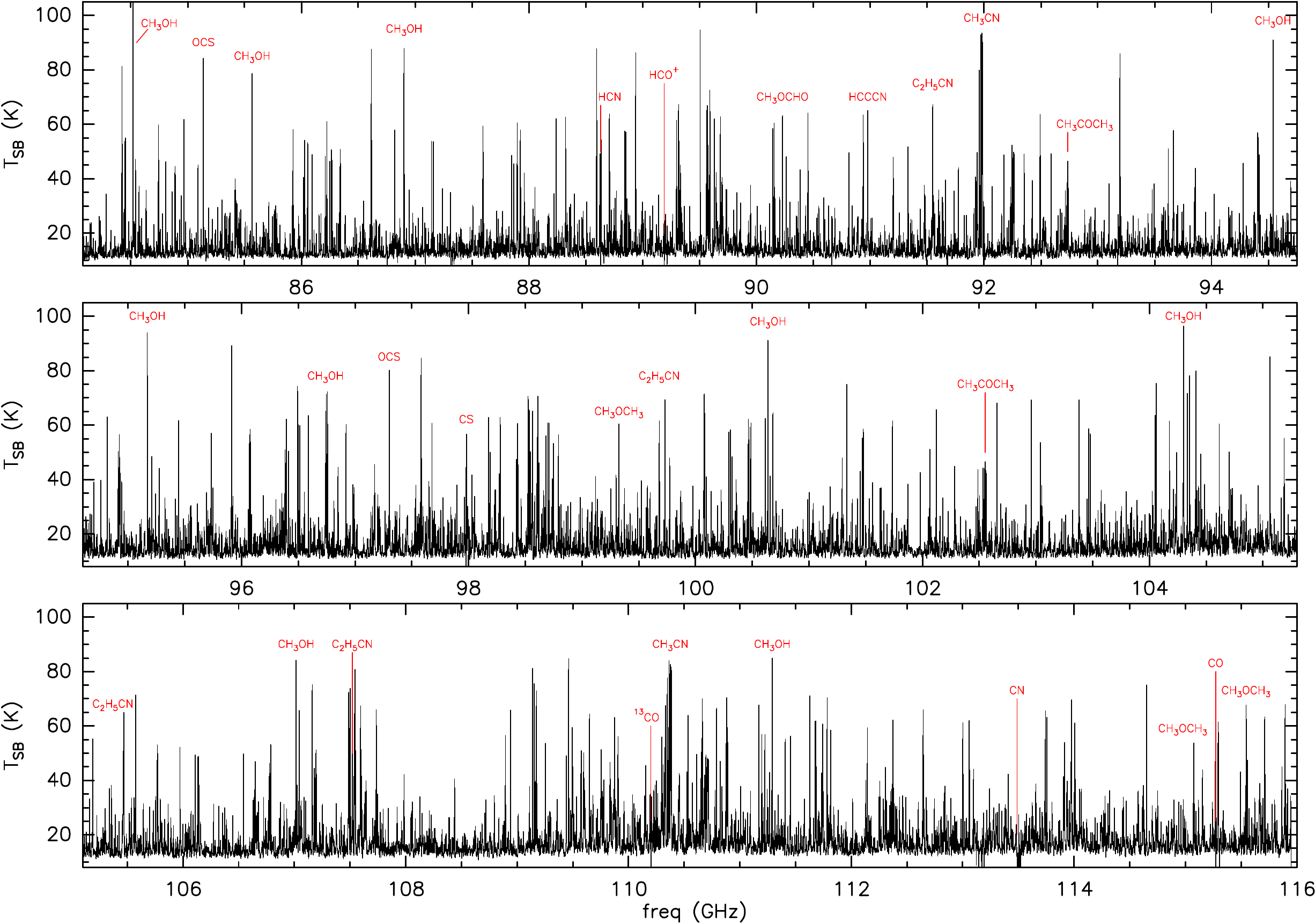}
\caption{Full final spectrum from 84 to 116 GHz. In red is reported the name of the molecular species associated to some of the most common or bright lines.}
\label{LIDtotalspectrum}

\end{sidewaysfigure*}
We converted the spectra to rest frequency, assuming a velocity of the source of 96.5 $\rm{km\,s^{-1}}$, and rebinned the spectra to a constant value of the bin width equal to 0.48840 MHz using HIPE\footnote{https://www.cosmos.esa.int/web/herschel/hipe-download} (\textit{Herschel Interactive Processing Environment}, \citealt{ottHIPE}), since the small differences in the bandwidth of different basebands, lead to differences in the channel widths (see Table \ref{SpecSet}) which span between 0.48837 and 0.48845 MHz (after the conversion to rest frequency). This was done also considering the possibility of analyzing the spectrum with different software, since some of the commonly used software to analyze spectra (e.g. CLASS or MADCUBA\footnote{Madrid Data Cube Analysis (MADCUBA) is a software developed in the Center of Astrobiology (Madrid) to visualize and analyze data cubes and single spectra \citep{martin2019}: http://cab.inta-csic.
es/madcuba/MADCUBA\_IMAGEJ/ImageJMadcuba.html}) need a unique  frequency width to import the data.\newline\indent We remark that the procedure we used in principle did not affect in any way the slope of the baseline in each baseband. Considered the artificial nature of the jumps in adjacent spectra, we expect the slope of the baseline not to be considerably changed from its true value. \newline\indent
The $rms$ on the level of the baseline, calculated as the root mean square of the differences between the original baseline in each baseband and the baseline in the final spectrum, is $1.2\,\mathrm{K}$ ($\sim12\%$ of the continuum level, see Sect. \ref{ancontinuum}). This 12\% error on the continuum level is included as additional error on the parameters derived from the fit to the spectrum. The $rms$ of the spectra has been derived from the $rms$ of the maps (see Table \ref{SpecSet}) and vary from 7 mK to 27 mK, for the different basebands, but we consider a conservative value of 27 mK for the entire final spectrum. The fluctuations in the $rms$ of the maps are the result of different atmospheric conditions in different days of observations and of the presence of bright lines in some of the basebands (therefore, a dynamic range effect). Figure \ref{LIDtotalspectrum} shows the total final spectrum.

\section{ Analysis}\label{analysis}
\subsection{Continuum}
\label{ancontinuum}
The map of the continuum (see Fig. \ref{FigContinuum}) shows the presence of two compact sources. The brightest source is our target, the HMC G31.41+0.31, that peaks at  R.A. $18^{\rm{h}}\,47^{\rm{m}}\,34.321^{\rm{s}}$ Dec. $-01^{\circ} 12\arcmin 45.977\arcsec$ (J2000). Located at a distance of $\sim 4\farcs5$ in the NE direction, there is a fainter continuum source that peaks at R.A. $18^{\rm{h}}\, 47^{\rm{m}}\, 34.56^{\rm{s}}$ Dec. -$01^{\circ}12\arcmin 43.35\arcsec$ (J2000) consistent with the position of the nearby UC HII region observed by \cite{cesa98} in the continuum at 1.3 cm with the VLA. The NE core, identified by \citet{beltran2018} with high resolution observations, is not resolved in our data at $1\farcs2$, resolution.   \newline\indent We have fitted a 2D gaussian to the continuum emission of the main core using the task \textit{imfit} of CASA, and obtained the source size of the HMC, reported in Table \ref{2Dgaussianresults}. The source is barely resolved, with the deconvolved size of the continuum being $1\farcs41\times1\farcs22$.
\newline\indent The flux of the continuum inside the area from which we extracted the spectrum is of 0.1 Jy at the mean frequency of 99.0 GHz (see Sect. \ref{excontinuum}). Assuming a dust temperature of $150\,\mathrm{K}$, $\beta = 2.0$, a dust opacity coefficient $k_0 = 0.8\,\mathrm{cm^2/g}$ at 220 GHz \citep{ossenkopf1994}, a mean molecular weight of 2.33 and a gas-to-dust ratio of 100 the column density of $\rm{H}_2$ is $N(\mathrm{H}_{2})=1.0\times10^{25}\,\mathrm{cm^{-2}}$. This value is consistent within a factor $\sim4$ with the previous estimate by \citet{rivilla2017a}.
\begin{table*}	
\caption[]{Parameter of the best fit for a 2D gaussian model to the map of the continuum and the maps of MF, AA and GA in Figs. \ref{figuremappe} and \ref{figuremappemean}, and difference between the mean diameter of each map with the mean diameter of the mean map.}
\label{2Dgaussianresults}
\centering
\begin{centering}
\begin{tabular}{cccccccc}
\hline
  & $\nu_{0}$ & $E_{\rm{u}}$ & R.A. (J2000) & Dec. (J2000) & $\theta_{conv}\times\theta_{conv}$  &  PA & $\bar{\theta}-\bar{\theta}_{\rm{mean}}$  \\	 
  &  \small{[GHz]} &  \small{[K]} & \small{[18h 47m s]} & \small{[-01$^{\circ}$ 12\arcmin \hspace{0.6mm} \arcsec]} & \small{[$\arcsec\times\arcsec$]} &  \small{[$^{\circ}$]}  & [$\arcsec$]\\
\hline
continuum & & & 34.321 & 45.976 & $1.85\times1.71$ &  29  & \\
\hline
$\mathrm{CH_{3}OCHO}$ & 88.2207 & 205 & 34.316 & 46.0414 & $1.61\times1.56$ & 44 & -0.14 \\
& 88.7707 & 208 &  34.316 & 46.030 & $1.63\times1.60$ & 27 & -0.11 \\
& 98.8153 & 215 & 34.320 & 45.949 & $1.65\times1.59$ & 75  & -0.10\\
& 103.4787& 25 &  34.318 & 46.069 & $1.84\times1.81$ & 60     &+0.10 \\
& 110.7765& 216 & 34.318 & 46.053 & $1.60\times1.58$  & 63  &-0.13 \\
& 111.4084 & 37 & 34.318 & 46.087 & $1.81\times1.79$ & 68  &+0.08 \\
& 111.4533 & 37 & 34.317 & 46.094 & $1.83\times1.82$ & 83  &+0.11 \\
& 111.6822 & 28 & 34.317 & 46.101 & $1.82\times1.81$ & 65 &+0.10 \\
 & mean & - & 34.318 & 46.054 & $1.73\times1.71$   & 55 & \\
\hline
$\mathrm{CH_{3}COOH}$ &85.319/.322&  181/207 & 34.320 & 46.098 & $1.53\times1.39$ & 6  & +0.06\\
& 85.6331 & 229 & 34.320 & 46.078 &  $1.48\times1.31$ & 8 & -0.00 \\
& 90.2462 & 20 &34.313 & 46.107 &  $1.46\times1.43$ & 19 &+0.05 \\
& 94.4995 & 58 &34.311 & 46.018 & $1.37\times1.35$ & 109 & -0.03\\
& 100.8554 & 26 &34.314 & 46.017 & $1.43\times1.36$ & 58  &-0.00 \\
&100.8975 &  25& 34.314 & 46.028 & $1.45\times1.37$  & 50 & +0.01\\
& 104.0780/.0786 & 80/266 &34.317 & 46.049 & $1.33\times1.31$ & 92  & -0.07\\
& 114.6380 & 75 &34.313 & 46.068 & $1.39\times1.35$ & 179  &-0.02 \\
& mean &-& 34.315 & 46.057 & $1.42\times1.37$    & 19 & \\
\hline
$\mathrm{CH_{2}OHCHO}$ & 88.5304 & 30 & 34.304 &46.163  & $1.43\times1.37$    & 19 & -0.05 \\
& 88.6912 & 49 & 34.306 &46.124  & $1.52\times1.46$    & 167 & +0.04 \\
& 93.0485 & 192 &34.310 &46.054  & $1.44\times1.36$    & 45  &-0.05 \\
& 93.0527 & 23 & 34.309 &46.066  & $1.50\times1.45$    & 42 & +0.03\\
& 95.7562 & 178 &  34.305 & 46.103 & $1.44\times1.37$  & 73  &-0.04 \\
&102.5729 & 64 &   34.314 & 46.049 & $1.51\times1.44$  & 79  &+0.03 \\
&102.6144 & 104 & 34.310 &46.098  & $1.41\times1.35$    & 77  &-0.07 \\
&103.6680 & 32 & 34.310 & 46.100 & $1.56\times1.51$  & 71   &+0.09 \\
 & mean &-& 34.309 &46.094  & $1.47\times1.42$    & 56 & \\
\hline

\end{tabular}
\end{centering}
\tablefoot{$\bar{\theta}-\bar{\theta}_{\rm{mean}}$ is calculated as $\sqrt{\theta_1\,\theta_2}-\sqrt{\theta_1^{\rm{mean}}\,\theta_2^{\rm{mean}}}$. }
\end{table*}
\newline\indent We have fitted by eye the continuum level of the spectrum assuming a power law,
\begin{equation}
\label{eqcontinuum}
 T_{\rm{SB}}(\nu)=T_{\rm{SB}}(\nu_{0})(\nu/\nu_{0})^\beta \,,
\end{equation}
where $\nu_{0}$ is the reference frequency, $T_{\rm{SB}}(\nu_{0})$ is the level of the continuum at the arbitrary reference frequency and $\beta$ is the dust emissivity spectral index. The parameters of the continuum are given in Table \ref{baselinefit}. The dust emissivity spectral index $\beta$ is related to $\alpha$, the flux spectral index ($S_{\rm{\nu}}\propto \nu^{\alpha}$), by
\begin{equation}
  \alpha = 2+\beta\,.
\end{equation}
The value of $\beta=0.4$ ($\alpha=2.4$) is smaller than the expected value for dust thermal emission from small grains ($\beta \in [1;2]$). However, it is consistent with the thermal emission by dust in the case of the presence of larger grains together with possibly differents grain chemical composition (see \citealt{beckwith1991,galametz2019,ysard2019} and references therein). Based on the fluxes at 7 mm and 1.3 cm of the two unresolved radio sources found within the HMC by \citet{cesa2010}, the contribution of free-free emission to the continuum flux should vary in a range of a few percent to a maximum value of 8\%, assuming a spectral index for the free-free emission of 0.6 and 1.3 (maximum spectral index found by \citealt{cesa2010}), respectively.
The vast majority of the flux of the continuum is thus due to thermal emission from dust grains, with possibly larger grains already present. 
\begin{table}	
\caption[]{Parameters of the continuum baseline: reference frequency $\nu_{0}$, level of the continuum at the reference frequency $T_{SB}(\nu_{0})$ and spectral index $\beta$. }
\label{baselinefit}
\begin{center}
\begin{tabular}{ccccc}
\hline
 $\nu_{0}$ & & $T_{\rm{SB}}(\nu_{0})$ & & $\beta$ \\	 
   \small{[GHz]} & & \small{[K]}  & &  \\
\hline
 84.079 & & 10.4 & & 0.4\\
\hline
\end{tabular}
\end{center}
\end{table}

\subsection{Spectral analysis}
The line identification and the fit to the $\mathrm{C_{2}H_{4}O_{2}}$ isomers have been obtained with the software XCLASS\footnote{https://xclass.astro.uni-koeln.de} (eXtended CASA Line Analysis Software Suite - \citealt{xclass}), which makes use of the Cologne Database for Molecular Spectroscopy\footnote{https://cdms.astro.uni-koeln.de} (CDMS, \citealt{cdms2001,cdms2005}) and Jet Propulsion Laboratory\footnote{https://spec.jpl.nasa.gov} (JPL, \citealt{jpl1998}) catalogs, via the Virtual Atomic and Molecular Data Centre (VAMDC, \citealt{endres2016}). The spectroscopic data of acetic acid, from the work of \citet{ilyushin2013}, were introduced in CDMS in May 2019 to perform the analysis presented in this work. A complete documentation of the spectroscopic works on which the entries of the catalogues for MF, AA and GA are based, is given in Appendix A.  
For each molecular species the software computes a synthetic spectrum assuming LTE (the software also allows non-LTE analysis) using as possible free parameters the excitation temperature, $ T_{\rm{ex}}$, the total column density, $N_{\rm{tot}}$, of the molecule, the line FWHM, $\Delta\mathrm{v}$, the velocity of the source, $V_{\rm{LSR}}$, and the angular diameter of the source $\theta_{\rm{s}}$. The LTE assumption is justified by the high density of hot cores; in fact, from the column density of H$_2$ derived in Sect. \ref{ancontinuum}, a rough estimate of the volume density of the core is $n\sim10^8\,\mathrm{cm^{-3}}$. The line analysis can be done without subtracting the continuum, which can be simulated simultaneously with the line emission. The fitting procedure varies the free parameters to minimize the $\chi^{2}$ between the observed spectrum and the synthetic one. It is possible to select the  frequency ranges for the fit, which allows to limit the fit to a selected number of transitions. 
In order to estimate the degree of contamination in the observed transitions of AA, GA and MF, due to emission by other species, we performed a preliminary line identification across the whole spectrum. Starting from the molecules responsible for the brightest lines, we carried out an investigation to identify as many species that can potentially contribute to the observed spectrum as possible.
For the species identified in this way, we adjusted the parameters of the synthetic spectra until a reasonable fit (by visual inspection) was obtained. This preliminary identification was used to identify the transitions of the isomers of $\mathrm{C_2H_4O_2}$, and among them the most unblended to use in the XCLASS fitting procedure.\newline\indent   
We followed three rules to identify a transition of one of the 3 $\mathrm{C_2H_4O_2}$ isomers: i) the intensity of the line should be higher than $3\times rms$; ii) blending of a line of another species, with clearly separated peak, should start at a distance larger than FWHM/2 from the peak; iii) in case of blending with separation <FWHM/2 the intensity of the line of the other species should be $<15\,\%$ of the peak.
In the case of AA, whose lines are more affected by blending, we included also lines with one of the two wings blended up to 60\% of the peak (with respect to the baseline). The MF spectrum presents 196 identified lines, with brightness temperatures up to $\sim70\,\,\mathrm{K}$. These include both ground state and vibrationally excited $\varv_{18}=1$ transitions, with upper state energies ($E_{\rm{u}}$) between $\sim10\,\mathrm{K}$ and $\sim400\,\mathrm{K}$. For GA we detected 20 lines, with brightness temperatures up to $\sim28\,\,\mathrm{K}$. 
All the transitions are in the ground vibrational state and have energies $E_{\rm{u}}$ between $\sim25\,\mathrm{K}$ and $\sim190\,\mathrm{K}$. Lastly, we identified 26 lines of AA, with brightness temperatures up to $\sim23\,\,\mathrm{K}$. These are ground state and vibrationally excited $\varv_{18}=1$ transitions,  with energies $E_{\rm{u}}$ between $\sim25\,\mathrm{K}$ and $\sim400\,\mathrm{K}$. 
To compare the best fit physical parameters of the 3 species in a consistent way, we performed the fit by selecting the most unblended transitions with energies $E_{\rm{u}}<250\,\mathrm{K}$. This energy constraint was due to GA not presenting identified unblended transitions of higher upper state energy.
\citet{beltran2018} found a decrease of  $ T_{\rm{ex}}$ of methyl formate with distance from the core center  from $ T_{\rm{ex}}\sim 497\,\mathrm{K}$ at the inner radius of $0\farcs22$ to $ T_{\rm{ex}}\sim 110\,\mathrm{K}$ at a radius of $1\farcs32$. From these data, the average value of $ T_{\rm{ex}}$ of MF inside the main core is $\sim250$\,K. Despite the presence of this  gradient in $ T_{\rm{ex}}$, we decided to perform a single temperature component analysis, to reduce the number of free parameters as much as possible. For each molecule we selected the lines for the fitting procedure choosing the most unblended transitions and balancing the presence of low $E_{\rm{u}}$ and high $E_{\rm{u}}$ transition, to avoid possible biases. For MF we selected 25 lines, for GA 12 and for AA 14.
We calculated the optical depth $\tau_{0}$ at the center of all the identified lines (see Appendix B) using the values listed in Table \ref{fitresults} and we find that all transitions are optically thin ($\tau_{0}\ll1$). The identified transitions of the 3 molecules are listed Appendix C in Tables C.1, C.2 and C.3, together with $\tau_0$. The transitions selected for the XCLASS fit are listed in boldface. As only exception to the $E_{\rm{u}}<250\,\mathrm{K}$ constrain, we have included in the fit the transition $24(14,11) A2 \,\rightarrow \,24(13,12) A1$ of AA, which has  $E_{\rm{u}}=266\,\mathrm{K}$, because it is blended with other transitions of the same species with $E_{\rm{u}}=80\,\mathrm{K}$. During the fit, the continuum, from Eq. (\ref{eqcontinuum}), the FWHM and $V_{\mathrm{LSR}}$ of the selected species were kept fixed to the values found by a preliminary fit (in which single transitions were selected and all the parameters were free to vary), while $ T_{\rm{ex}}$ and $N_{\rm{tot}}$ were free to vary. 

\section{Results}
\begin{figure*}
\centering
\includegraphics[width=15cm]{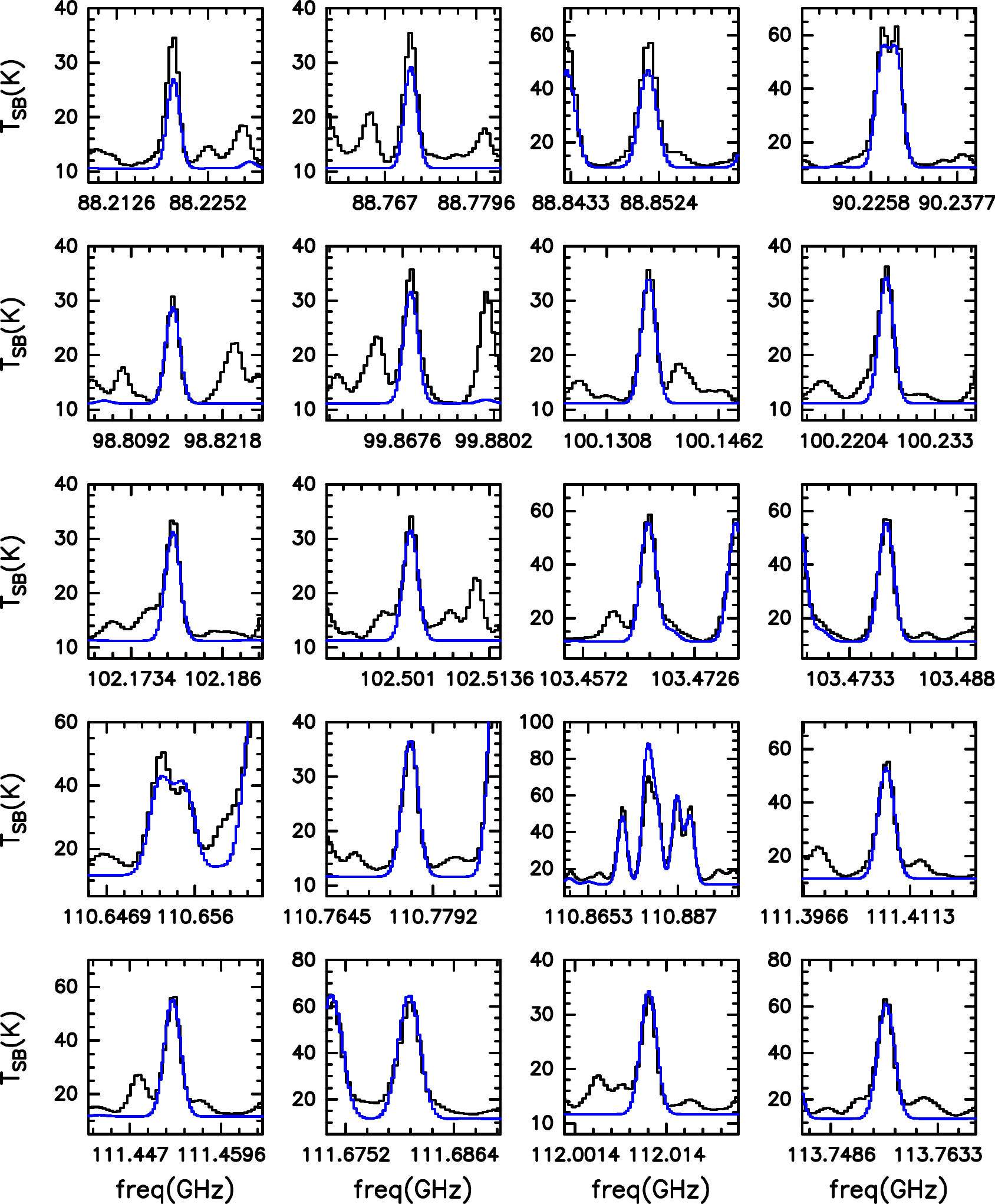}
\caption{Transitions used to fit methyl formate, with the LTE synthetic spectra simulated using the best fit parameters given in Tab. \ref{fitresults} in blue.}
\label{figsingleMF}
\end{figure*}
\begin{figure*}
\centering
\includegraphics[width=14cm]{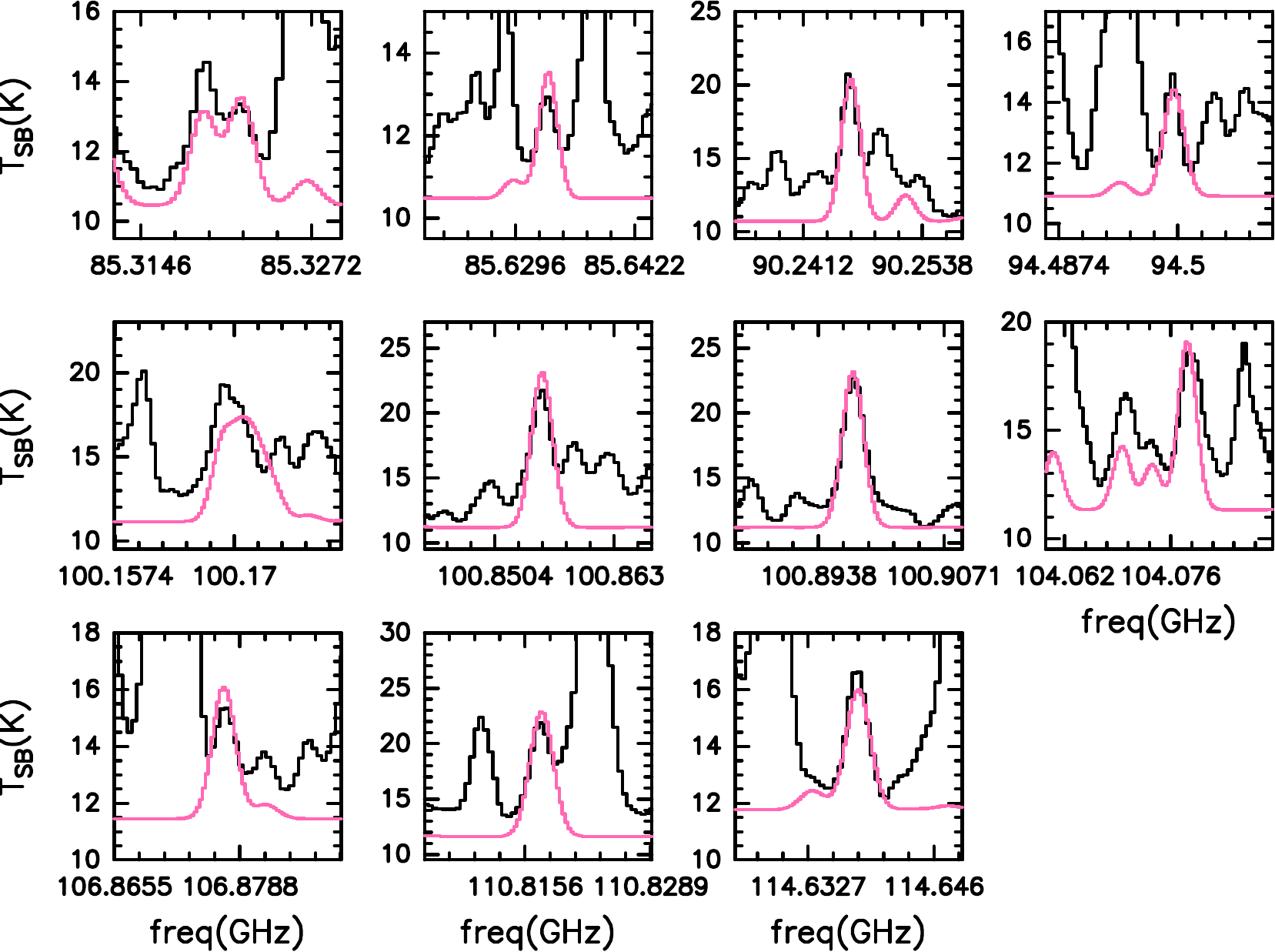}
\label{figsingleAA}
\caption{Transitions used to fit acetic acid, with the LTE synthetic spectrum simulated using the best fit parameters given in Table \ref{fitresults} in pink.}

\vspace*{\floatsep}

\includegraphics[width=14cm]{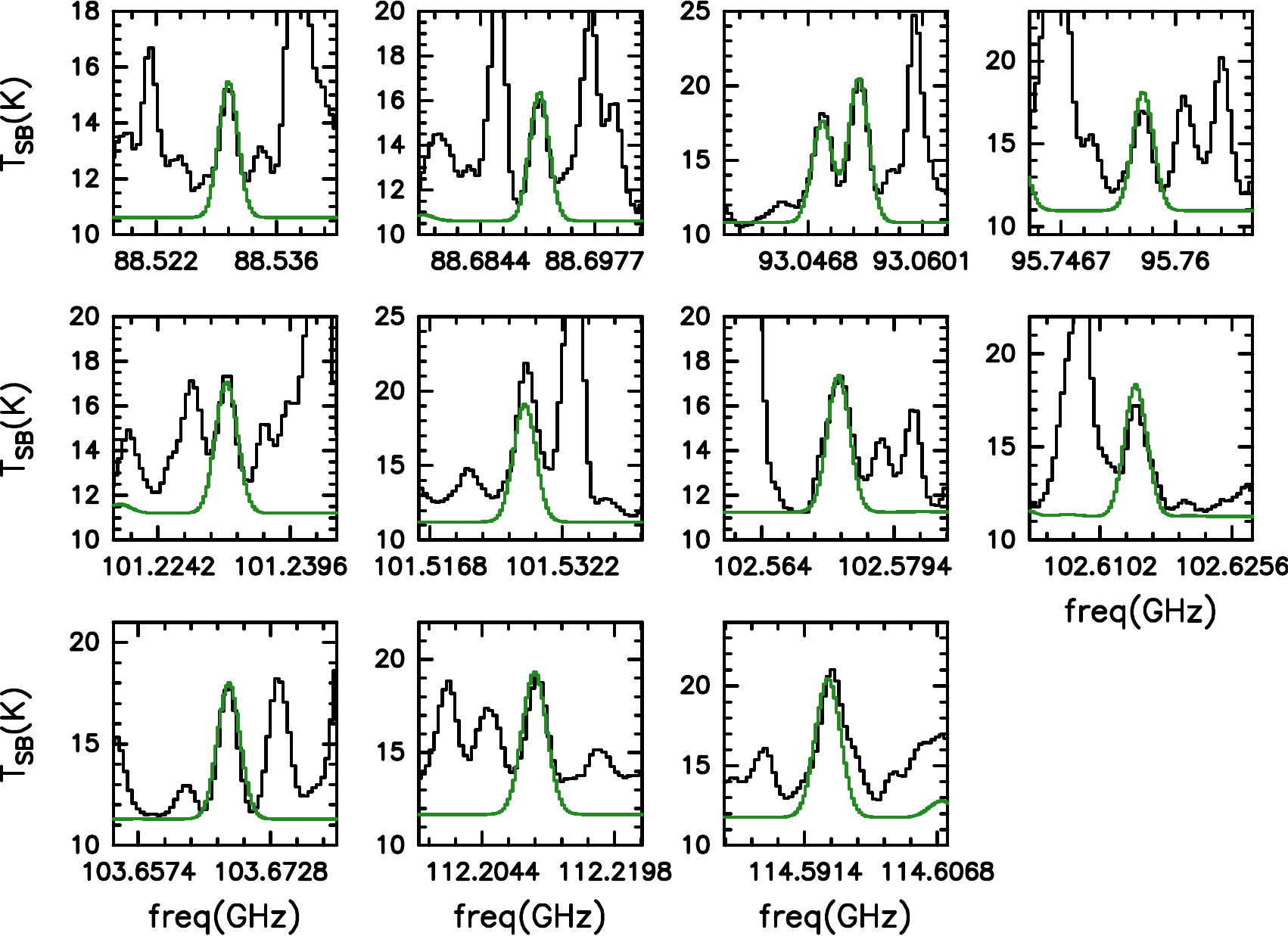}
\label{figsingleGL}
\caption{Transitions used to fit glycolaldehyde, with the LTE synthetic spectra simulated using the best fit parameters given in Table \ref{fitresults} in green.}

\end{figure*}
\begin{sidewaysfigure*}
\centering
\includegraphics[width=23cm  ]{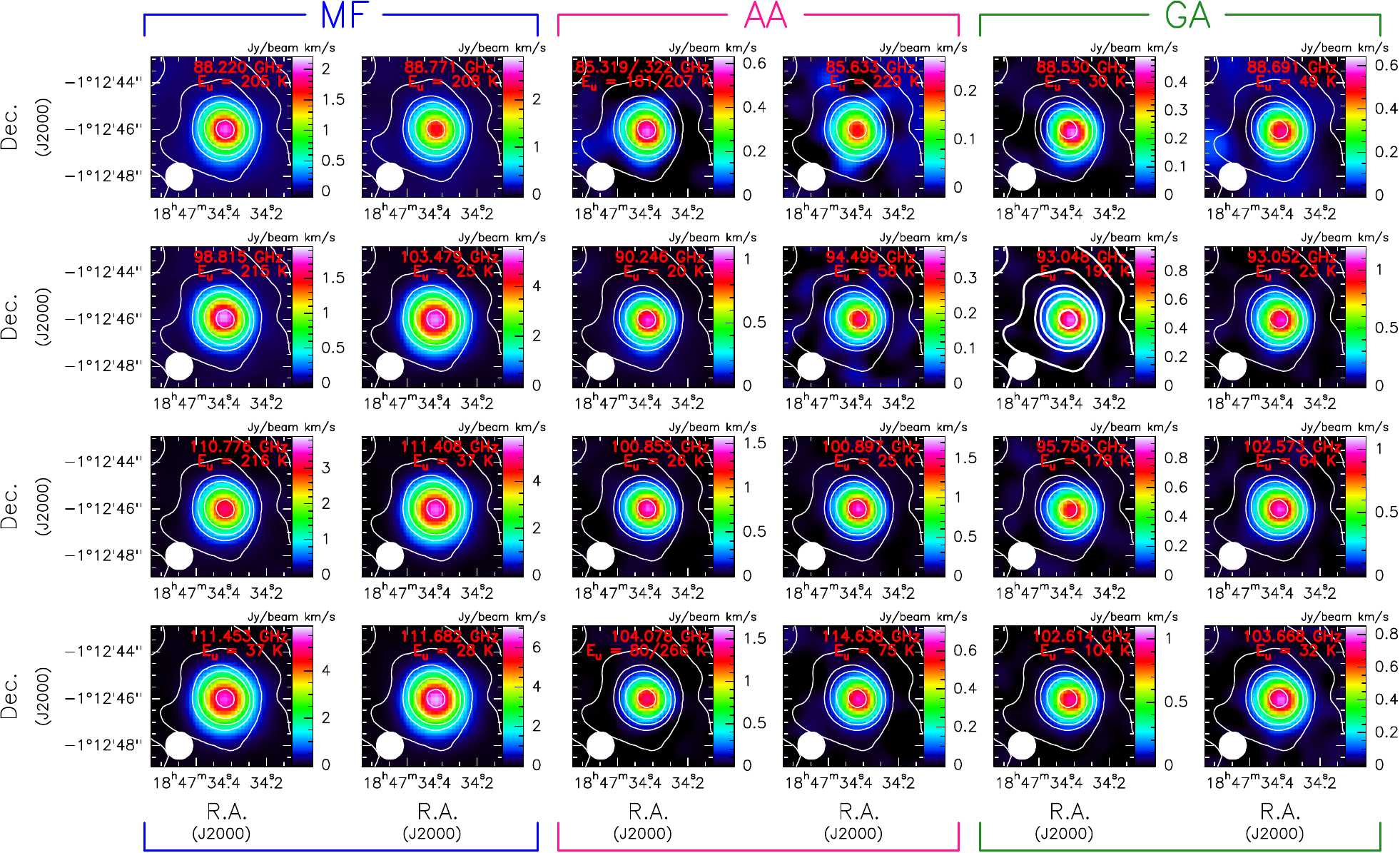}\\
\caption{Maps of integrated emission of 8 transitions with different $E_{u}$ (from $\sim25\,\mathrm{K}$ to $\sim200\,\mathrm{K}$) for MF (\textit{left panels}), AA (\textit{middle panels}) and GA (\textit{right panels}) in color scale. Continuum emission is shown in withe contours, as in Fig. \ref{FigContinuum}, and the beam is indicated in the left-bottom corner of the maps.
}
\label{figuremappe}
\end{sidewaysfigure*}
\begin{figure*}
\centering
\includegraphics[width=\hsize, angle=0 ]{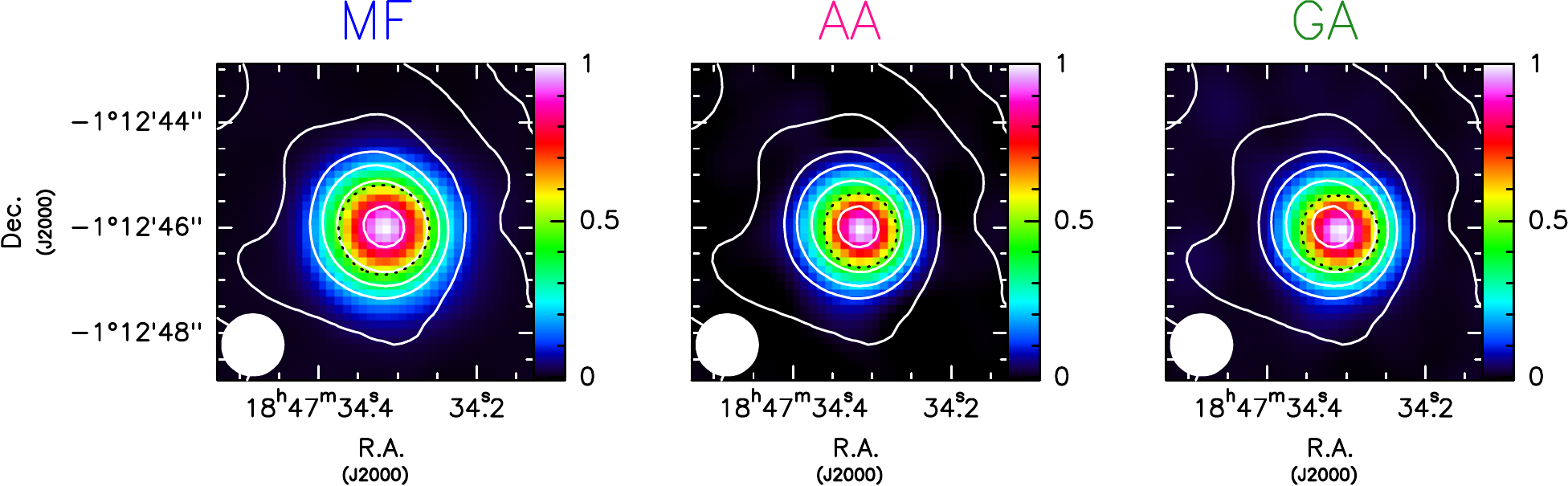}\\
\caption{Mean map for each molecule, obtained from the average of the 8 maps at different $E_{u}$ in Fig. \ref{figuremappe}, rescaled to a peak of intensity 1 before averaging. The black dashed line is the contour where the emission reaches the half the peak intensity.}
\label{figuremappemean}
\end{figure*}
\begin{figure*}
\centering
\includegraphics[width=17cm]{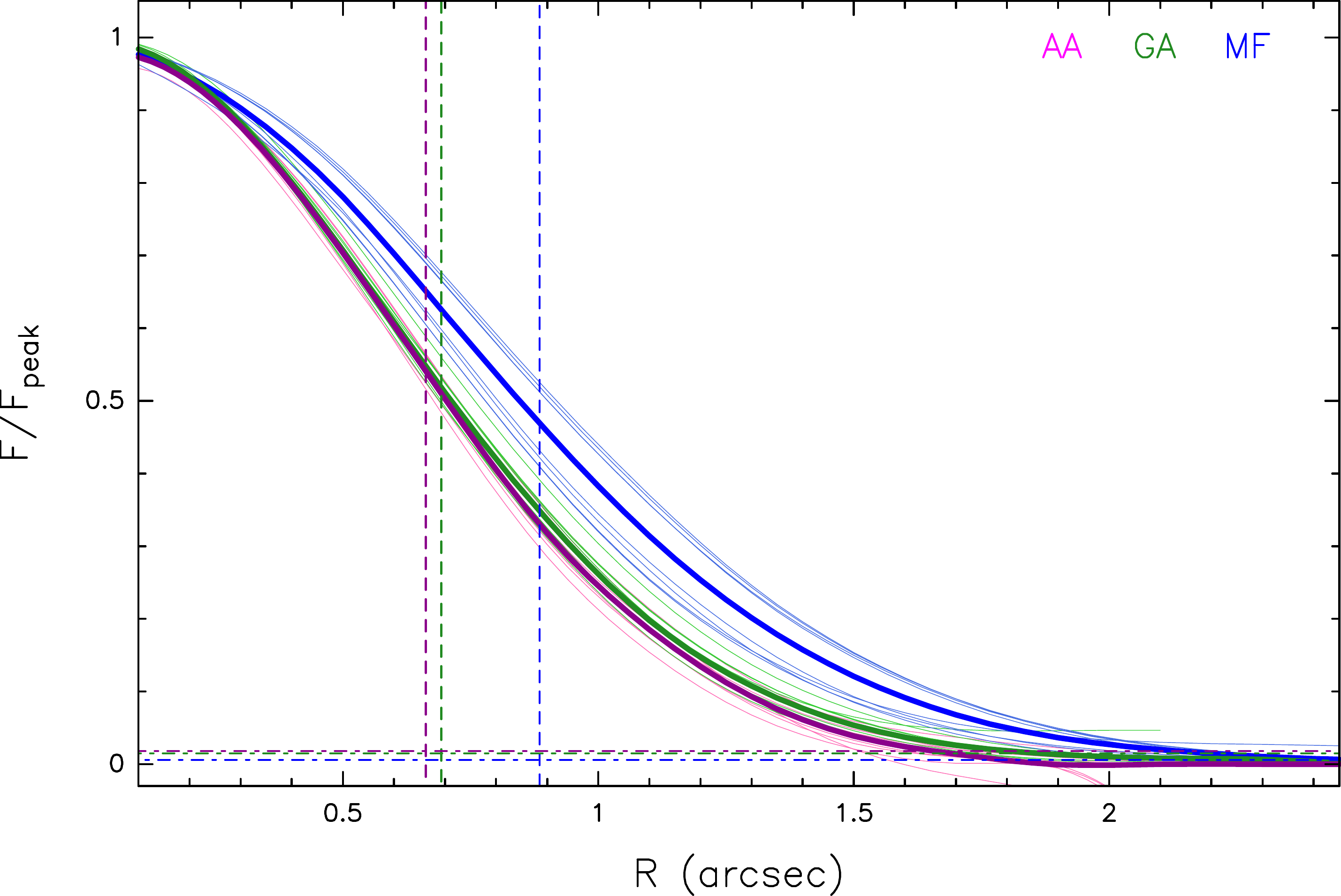}
\caption{Relative flux of emission maps shown in Fig. \ref{figuremappe} (thin lines) and \ref{figuremappemean} (bold lines) as a function of R. The dashed vertical lines indicate the mean value between the two semiaxes of the elliptic 2D fit of the mean maps, calculated from the values given in Table \ref{2Dgaussianresults}. At these radii the emission of the mean maps differs slightly from 0.5 due to the fact that the fluxes plotted are calculated in circular rings, for simplicity. The horizontal dashed lines show the $3\times rms$ level for the mean emission maps.} 
\label{figFvsR}
\end{figure*}
\begin{figure*}[h!]
\centering
\vspace{-4pt}
\includegraphics[width=14cm]{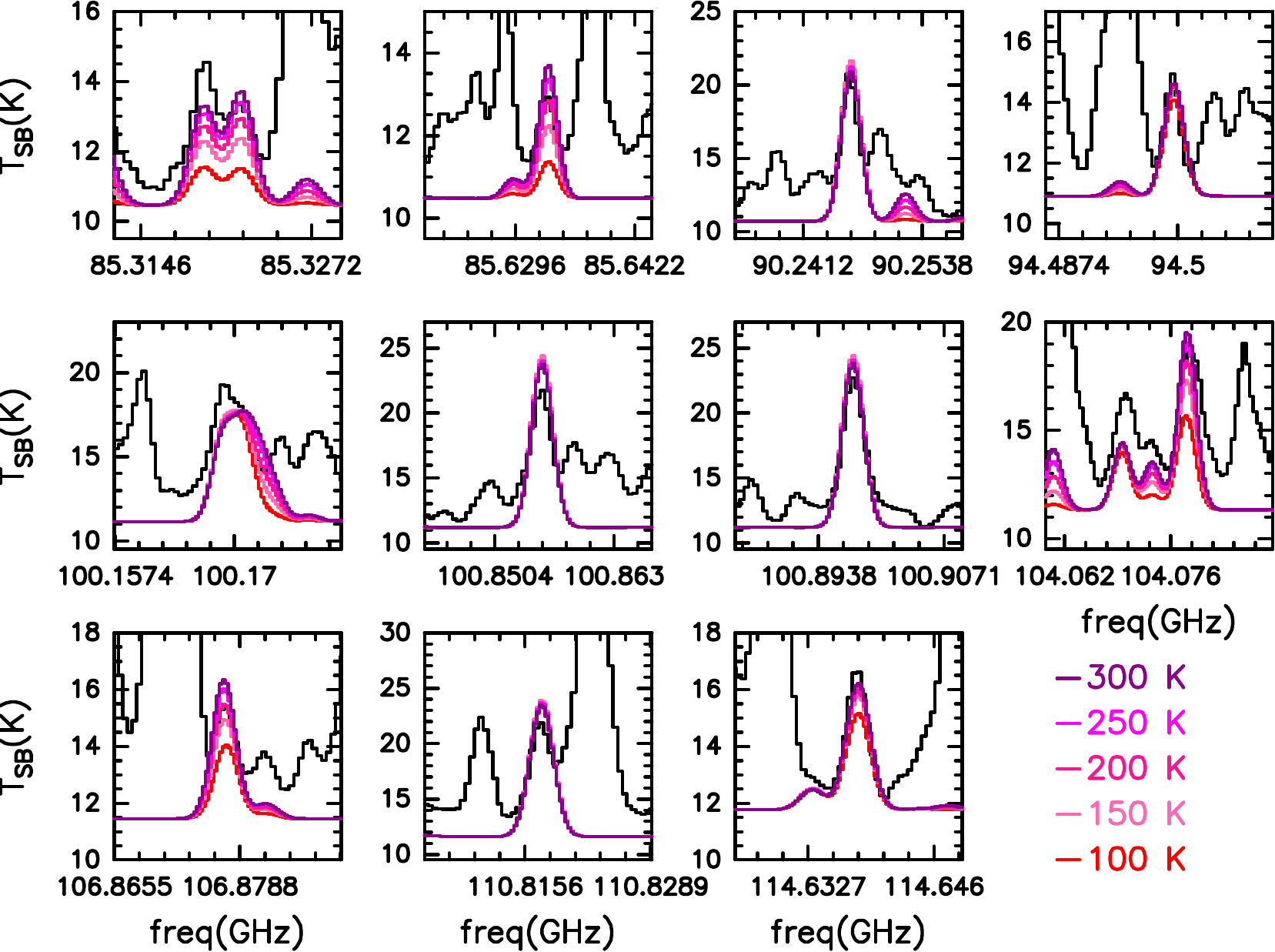}
\caption{Transitions used to fit acetic acid, with the LTE synthetic spectra simulated assuming different excitation temperatures, and leaving as free parameters only the column density. The value used are given in Table \ref{fitresults_GRIDAA}.}
\label{figsingleAAgrid}

\vspace*{5pt}
\includegraphics[width=14cm]{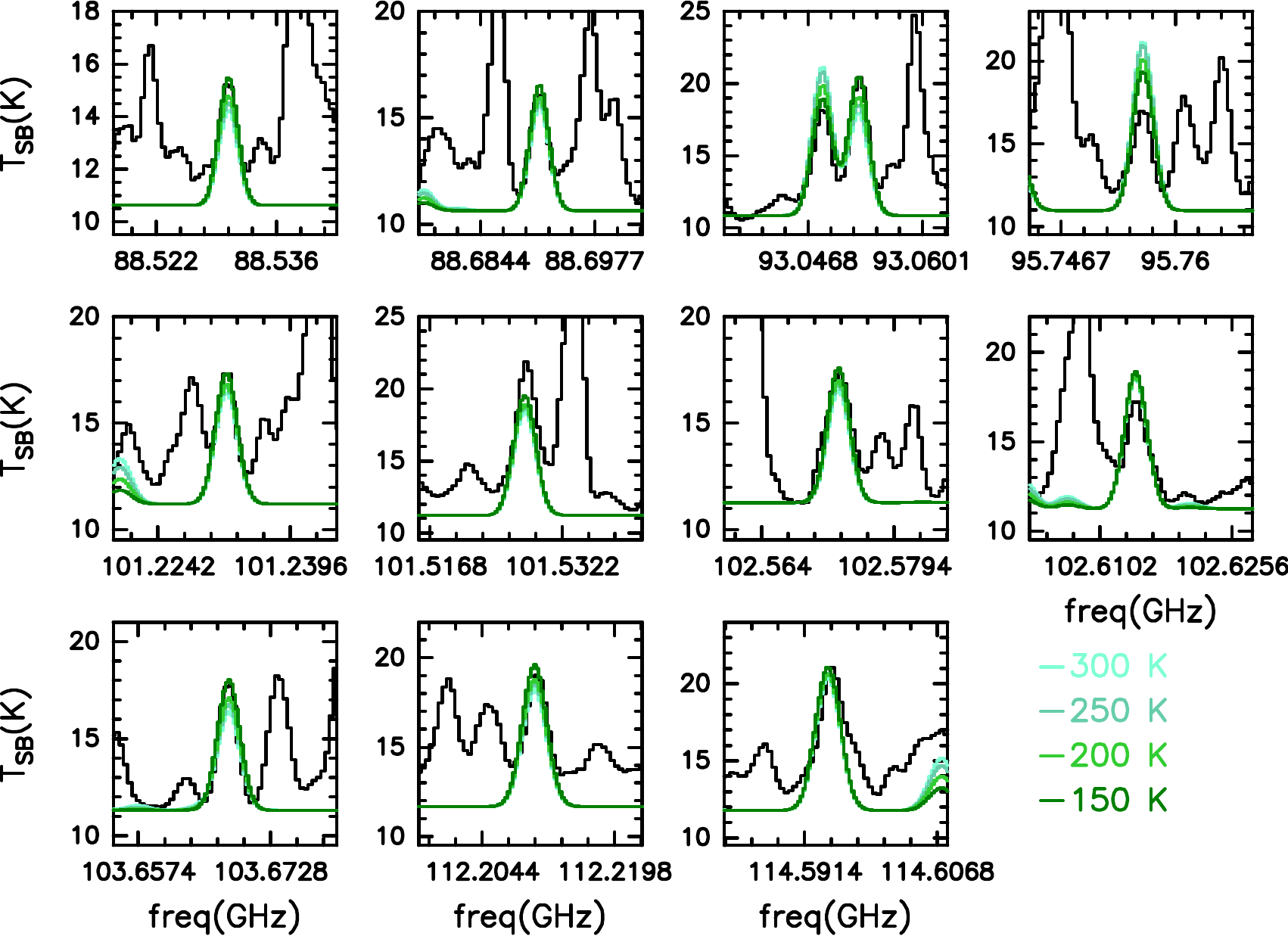}
\caption{Transitions used to fit  glycolaldehyde, with the LTE synthetic spectra simulated assuming different excitation temperatures, and leaving as free parameters only the column density. The value used are given in Table \ref{fitresults_GRIDGL}.}
\label{figsingleGL_GRID}
\end{figure*}
\begin{table*}	
\caption[]{Parameter of the best fit and abundances for MF, AA and GA.}
\label{fitresults}
\begin{centering}
\begin{tabular}{cccccc}
\hline
   	& FWHM & $V_{\rm{LRS}}$	&	$T_{\rm{ex}}$ &	 $N_{\rm{tot}}$ & $X$ \\	 
     & \small{[$\mathrm{km\,s^{-1}}$]} & \small{[$\mathrm{km\,s^{-1}}$]} & \small{[K]} &   \small{[$10^{17}\,\mathrm{cm^{-2}}$]}  & \\
\hline
$\mathrm{CH_{3}OCHO}$  & $6.8$ & $1.1$ & $221\pm27$   & $20\pm4$      & $(2.0\pm0.6)\times10^{-7}$  \\
$\mathrm{CH_{3}COOH}$  & $7.8$ & $0.0$ & $299\pm42$  & $8.4\pm1.4$       & $(8\pm3)\times10^{-8}$  \\
$\mathrm{CH_{2}OHCHO}$ & $8.8$ & $0.0$ &$128\pm17$    & $0.50\pm0.09$ & $(5.0\pm1.4)\times10^{-9}$  \\
\hline

\end{tabular}
\tablefoot{$N_{\rm{tot}}$: column density inside $1\farcs2$ beam;  $X$: abundances inside a beam calculated as $N_{\rm{tot}}/N_{\rm{H2}}$, where $N_{\rm{H_2}}=(1.0\pm0.2)\mathrm{\times10^{25}\,cm^{-2}}$. FWHM and $V_{\rm{LRS}}$ parameters were kept fixed in the fitting procedure. }
\end{centering}
\end{table*}

\begin{table}	
\caption[]{$N_{\rm{tot}}$ of the best fit for AA given a grid of fixed temperatures, with FWHM = 7.8$\,\mathrm{km\,s^{-1}}$  and $V_{\rm{LRS}}$ = 0.0 $\mathrm{km\,s^{-1}}$. }
\label{fitresults_GRIDAA}
\centering
\begin{tabular}{ccc}
\hline
  	&	$T_{\rm{ex}}$ &	 $N_{\rm{tot}}$	 \\	 
 & \small{[K]} &   \small{[$10^{17}\,\mathrm{cm^{-2}}$]}     \\
\hline

$\mathrm{CH_{3}COOH}$    & 100 & $1.3$  \\
   & 150 & $2.6$  \\
   & 200 & $4.3$  \\
   & 250 & $6.3$  \\
   & 300 & $8.5$  \\
\hline
\end{tabular}
\end{table}

\begin{table}	
\caption[]{$N_{\rm{tot}}$ of the best fit for GA given a grid of fixed temperatures, with FWHM = 8.8$\,\mathrm{km\,s^{-1}}$  and $V_{\rm{LRS}}$ = 0.0 $\mathrm{km\,s^{-1}}$. }
\label{fitresults_GRIDGL}
\centering
\begin{tabular}{ccc}
\hline
  	&	$T_{\rm{ex}}$ &	 $N_{\rm{tot}}$ 	 \\	 
 & \small{[K]} &   \small{[$10^{17}\,\mathrm{cm^{-2}}$]}    \\
\hline

$\mathrm{CH_{2}OHCHO}$    & 150 & $0.6$ \\
    & 200& $0.9$  \\
   & 250 & $1.3$  \\
   & 300 & $1.7$  \\
\hline

\end{tabular}

\end{table}
In the following we will analyze the  best fit physical parameters obtained from the line fitting performed with XCLASS for the three different isomers. The results of the fit are given in Table \ref{fitresults}, while in Figs. \ref{figsingleMF}, 6 and 7 the transitions used to constrain the fit for MF, AA and GA are reported, with the synthetic spectra obtained using the best fit parameters overlaid on the data. The total spectrum with the synthetic spectra of the three isomers obtained using the best fit parameters is given in Appendix D. The errors on $ T_{\rm{ex}}$ and $N_{\rm{tot}}$ take into account a $12\%$ error, from the $rms$ of the continuum baseline determination (see Sect. \ref{onefinalspectrum}). To determine the abundances of the molecules we used the value of the column densities inside the $1\farcs2$ beam, divided by the value of column density of $\rm{H}_2$ found in  Section \ref{ancontinuum}, extracted from the same area, considering a 20\% uncertainty.\\
\indent Figure \ref{figuremappe} shows the integrated maps of 8 of the most unblended transitions of MF, AA and GA with different values of upper state energy $E_{\rm{u}}$  (from $\sim 25\,\mathrm{K}$ to $\sim 200\,\mathrm{K}$). As seen in this figure, the emission of MF, AA and GA comes only from the HMC  and has a nearly circular shape in all the emission maps. 
To obtain a mean value of diameter of the emitting region of each molecule, the 8 maps of a given molecule have been normalized with respect to the peak intensity, and averaged together. The result is shown in Fig. \ref{figuremappemean}. The task  \textit{imfit} of CASA has been used to fit a 2D gaussian to these maps and thus obtain an estimate of the angular diameters of the emission, reported in Table \ref{2Dgaussianresults}. 
The column densities of the 3 isomers have been derived assuming a beam filling factor of 1. 
\newline\indent The decrease of the normalized flux ($F/F_{\rm{peak}}$) in the emission maps, as a function of the radius from the peak $R$, is shown in Fig. \ref{figFvsR} for the mean maps (\textit{bold lines}) and for the 8 selected transitions (indicated in Table \ref{2Dgaussianresults}) at different values of $E_{\rm{u}}$. 
These data were obtained by averaging the normalized flux inside circular rings centered on the mean map peak. As shown in Fig. \ref{figFvsR}, the emission of MF is the most extended among the 3 isomers, while AA is the most compact, even if with a small difference with respect to GA.\newline
\indent In addition, we performed the fit of AA and MF including also the most unblended transitions of high-energy ($E_{\mathrm{u}}>250\,\mathrm{K}$) and discuss the results in Sect. \ref{alsohighE}.

\subsection{Methyl Formate}
 For methyl formate we found an excitation temperature of $ T_{\rm{ex}}=221\pm27\,\mathrm{K}$, a column density of $N_{\rm{tot}}=\mathrm{(2.0\pm0.4)\times10^{18}\,cm^{-2}}$ and an abundance $X=(2.0\pm0.6)\times10^{-7}$. The $T_{\rm{ex}}$ is consistent with $\mathrm{250\,K}$, the mean temperature of MF in the main core calculated using the temperature gradient found by \citet{beltran2018}. From the total spectrum in Appendix D, we see that there are transitions of MF, not used for the fit, that are under-reproduced by more than 20\% by the best fit synthetic spectrum. About a third of these transitions are indeed blended with other molecular species, such as NS, CH$_3$CN, $^{13}$CH$_3$CN, C$_2$H$_5$CN and more. The others are mostly low energy transitions, which would be better reproduced by a lower temperature component of the fit. However, a two-temperature component fit does not properly converge and does not significantly improve the fit, either in terms of $\chi^{2}$ or of visual comparison with the observed spectrum. These transitions will likely be better reproduced by a fit that takes into account the temperature gradient observed by the previous study of \citet{beltran2018}. However, this is beyond the scope of this paper. In addition, we cannot exclude that some of these transitions may also be blended with still unidentified molecular species, since the preliminary line identification does not explain the totality of the transitions in the spectrum.\newline\indent
The mean value of column density of MF inside $1\farcs1$ from the data of \citet{beltran2018} is of $\sim1.5\times10^{18}\,\mathrm{cm^{-2}}$, consistent with our results. The column density derived by us differs by only a factor 2.5 from the value of $\mathrm{5\times10^{18}\,cm^{-2}}$ found by \citet{rivilla2017a} from IRAM 30m data, considering a source size of $1\farcs7$, and is also consistent with the estimate of $\mathrm{3.4\times10^{18}\,cm^{-2}}$ of \citet{Calcutt2014}, when rescaled to the same source size of this work. 
The distance between the peak of the continuum emission and the MF emission is of only $0\farcs09$, smaller than the pixel size ($0\farcs15$).
\subsection{Acetic Acid}
\label{ResultsAA}
 The best-fit parameters obtained with XCLASS for acetic acid are $ T_{\rm{ex}}=299\pm42\,\mathrm{K}$, $N_{\rm{tot}}=(8.4\pm1.4)\mathrm{\times10^{17}\,cm^{-2}}$ and an abundance of $X=(8\pm3)\times10^{-8}$. The temperature is consistent within the errors with the mean temperature in the main core of $\sim 250\,\mathrm{K}$ found from the relationship of $ T_{\rm{ex}}(\mathrm{MF})$ given in \citet{beltran2018}. Nevertheless, the values of $T_{\rm{ex}}$ for the three isomers cover a large range from $\sim130\,\mathrm{K}$ to $\sim300\,\mathrm{K}$ (see Table \ref{fitresults}). For this reason, to understand if the value of $T_{\rm{ex}}$ of AA could be overestimated, we performed the fit to AA for a grid of fixed temperatures, ranging from $100\,\mathrm{K}$ to  $300\,\mathrm{K}$. The only free parameter was the column density $N_{\rm{tot}}$.
 The results are given in Table \ref{fitresults_GRIDAA} and shown in Fig. \ref{figsingleAAgrid}. From the plots it is possible to exclude $ T_{\rm{ex}}\leq 150\,\mathrm{K}$, but it is also clear that for 7 transitions out of 14, there is a complete degeneracy between $ T_{\rm{ex}}$ and $N_{\rm{tot}}$, for $ T_{\rm{ex}}\geq 200\,\mathrm{K}$. From this plot we conclude that the excitation temperature of AA can be in the range $\mathrm{200\,-\,299\,K}$ (i.e. the value of best fit), leading to a range in column density of $\mathrm{4.3-8.4\times10^{17}\,cm^{-2}}$, with an uncertainty on $N_{\rm{tot}}$ of a factor $\sim2$. This implies a range in abundance $X$ between $4.3\times10^{-8}$ and $8\times10^{-8}$.  
The distance between the peak of the continuum emission and the AA emission is of $0\farcs12$, slightly larger with respect to MF but still smaller than the pixel size ($0\farcs15$).
\subsection{Glycolaldehyde}
 For GA we found an excitation temperature of $ T_{\rm{ex}}=128\pm17\,\mathrm{K}$, the lowest of the 3 isomers, a column density  $N_{\rm{tot}}=\mathrm{(5.0\pm0.7)\times10^{16}\,cm^{-2}}$ and an abundance $X=(5.0\pm1.4)\times10^{-9}$. The low $ T_{\rm{ex}}$ can be due to the small number of detected transitions with $E_{\rm{u}}>150\,\mathrm{K}$ (see Table B.3), and consequently the presence of only 2 high-energy transitions used for the fit, with $E_{\rm{u}}= 192\,\mathrm{K}$ and $178\,\mathrm{K}$, respectively. To understand if a higher $ T_{\rm{ex}}$ is compatible with the observed spectra, we performed the fit of GA for fixed values of $ T_{\rm{ex}}$ in the range $150\,\mathrm{K}$ to $300\,\mathrm{K}$, varying only the column density. The results of the fits are given in Table \ref{fitresults_GRIDGL} and shown in Fig. \ref{figsingleGL_GRID}. This figure shows that the transitions most sensitive to the value of $ T_{\rm{ex}}$ are those at frequency 93.048 GHz, 93.053 GHz and 95.756 GHz. From the comparison of the spectra simulated and observed at these frequencies  (Figs. 6 and \ref{figsingleGL_GRID}), the best fit seems to be compatible with $ T_{\rm{ex}}\leq150K$, confirming the results of the XCLASS fit. The values of the column density found in G31 by \citet{rivilla2017a} and \citet{Calcutt2014} differ by a factor $\sim 3.5$ and $\sim0.8$, respectively. 
 GA shows the largest shift in position with respect to the peak of the continuum emission, with a distance of $0\farcs22$, still small compared to the angular resolution of the data.
\begin{figure*}[h!]
\vspace{10pt}
\centering
\includegraphics[height=6cm]{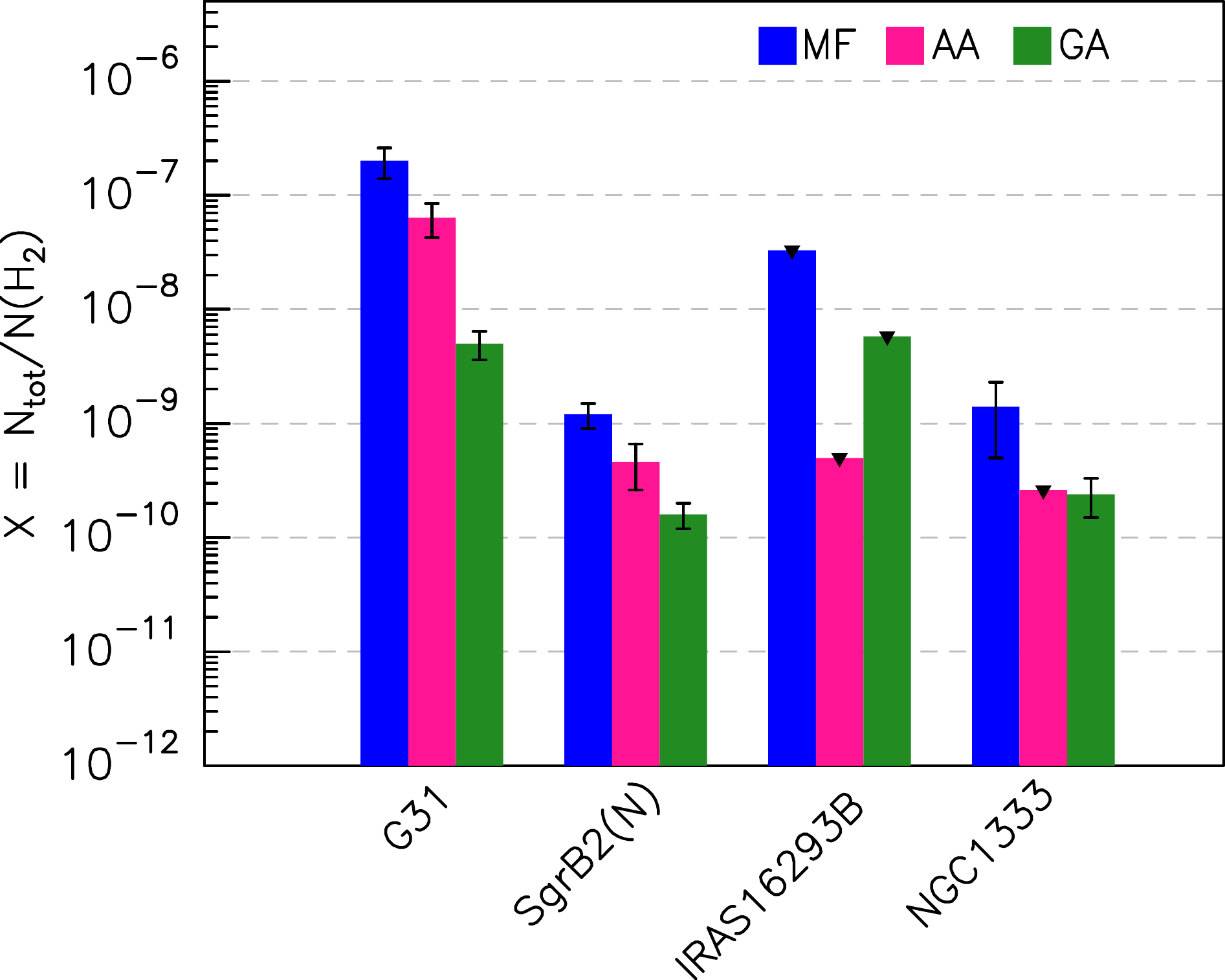}
\includegraphics[height=6cm]{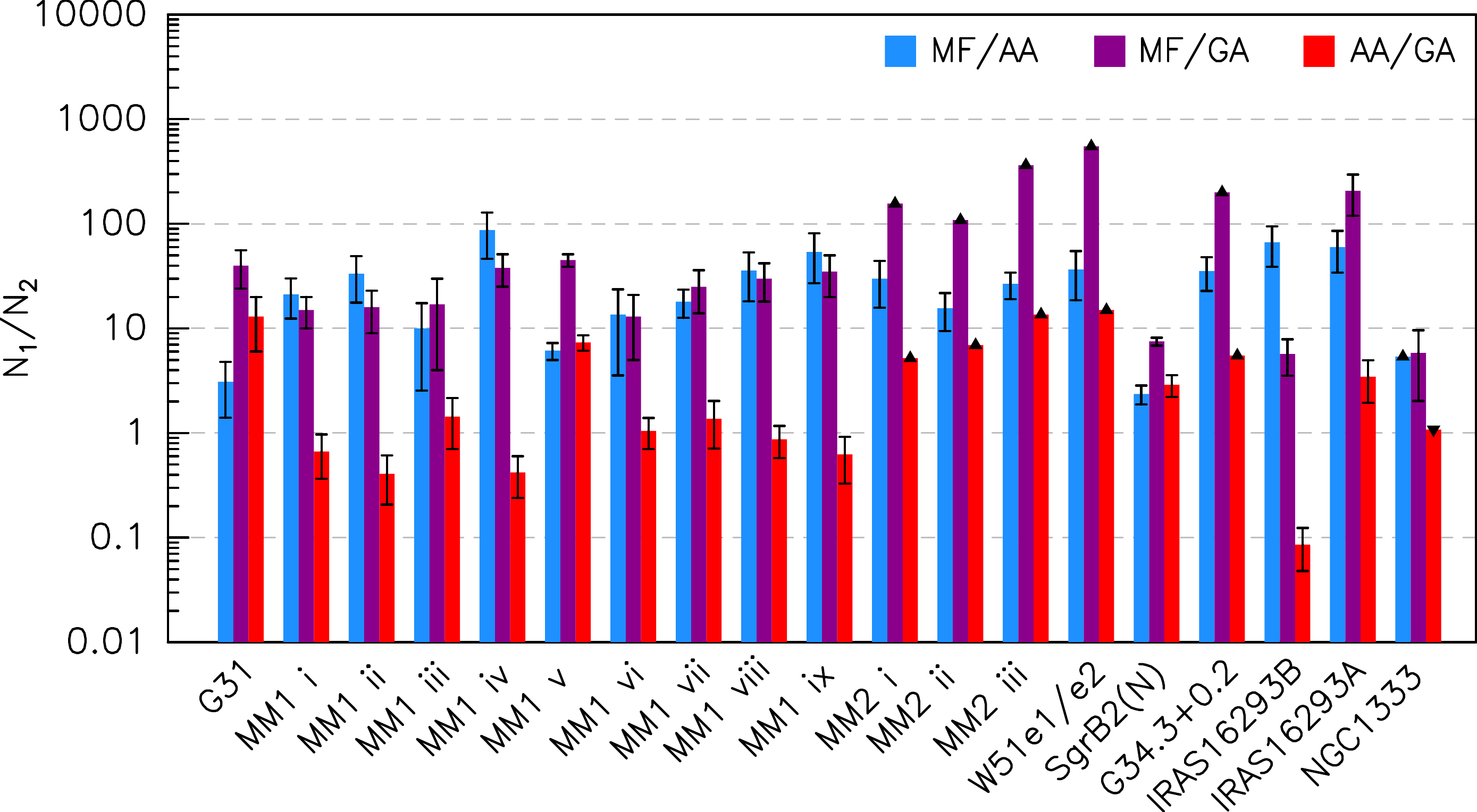}
\caption{\textit{Upper panel:} abundances wrt H$_{2}$ for the sources with an estimate in Table \ref{ratiolitterature}. For G31 the values of X(AA) given in figure has been calculated from the mean value of the observed range in column density $4.3\times10^{17} - 8.4\times10^{17}\mathrm{cm^{-2}}$, while the error bars cover the entire range. For IRAS16293 B the value of $N(\rm{H_2})$ is a lower limit, leading to upper limits on the abundances, but not affecting the ratios between abundances. 
\textit{Lower panel:} ratio of the column densities of the 3 isomers of C$_2$H$_4$O$_2$ for the sources listed in Table \ref{ratiolitterature}. For G31 the values of MF/AA and AA/GA have been calculated using the mean value of the observed range $4.3\times10^{17} - 8.4\times10^{17}\mathrm{cm^{-2}}$ for AA, while the error bars cover the ratios given by entire range. The sources labeled as MM1 i-viii and MM2 i-iii refer to NGC6334I MM1 and NGC6334I MM2, while the source labeled as NGC1333 is NGC1333 IRAS 4A.}
\label{figistogrammi}
\end{figure*}
\begin{figure*}[h!]
\centering
\vspace{0.1cm}
\includegraphics[width=8cm]{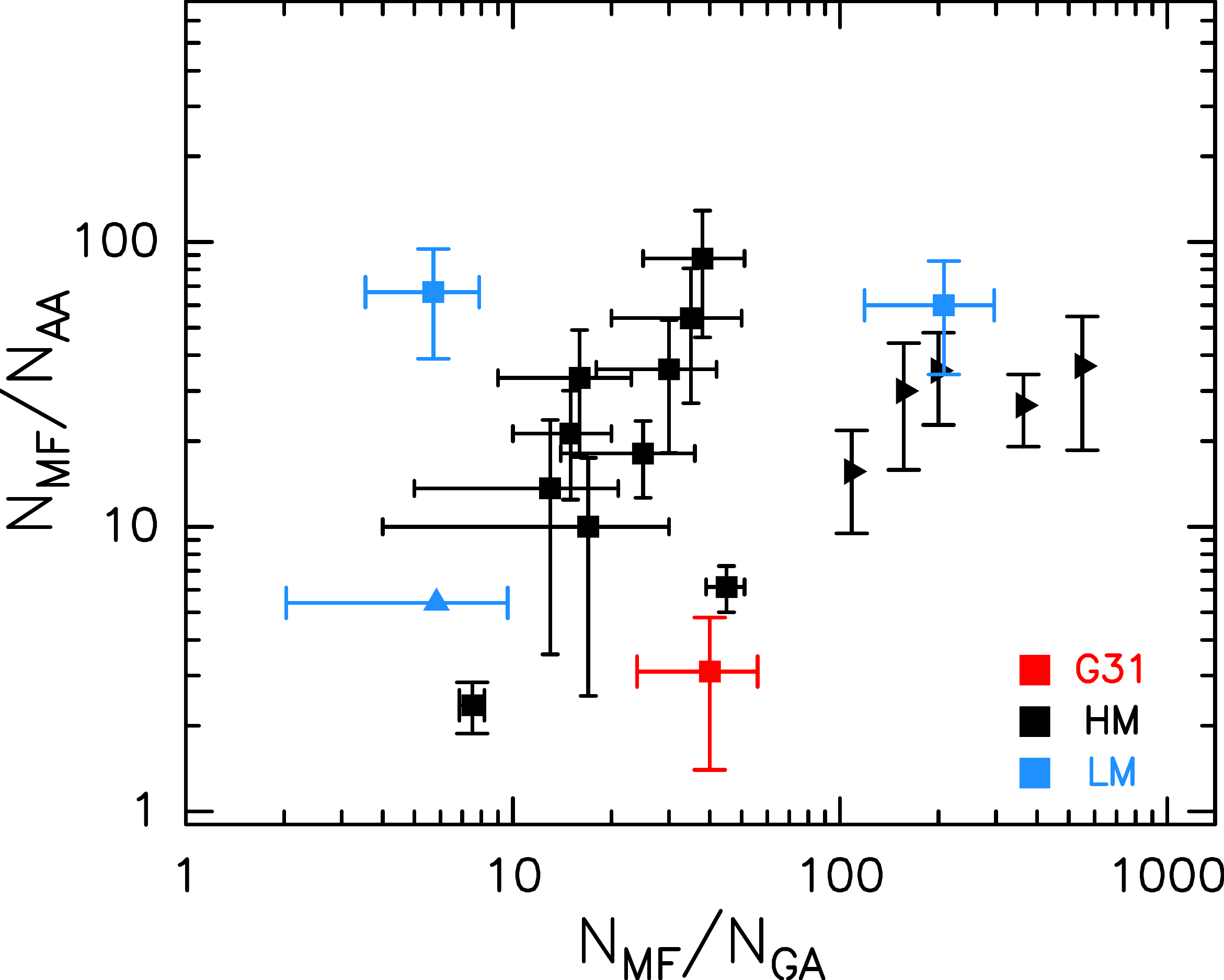}
\includegraphics[width=8cm]{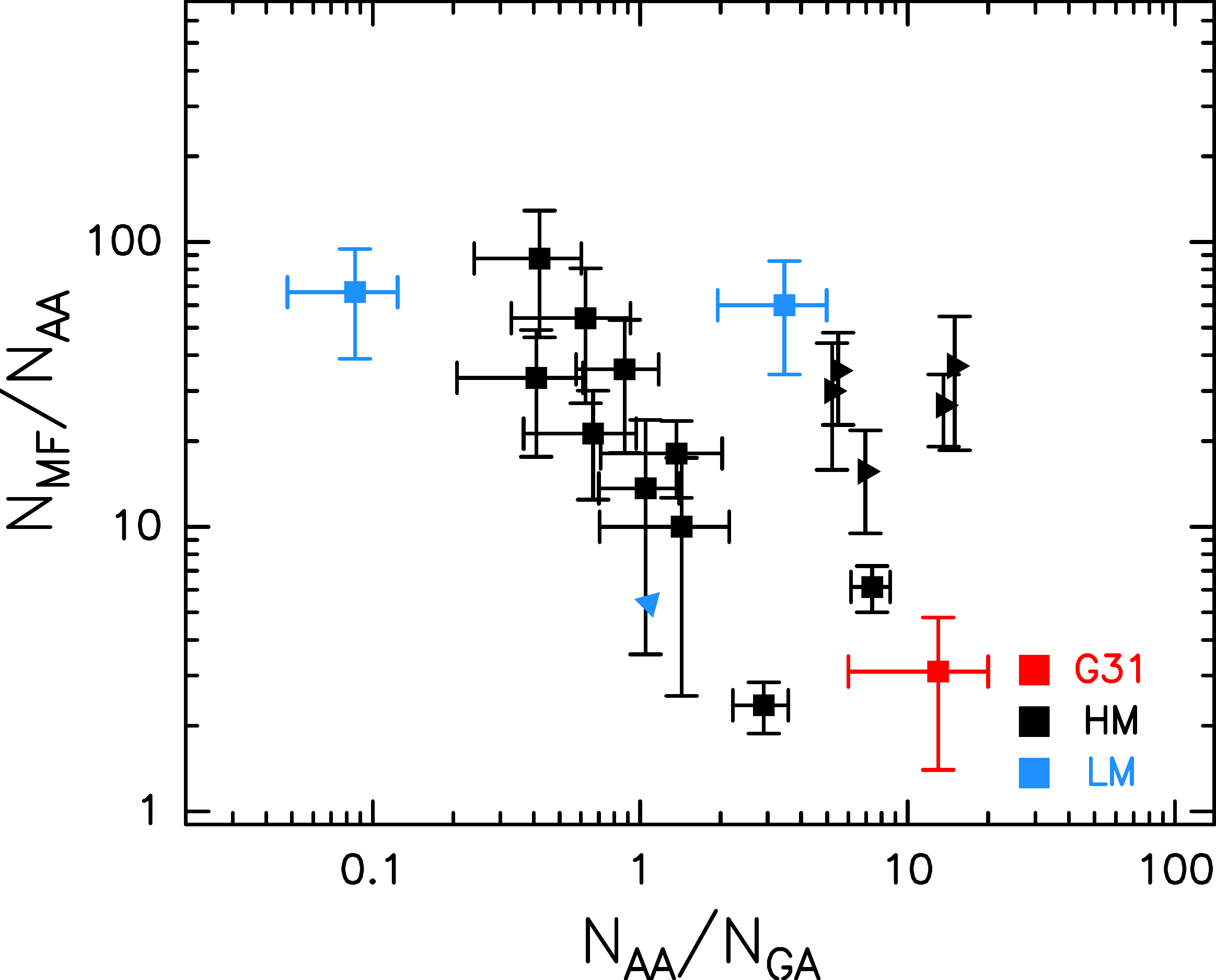}\\
\vspace{0.3cm}
\includegraphics[width=8cm]{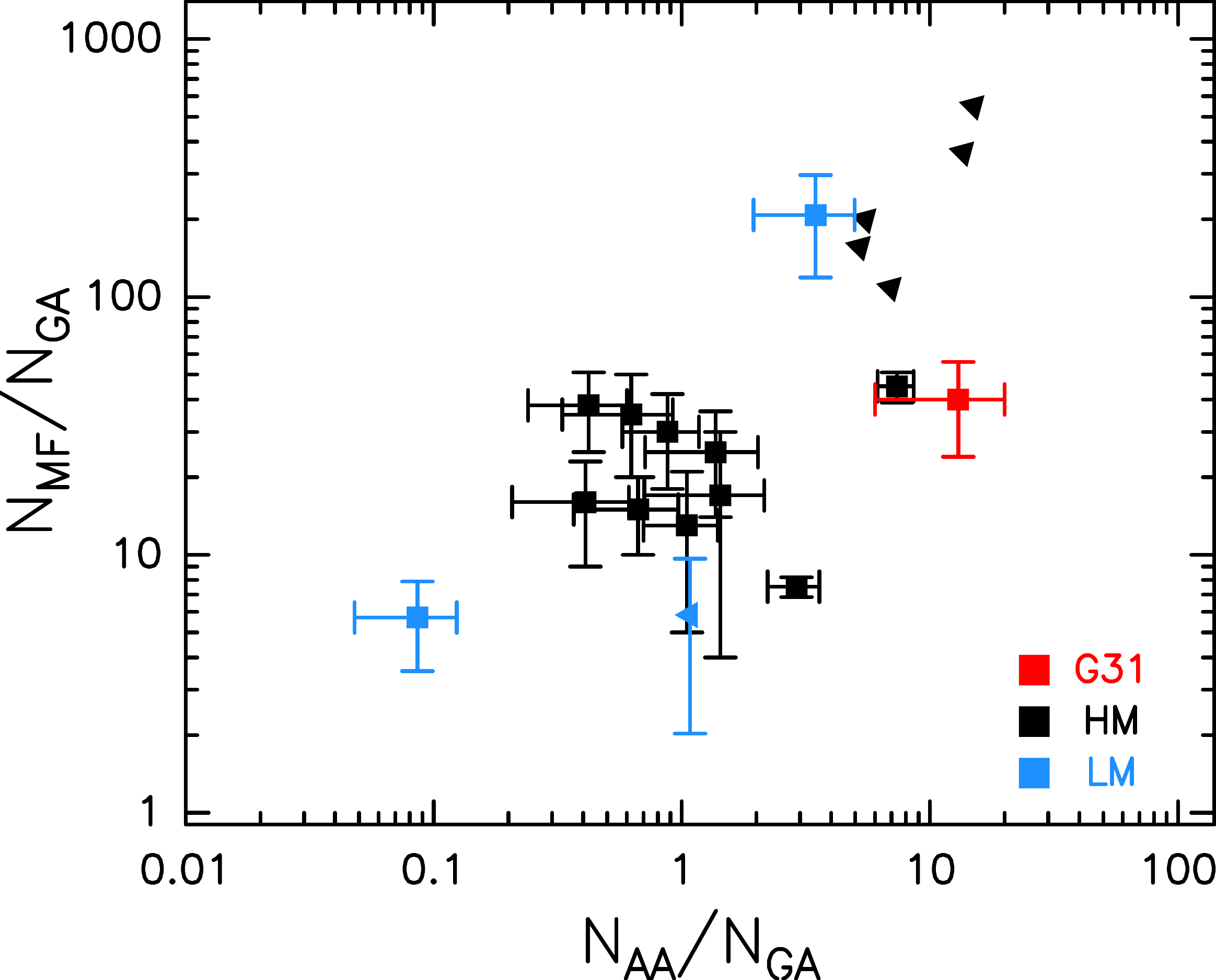}
\caption{Comparison of the ratios between the 3 isomers of C$_2$H$_4$O$_2$ for the sources listed in Table \ref{ratiolitterature}. For G31 the values of MF/AA and AA/GA given in the figure are calculated using the mean value of the observed range $4.3\times10^{17} - 8.4\times10^{17}\mathrm{cm^{-2}}$ for AA, while the error bars have been estimated from the entire range.}
\label{figconfronti}
\end{figure*}
\subsection{Methyl formate and acetic acid fit including $E_{\rm{u}}>250\,\mathrm{K}$ }
\label{alsohighE}
In Sects. 4.1 and 4.2 we reported the fit of MF and AA taking into account only the most unblended transitions with $E_{\rm{u}}<250\,\mathrm{K}$. We imposed this restriction to trace the same gas, with the same physical conditions, for all the 3 isomers. We thus not included high energy transitions (higher than the highest energy transition of GA available), that are in principle possibly emitted by a warmer gas. For completeness we present the fit of AA and MF including also the most unblended transitions of energy $E_{\rm{u}}>250\,\mathrm{K}$. We selected 4 transitions for AA ranging from 260 K to 510 K, and 7 for MF ranging from 250 K to 420 K, to add to the previously selected transitions for the fit. These new transitions are indicated by an asterisk in Table C.1 and C.2.\newline The results of the new fit for AA are an excitation temperature $T_{\rm{ex}}=290\pm40\,\mathrm{K}$ and a column density of $N_{\rm{tot}}= (8.0\pm1.2)\times10^{17}\,\mathrm{cm^{-2}}$, while for MF we find $T_{\rm{ex}}=229\pm28\,\mathrm{K}$ and a column density of $N_{\rm{tot}}= (2.1\pm0.3)\times10^{18}\,\mathrm{cm^{-2}}$. From a comparison with the values in Table \ref{fitresults}, the results are well consistent within the errors. This indicates that for MF and AA in G31 even the highest energy transitions have the same excitation conditions of lower energy transitions, and are thus emitted by gas with the same physical conditions.
\newline\indent
\section{Discussion}
\label{discussion}
\subsection{Comparison with other sources}
Figure \ref{figistogrammi} shows the abundances with respect to H$_2$ (upper panel) and the column density ratios (lower panel) of the 3 isomers of C$_2$H$_4$O$_2$ in G31 and in the sources in literature for which all three isomers have been observed, and at least two of them detected. The upper panel shows a very restricted sample, with respect to the lower panel, because for only 4 sources we found an estimate of $N(\rm{H}_2)$ on an angular scale comparable to the size of the region over which the column densities of the 3 isomers have been derived. The column density ratios, the abundances and the $N(\rm{H}_2$) adopted, and the corresponding references are given in Table \ref{ratiolitterature}. G31 has the highest values of abundance of MF and AA among the 4 sources shown, with a difference of 1 and 2 orders of magnitude, respectively, from the second highest values in the sample. The abundances in G31 are calculated using $N(\rm{H}_2)=(1.0\pm0.2)\times10^{25}\,\rm{cm^{-2}}$ (see Sect. \ref{ancontinuum}). Even considering this value as a lower limit - \citet{rivilla2017a} found a value 4 times larger - and adopting a value one order of magnitude higher, the abundance of AA would still be the largest in the sample, at least one order of magnitude above the others. From the upper panel, it does not seem to exist a unique behavior for the three isomers in both high-mass and low-mass sources. This is confirmed by the column density ratios in the extended sample showed in the lower panel. For IRAS 16293-2422A, \citet{rivilla19} found a column density of GA towards the peak one order of magnitude larger than the one in Table \ref{ratiolitterature} and in Fig. \ref{figistogrammi} by \citet{manigand2020}. \citet{manigand2020} selected an off-peak position to extract the spectrum, shifted by an amount comparable (by eye) with the emission size of AA and GA, thus possibly leading to underestimated values of these column densities.\\ \indent The lower panel in Fig. \ref{figistogrammi} confirms the large fraction of acetic acid in G31, with respect to the extended sample. In fact, in only 3 sources (4 including the lower limit in NGC1333 IRAS 4A) MF/AA<10: G31, SgrB2(N) and  the position "v" in NGC6334I MM1. The column density ratios of the isomers are plotted in Fig. \ref{figconfronti}, to search for possible
correlations. The first plot (\textit{upper-left}) does not show a clear
correlation between MF/AA and MF/GA, however, although the large scatter, a hint of the dual distribution firstly seen by \citet{el2019} in the space $N_{\mathrm{MF}}$ - $N_{\mathrm{GA}}$  is still visible. On the other hand, there seems to be an anti-correlation (Pearson's $\chi^{2} = 0.73$) between MF/AA and AA/GA (\textit{upper-right}), where we can also possibly see the same dual distribution - although not as clearly -, and a faint correlation (Pearson's $\chi^{2} = 0.51$) between MF/GA and AA/GA (\textit{lower panel}), even though with some spread. The bimodal distribution is not driven by a difference between high-mass sources and low-mass sources, but the origin must be related to particular physical or chemical conditions inside some sources (both high-mass and low-mass), as already stated by \citet{el2019}.\newline
\indent Lastly, from our work we derived the size of the emission of the 3 isomers, confirming a difference in the extension and showing as AA and GA are more compact than MF. This result is consistent with what shown in the maps towards other sources by  \citet{remijian2002}, \citet{Jorge2016}, \citet{xue2019}, \citet{el2019} and \citet{manigand2020}. Towards G31, AA shows the most compact emission among the 3 isomers. 
\subsection{Comparison with chemical models and laboratory studies}
For several years, the only efficient chemical pathways leading to the formation of the three isomers of $\mathrm{C_2H_4O_2}$ were found to be on the surface of dust grains \citep{garrod2006,garrod2008,beltran2009,laas2011,woods2012,garrod2013,woods2013,coutens2018, rivilla19}. When temperatures are sufficiently high, radicals have enough energy to diffuse across the surface and react to form COMs. The main routes on warm dust grains for the formation of the 3 isomers from the aforementioned studies are:
\begin{equation}
\mathrm{HCO + CH_3O \longrightarrow 
CH_3OCHO}
\label{eqstart}
\end{equation}
\begin{equation}
\mathrm{CH_3CO + OH\, \longrightarrow\, CH_3COOH}
\end{equation}
\begin{equation}
\mathrm{HCO + CH_2OH\, \longrightarrow\, CH_2OHCHO}
\end{equation}
\begin{equation}
\mathrm{CH_3OH + HCO\, \longrightarrow\, CH_2OHCHO + H}
\end{equation}
\begin{equation}
\label{eqend}
\mathrm{2HCO + 2H \,\longrightarrow\, CH_2OHCHO}.
\end{equation}
Laboratory studies also investigated possible routes for the formation of MF and GA on the surface of dust grains at low temperatures (\citealt{fedoseev2015, chuang2016}), showing that the reactions can take place even at temperatures as low as $\sim10\,\mathrm{K}$. This has also been confirmed by \citet{simons2020} using microscopic kinetic Monte Carlo simulations to simulate ice chemistry. Nevertheless, the increase of temperature in more evolved sources, like G31, increases the mobility of radicals on the surface leading to higher efficiency of the routes (\ref{eqstart})-(\ref{eqend}) than those proposed in the experiment at low temperature.\\
\indent Gas phase chemical routes, might be also efficient at low temperatures, and therefore able to explain the detection of COMs in cold environment \citep{Ob2010,bac2012,jimenez2016}. 
In the model by \citet{vasy13} the most efficient gas-phase pathway to the formation of MF in a cold environment ($\mathrm{T\sim10\,K}$) is given by the radiative association reaction \citep{horn2004}
\begin{equation}
\mathrm{H_{2}CO^{+} + H_2CO \longrightarrow 
H_{2}COHOCH_{2}^{+}+ h}\nu\,,
\label{eqchangestructure}
\end{equation}
followed by a dissociative recombination and an isomerization, that leads to the formation of MF. 
\citet{balucani2015} proposed the following chain of reactions for the production of MF
\begin{equation}
\mathrm{ CH_3OH + OH \,\longrightarrow\, CH_3O + H_2}
\end{equation}
\begin{equation}
\mathrm{ CH_3O + CH_3 \,\longrightarrow\, CH_3OCH_3}
\end{equation}
\begin{equation}
\label{fluoroCloro}
\mathrm{ CH_3OCH_3 + F/Cl\,\longrightarrow\, CH_3OCH_2 + HF/HCl }
\end{equation}
\begin{equation}
\centering
\label{lastbalucani}
\mathrm{ CH_3OCH_2 + O \,\longrightarrow\, CH_3OCHO + H}\,,
\end{equation}
starting from the non-thermal desorption of methanol at low temperature.
\citet{vas2017} included these new gas-phase reactions in their network to model abundances of complex organic species in the prestellar core L1544 including MF. Reaction (\ref{lastbalucani}) of the chain proposed by \citet{balucani2015} is indeed the most efficient pathway for the formation of MF in this cold environment, accounting for two-thirds of the total production. Nevertheless, the production of the reactant $\mathrm{ CH_3OCH_2}$ is mainly dominated by the reaction studied by \citet{Shannon2014}:
\begin{equation}
\centering
\mathrm{ CH_3OCH_3 + OH \,\longrightarrow\, CH_3OCH_2 + H_2O}\,,
\end{equation}
rather than reaction (\ref{fluoroCloro}), due to the low abundances of F and Cl. Approximately  the  remaining one third of total MF is generated by the ion-neutral reaction 
\begin{equation}
\mathrm{CH_3OH_{2}^{+} + HCOOH \,\longrightarrow\, HC(OH)OCH_{3}^{+}+H_2O}\,.
\end{equation}
and the subsequent dissociative recombination of $\mathrm{HC(OH)OCH_{3}^{+}}$. The abundances of COMs such as MF, dimethyl ether, formamide and acetaldehyde are of the order of $10^{-12}-10^{-10}$, thus gas-phase reactions can reproduce the observed abundances in cold low-mass objects. However, the abundance of MF derived in e.g. L1544 or L1689B \citep{bac2012,jimenez2016} is 2-3 orders of magnitude below that obtained for G31. \citet{coutens2018} report that in their models for star-forming regions hosting a proto-star with mass from 1 to $60\,M_{\odot}$ MF reaches abundances in the range $7\times10^{-10}-2.5\times10^{-8}$, when the temperatures are high ($100-300\,\mathrm{K}$). These results are obtained with a network considering only gas-phase reactions for MF, including the ones in \citet{balucani2015}. They found that the peak of abundance of MF from gas-phase reactions only is a factor 2 smaller than the value estimated in G31 by \citet{rivilla2017a}, and the inclusion of route (4) on grain-surface is able to bridge this discrepancy. However, the new estimate presented in this work is a factor 5 larger, implying that grain surface reactions are likely responsible for the majority of the abundance of MF detected in G31. \citet{garrod2013} found that reaction (4) can efficiently lead to abundances $\sim1\times10^{-7}$. From this, MF would behave in a similar way as other COMs such as formamide (see \citealt{quenard2018}), whose production at low T is dominated by gas-phase chemistry, while its formation is governed by grain surface reactions in warm/hot sources such as hot corinos/hot cores.
 \\\indent For GA and AA, \citet{skouteris2018} introduced a new chain of reactions that leads to the formation of these C$_2$H$_4$O$_2$ isomers:
\begin{equation}
\begin{aligned}
\label{skoutOH}
\mathrm{CH_3CH_2OH + OH/Cl \,\longrightarrow\,} & \mathrm{CH_2CH_2OH + H_2O/HCl}\\
    & \mathrm{CH_3CHOH + H_2O/HCl}
\end{aligned}
\end{equation}
\begin{equation}
\mathrm{CH_3CHOH + O \,\longrightarrow\, CH_3COOH + H }
\end{equation}
\begin{equation}
\mathrm{CH_2CH_2OH + O \,\longrightarrow\, CH_2OHCHO + H}\,,
\end{equation}
starting from the sublimation of ethanol in gas-phase. The GA abundance prediction of the model that simulates the chemistry in a hot corino, varies over a broad range $\sim10^{-10}-10^{-8}$ and shows a strong dependence on the hydroxyl radical OH, on the rate coefficient of the reaction (\ref{skoutOH}), and on the initial abundance of ethanol (CH$_3$CH$_2$OH). Assuming the most favorable rate coefficient and an abundance of ethanol of $\sim6\times10^{-8}$ (wrt to H$_2$), the model of \citet{skouteris2018} predicts an abundance of GA close to the one derived in this work, but the abundance of AA is $\sim1$ order of magnitude lower than the new estimate towards G31. These values are derived before sublimated ethanol is fully consumed for a gas with temperature equal to 100 K and H nuclei density $2\times10^8\,\mathrm{cm^{-3}}$. Nevertheless, \citet{coutens2018} have shown that GA can efficiently form from reactions (6) and (8) on dust grains and reproduce the abundances in G31 observed by \citet{rivilla2017a} and the new estimate presented in this work. It has to be noted that these calculations have been made assuming a mass of $25\,M_{\odot}$ (from \citealt{oso2009}) for the proto-star embedded in G31, derived using the old estimates of distance of 7.9kpc. The new estimate of the distance to G31, which affects the estimate of the mass and luminosity of the proto-star(s) embedded in G31, might lead to a change in the modeled abundances. Similarly to \citet{coutens2018}, \citet{rivilla19} have found that the main formation route for GA in IRAS 16293-2422 proto-stars is reaction (6), with possible contributions also from reaction (8). The formation of GA through grain-surface reactions in \citet{rivilla19} is strengthen by the fact that their grain-surface network reproduces both GA and that of its precursor HCO abundances.\\\indent Unlike GA, the abundance of AA from gas-phase reactions as a function of ethanol abundance and the different rate coefficient possibilities is not given in \citet{skouteris2018}. If we assume that the ratio AA/GA$\sim10$ (found in the aforementioned specific case of ethanol abundance $\sim6\times10^{-8}$), is conserved, in principle the abundance of AA in this work could be covered by the wide range derived from multiplying for a factor 10 the given range $\sim10^{-10}-10^{-8}$  of GA in \citet{skouteris2018}. However, this relies on many 
assumptions that need to be further tested.  The observed abundance estimated this work cannot be reproduced with reaction (5) included in the chemical network of \citet{garrod2013}. Further investigation on reaction (\ref{skoutOH}) and on the abundance of ethanol in G31 is needed to constrain the formation route of AA in G31, which might also lead to completely discard the proposed gas-phase pathways (16)-(18).\\ 
\indent The fact that the rate coefficients of reactions (17) and (18) have a dependence on temperature implies that both gas-phase and grain-surface reactions can possibly explain the more extended emission of MF, with respect to AA and GA. In fact, the gas-phase reactions for AA and GA would be more efficient closer to the center of the core, due to the increase in temperature. On the other hand for grain-surface reactions, \citet{burke2015} found that MF desorbs at lower temperature with respect to AA and GA, which follow the desorption of water ice.\newline
\indent Recently, \citet{ahmad2020} proposed a new pathway for the formation of the three isomers both in gas-phase and on grain surface, involving formaldehyde:
\begin{equation}
\begin{aligned}
\label{ahmadreact}
\mathrm{H_2CO + H_2CO \,\longrightarrow\,} & \mathrm{CH_2OHCHO}\\
    & \mathrm{CH_3OCHO}\\
    & \mathrm{CH_3COOH}
\end{aligned}
\end{equation}
All the 3 channels show high energy barriers, that can be crossed thanks to quantum tunneling or thermal hopping. The study also reports that reaction (19), in all the three cases, is more efficient on grain-surface. However, no calculations have been made to predict the abundances that can be reached thanks to this pathway on grain surface in star-forming regions. \citet{silva2020} simulated the channel of reaction (19) that leads to GA in the condition of a high-mass protostellar object in gas-phase, and the maximum abundance is of $\sim10^{-12}$, orders of magnitudes below the value observed in G31 as well as SgrB2(N).\newline
\indent A summary of the reaction routes and the predictions of the different models is given in Table (\ref{table_recap}). In conclusion, for MF grain-surface reactions are likely responsible for its high abundance in G31. The results from GA and its precursor HCO in IRAS 16293-2422 A \& B by \citet{rivilla19} could suggest a grain-surface formation for GA in hot sources like G31. However, the case of GA is not of straightforward interpretation as both gas-phase and grain-surface reactions could be at play considered the uncertainties in the chemical models aforementioned, and, as a results,both scenarios are able to explain the more compact morphology of GA with respect to MF. For AA, the routes included in \citet{garrod2013} are not able to reproduce the high-abundance found towards G31, while the gas-phase route proposed by \citet{skouteris2018} could not be excluded, but need to be further tested in order to reduce the large uncertainties on their predictions. In case of future discrepancy with the narrowed prediction of this route and the abundance found in G31, the main pathway(s) responsible for the high abundance of AA in G31 should be further investigated. Similarly to GA, both gas-phase and grain-surface reactions could explain the more compact morphology of AA with respect to MF. 
\begin{sidewaystable*}
\centering
\caption{Column density ratios between the 3 isomers of $\mathrm{C_2H_4O_2}$ and abundances from star-forming regions in this work and in literature.}
\label{ratiolitterature}
     \begin{tabular}{lccccccccr}
         \hline
         Source & & AA/GA & MF/GA & diameter &X(MF) & X(AA) & X(GA) & N(H$_2$)& References \\
         & & & &[au] & & & & [cm$^{-2}$] & \\
         \hline
         \multicolumn{10}{c}{\textit{High-Mass Sources
          }}\\
         \textbf{G31} & &  \textbf{9-17}  &\textbf{40$\mathbf{ \pm16} $}  & \textbf{4400} & $(2.0\pm0.6)\times10^{-7}$ &$(4.3-8)\times10^{-8}$ & $(5.0\pm1.4)\times10^{-9}$ & $1.0\times10^{25}$ &\textbf{this work}\\
         \hline
          
         SgrB2(N)-LMH    & &  $2.9 \pm0.7 $   & $7.5 \pm0.7 $  & 17000  &$(1.2\pm0.3)\times10^{-9}$ &$(4.6\pm1.4)\times10^{-10}$ & $(1.6\pm0.4)\times10^{-10}$ & $1.0\times10^{26\,a)}$  &(1) \\
        \hline
         NGC 6334I MM1 & I &  $0.7 \pm0.3 $ & $15 \pm5 $   &  350     & - & - & - & - & (2) \\
         & II & $0.4 \pm0.2 $   & $16 \pm7 $   &  & - & - & - & - &  \\
         & III & $1.4 \pm0.8 $   & $17 \pm13 $    &  & - & - & - & - & \\
         & IV &  $0.42 \pm0.19 $   & $38 \pm13 $  &  & - & - & - & - &  \\
         & V &  $7.4 \pm1.2 $   & $45 \pm6 $  &  & - & - & - & - &  \\
         & VI &  $1.0 \pm0.4 $   & $13 \pm8 $  &  & - & - & - & - &\\
         & VII & $1.4 \pm0.7 $   & $25 \pm11 $  &  & - & - & - & - &\\
         & VIII &  $0.9 \pm0.3 $   & $30 \pm12 $  &  & - & - & - & - & \\
         & IX &  $0.6 \pm0.3 $   & $35 \pm15 $  &  & - & - & - & - &\\
         \hline
         NGC 6334I MM2 & I &  $>5^{\ast }$   & $>158^{\ast }$    & 350     & - & - & - & - &(2) \\
         & II &  $>7^{\ast }$   & $>110^{\ast }$&   & - & - & - & - & \\
         & III &   $>14^{\ast }$   & $>373^{\ast }$  &  & - & - & - & - & \\
         \hline
         W51e1/e2 & & $>15^{\ast }$   & $>550^{\ast }$  &  13000     & - & - & - & - & (3),(4) \\
         G34.3+0.2 & & $>5^{\ast }$   & $>200^{\ast }$ & 29000      & - & - & - & - & (3),(5)\\
         \hline
         \multicolumn{10}{c}{\textit{Low-Mass Sources}}\\
         IRAS 16293-2422B  & &  $0.09 \pm0.04 $   & $6 \pm3 $  &  60   & $<3.3\times10^{-8}$  & $<5.0\times10^{-10}$ &$<5.8\times10^{-9}$ &$>1.2\times10^{25\,b)}$ & (6),(7)\\
         IRAS 16293-2422A  & & $3.5 \pm1.5 $   & $210 \pm90 $  & 60     & - & - & - & - & (8)\\
         NGC 1333 IRAS 4A & & $<1.1^{\ast\ast }$   & $6 \pm4^{\ast\ast } $ & 120   &$(1.4\pm1.0)\times10^{-9}$ &$<2.6\times10^{-10}$ & $(2.4\pm1.1)\times10^{-10}$ & $3.7\times10^{25\,c)}$&(5),(9)\\
        
         \hline
\end{tabular}
      
     \tablefoot{Sources in literature where all three isomers have been detected, or only one isomer not detected with an estimate on the upper limit of the column density; column density ratio  AA/GA; column density ratio  MF/GA; linear scale to which the ratio refers; abundance of MF; abundance of AA; abundance of GA; $N(\rm{H_2})$ used to derive the abundances; references for the column density values of the 3 isomers of $\mathrm{C_2H_4O_2}$. The high-mass sources are listed in the upper part of the table, while the low-mass sources are listed in the lower part. $\ast$) the column density of GA is an upper limit, so the ratio with MF and AA are lower limits; $\ast\ast$) estimate of MF and GA inside $0\farcs5$ from \citet{taquet2015} and upper limit of AA rescaled to the same source size ($0\farcs5$) from the value in \citet{remijan2003}. Since the column density of AA is an upper limit, the ratio AA/GA is an upper limit too. Abundances have been calculated only for sources for which an estimate of $N(\rm{H_2})$ is available within an angular diameter close to the one of the derived column densities of the 3 isomers. References to the works on the 3 isormers: (1) \citet{xue2019}; (2) \citet{el2019}; (3) \citet{lykke2015}; (4) \citet{remijian2002}; (5) \citet{remijan2003}; (6) \citet{Jorge2012}; (7) \citet{Jorge2016}; (8) \citet{manigand2020}; (9) \citet{taquet2015}. Reference to the $N(\rm{H_2})$ values a) Mean value for position AN01 from \citet{alvaro2017}; b) lower limit from (7); c) derived from emission at 145GHz from (9). 
     }

\end{sidewaystable*}
\begin{sidewaystable*}	
\caption[]{Summary of the chemical routes to the formation of the 3 isomers of C$_2$H$_4$O$_2$, and related reactions.}
\label{table_recap}
\centering
\begin{centering}
\begin{tabular}{p{1.3cm}p{7.6cm}p{0.7cm}p{12.2cm}}
\hline
$\#$ in text & Reaction & type & comments \\
\hline
\multicolumn{4}{c}{\textbf{MF}}\\
(4) & $\mathrm{HCO + CH_3O \longrightarrow 
CH_3OCHO}$ & gs & efficient pathway for the formation of MF from (a) and (b).  \\
\hline
(9) & $\mathrm{H_{2}CO^{+} + H_2CO \longrightarrow 
H_{2}COHOCH_{2}^{+}+ h}\nu\,$ & \multirow{2}{*}{gp} & \multirow{2}{335pt}{efficient for the formation of MF in cold environment (c). Not efficient to reproduce the observed abundance in G31 (b).} \\
 & $\mathrm{ 
H_{2}COHOCH_{2}^{+} + e^{-}} \longrightarrow \mathrm{CH_3OCHO + H}$  \\
\hline
(10) & $\mathrm{ CH_3OH + OH \,\longrightarrow\, CH_3O + H_2}$ & \multirow{4}{*}{gp} & \multirow{4}{335pt}{efficient for the formation of MF in cold environment (d,e). (e) found react. (14) more efficient than (12). Not efficient to reproduce the observed abundance in G31 (b).}\\
(11) & $\mathrm{ CH_3O + CH_3 \,\longrightarrow\, CH_3OCH_3}$ &  \\
(12) & $\mathrm{ CH_3OCH_3 + F/Cl\,\longrightarrow\, CH_3OCH_2 + HF/HCl }$ &  \\
(13) & $\mathrm{ CH_3OCH_2 + O \,\longrightarrow\, CH_3OCHO + H}$ &  \\
\hline
(15) & $\mathrm{CH_3OH_{2}^{+} + HCOOH \,\longrightarrow\, HC(OH)OCH_{3}^{+}+H_2O}$ &\multirow{2}{*}{gp}&\multirow{2}{335pt}{second most efficient reaction in cold environment (e), after (13). Not efficient in reproduce the observed abundance in G31 (b).}  \\
 & $\mathrm{HC(OH)OCH_{3}^{+} + e^{-} \,\longrightarrow\, CH_3OCHO + H} $ & \\
 \hline
(19) & $\mathrm{H_2CO+H_2CO \,\longrightarrow\, CH_3OCHO}$ & $\ast$ & proposed by (f), presents high energy barrier, and is more efficient on grain-surface but no calculation made to predict the abundances in star-forming regions\\
\hline\hline
\multicolumn{4}{c}{\textbf{AA}}\\
(5) & $\mathrm{CH_3CO + OH\, \longrightarrow\, CH_3COOH}$ & gs & main route on grain-surface, not efficient enough to reproduce the observed abundance in this work (a).\\
\hline
(16) & $\mathrm{CH_3CH_2OH + OH/Cl \,\longrightarrow\,} \mathrm{CH_2CH_2OH + H_2O/HCl}$ & \multirow{3}{*}{gp} &  \multirow{4}{335pt}{predictions by (g) span in a broad range for GA, and may cover the abundance observed in this work for AA. Strong dependence on the rate coefficient of reaction (16) and on abundances of reactant OH and ethanol, which should be constrained to reduce uncertainties.}\\
   & $\qquad\qquad\qquad\qquad\qquad\quad\mathrm{CH_3CHOH + H_2O/HCl}$  & \\
(17) & $\mathrm{CH_3CHOH + O \,\longrightarrow\, CH_3COOH + H }$ & \\
 & & \\
 \hline
 (19) & $\mathrm{H_2CO+H_2CO \,\longrightarrow\, CH_3OOCH}$ & $\ast$ & proposed by (f), presents high energy barrier, and is more efficient on grain-surface but no calculation made to predict the abundances in star-forming regions\\
\hline\hline
\multicolumn{4}{c}{\textbf{GA}}\\
(6) & $\mathrm{HCO + CH_2OH\, \longrightarrow\, CH_2OHCHO}$ & gs & efficient to reproduce the abundance in this work together with (8) by (b) and in IRAS 16293-2422 A \& B by (h). The estimate by (b) might change due to the new estimate of the distance to G31.\\
\hline
(7) & $\mathrm{CH_3OH + HCO\, \longrightarrow\, CH_2OHCHO + H}$ & gs &  efficient route for the formation of GA (i), not tested in (b) and (h). \\
\hline
(8) & $\mathrm{2HCO + 2H \,\longrightarrow\, CH_2OHCHO}$ & gs &efficient to reproduce the abundance in this work together with (6) by (b) and possibly contributes in IRAS 16293-2422 A \& B by (h). The estimate by (b) might change due to the new estimate of distance to G31. \\
\hline
(16) & $\mathrm{CH_3CH_2OH + OH/Cl \,\longrightarrow\,} \mathrm{CH_2CH_2OH + H_2O/HCl}$ & \multirow{3}{*}{gp} & \multirow{3}{335pt}{predictions by (g) span in a broad range, that cover the abundances observed in this work. Strong dependence on the rate coefficient of reaction (16) and on abundances of reactant OH and ethanol, which should be constrained to reduce uncertainties.} \\
   & $\qquad\qquad\qquad\qquad\qquad\quad\mathrm{CH_3CHOH + H_2O/HCl}$  & \\
(18) & $\mathrm{CH_2CH_2OH + O \,\longrightarrow\, CH_2OHCHO + H}$ & \\
\hline
(19) & $\mathrm{H_2CO+H_2CO \,\longrightarrow\, CH_3OCHO}$ & $\ast$ & proposed by (f), presents high energy barrier, and is more efficient on grain-surface but no calculation made to predict the abundances in star-forming regions on grains surface. (j) simulated the abundance in gas-phase in high-mass protostellar objects, and it is not efficient to reproduce G31 abundance.\\
\hline\hline
(14) &  $\mathrm{ CH_3OCH_3 + OH \,\longrightarrow\, CH_3OCH_2 + H_2O}$ & gp & more efficient than reaction (12) to produce CH$_3$OCH$_2$ (e) needed for reaction (13). \\
\hline
\end{tabular}
\tablefoot{gs stands for grain-surface reactions; gp stands for gas-phase reactions; $\ast$ stands for both grain-surface and gas-phase reactions. References: a) \citet{garrod2013}; b) \citet{coutens2018}; c) \citet{vasy13}; d) \citet{balucani2015}; e) \citet{vas2017}; f) \citet{ahmad2020} ; g) \citet{skouteris2018}; h) \citet{rivilla19}; i) \citet{woods2012}; j) \citet{silva2020}}.  
\end{centering}
\end{sidewaystable*}
\section{Conclusions}
We have presented the data of the GUAPOS survey, an unbiased spectral survey performed with ALMA towards G31.41+0.31, one of the most chemically-rich HMC known, located outside the GC. The survey covered $\sim32\,\mathrm{GHz}$ bandwidth with a spectral resolution of 0.488 MHz and an angular resolution of $1\farcs2$.\newline\indent We have detected for the first time all the 3 isomers of $\mathrm{C_2H_4O_2}$ towards this HMC. This increases to a total number of 3 the number of high-mass sources, outside the GC, where the 3 isomers have been detected. The emission of all the isomers is compact towards the HMC and the most extended emission is from MF, while AA is the most compact. MF is the most abundant of the isomers, confirming what seen in the other sources. G31 shows the largest abundance of AA with respect to the other sources. From the comparison with literature works, there seems to not exist a unique behavior of the 3 isomers both in high- and low-mass sources. \newline\indent
The comparison with chemical models suggests that the MF abundance found in G31 seems not reproducible using only gas-phase routes, thus revealing the need for grain-surface reactions. On the other hand, the scenario for AA and GA is not of straightforward interpretation. Both gas-phase and grain-surface reactions are able to reproduce the observed abundance of GA, even though with uncertainties. In the case of the gas-phase reactions proposed by \citet{skouteris2018} the wide range of predicted abundances for GA is strictly related to the uncertainties in the reaction of $\mathrm{CH_3CH_2OH}$ with the hydroxyl radical, while the predictions from grain-surface reactions in \citet{coutens2018} might change adopting the new estimate of the distance to G31. For AA, the chemical model by \citet{garrod2013} including route (5) on grain-surface is not able to reproduce the high abundance found in G31. On the other hand, the gas-phase route by \citet{skouteris2018} should be further tested to reduce the uncertainties. Moreover, the more compact morphology of these two isomers, with respect to MF, can be explained in both scenarios (grain-surface and gas-phase) as well. To remove some of the uncertainties, laboratory and theoretical studies on reaction (16) are needed, together with an estimate of the abundance of the reactant $\mathrm{CH_3CH_2OH}$ in G31. If as a result, this route is not efficient enough to reproduce the abundance here presented, new formation routes should be investigated. To better understand and constrain whether the main chemical routes that lead to GA are in gas-phase or in grain-surface chemistry, or whether both reactions are needed, a systematic study including the new routes by \citet{skouteris2018}, like the one performed in \citet{quenard2018} for formamide in which the two types of chemistry were switched-off alternatively, is suggested.

\begin{acknowledgements}
The authors thank the Referee Brett A. McGuire for the constructive comments
which helped to improve the quality and readability of the paper.
      The authors wish to thank Holger P.S. Mueller and Christian Endres for inserting acetic acid in the CDMS catalogue.  C.M. and L.C. acknowledge support from the Italian Ministero dell'Istruzione, Universit\'a e Ricerca through the grant Progetti Premiali 2012 - iALMA (CUP C52I13000140001). I.J.-S. has received partial support from the Spanish FEDER (project number ESP2017-86582-C4-1-R). V.M.R. has received funding from the European Union's Horizon 2020 research and innovation programme under the Marie Sk\l{}odowska-Curie grant agreement No 664931. This paper makes use of the following ALMA data: ADS/JAO.ALMA\#2017.1.00501.S. ALMA is a partnership of ESO (representing its member states), NSF (USA) and NINS (Japan), together with NRC (Canada), MOST and ASIAA (Taiwan), and KASI (Republic of Korea), in cooperation with the Republic of Chile. The Joint ALMA Observatory is operated by ESO, AUI/NRAO and NAOJ.
\end{acknowledgements}
\bibliographystyle{aa}
\bibliography{biblio.bib}
\clearpage

\begin{appendix}
\section{Catalogue entries documentation for the three isomers of C$_2$H$_4$O$_2$}
\subsection{Methyl Formate}
The data set used in JPL catalogue is based on \citet{ilyushin2009} and includes data from \citet{brown75, bauder1979, demaison1983, plummer1984, plummer1986, oesterling1999,karakawa2001,odashima2003, ogata2004, carvajal2007, maeda2008, maeda2008b}. More information are available at 
\textbf{https://spec.jpl.nasa.gov/ftp/pub/catalog/doc/d060003.pdf}

\subsection{Acetic Acid} 
The data set used in CDMS catalogue is based on \citet{ilyushin2013} and includes data from \citet{krisher1971, vanEijck1981,demaison1982, demaison1983, vanEijck1983,wlodarczak1988, ilyushin2001, ilyushin2003}. More information are available at 
\textbf{https://cdms.astro.uni-koeln.de/cdms/portal/catalog/1605/\#}

\subsection{Glycolaldehyde}
The data set used in JPL catalogue is based on \citet{marstokk1970, marstokk1973, butler2001, widicus2005, carroll2010}. More information are available at 
\textbf{https://spec.jpl.nasa.gov/ftp/pub/catalog/doc/d060006.pdf}, 
\section{Optical depth of the lines}
Eq. (B.28) of \cite{xclass} gives the optical depth of a line as a function of frequency:
\begin{equation}
\label{inittau}
\tau_{\rm{\nu}} = \frac{c^{2}}{8\pi\nu_{\rm{u,l}}^{2}} A_{\rm{u,l}} N_{\rm{u}} \bigl(\rm{e}^{(E_{\rm{u}}-E_{\rm{l}})/k_{\rm{B}}T_{\rm{ex}}} -1\bigr) \phi_{\rm{u,l}}(\nu)\,,
\end{equation}
where $c$ is the speed of light, $\nu_{\rm{u,l}}$ is the frequency of the transition, $A_{\rm{u,l}}$ is the Einstein's coefficient of the transition, $N_{\rm{u}}$ is the column density of the molecule in the upper state of the transition, $E_{\rm{u}}$ and $E_{\rm{l}}$ are the energy level of the upper and lower states of the transition and $\phi_{\rm{u,l}}(\nu)$ is the normalized line profile (i.e. $\int \phi_{\rm{u,l}}(\nu)\, \rm{d}\nu = 1$). $N_{\rm{u}}$ is related to the total column density of the molecule, $N_{\rm{tot}}$, by the Boltzmann distribution
\begin{equation}
\label{Nu}
N_{\rm{u}} = \frac{g_{\rm{u}}}{Q(T_{\rm{ex}})} N_{\rm{tot}} \mathrm{e}^{-E_{\rm{u}}/k_{\rm{B}}T_{\rm{ex}}} \,,
\end{equation}
where $g_{\rm{u}}$ is the degeneracy of the level \textit{u}, $E_{\rm{u}}$ the corresponding and $Q(T_{\rm{ex}})$ the partition function of the molecular species, calculated at the temperature of excitation $T_{\rm{ex}}$. 
If we assume that $\tau_{\rm{\nu}}$ has a gaussian profile of the type
\begin{equation}
\label{tauprofile}
\tau_{\rm{\nu}} = \tau_{0} \rm{e}^{-\frac{4\ln{2}(\nu-\nu_{\rm{u,l}})^2}{\Delta \nu^2}}    
\end{equation}
where $\tau_{0}$ is the opacity at the line center and $\Delta\nu$ is the FWHM, then the integral of the optical depth over frequency will be related to the central opacity of the line by
\begin{equation}
\label{integraltau}
\int \tau_{\rm{\nu}} \, \rm{d}\nu = \frac{\sqrt{\pi}}{2\sqrt{\ln{2}}}\tau_{0} \Delta\nu\,.
\end{equation}
Integrating Eq. (\ref{inittau}) on $\nu$ and using Eq. (\ref{Nu}) and (\ref{integraltau}) we obtain the espression for the optical depth at the center of the line
\begin{equation}
\label{taufinal}
\tau_{0} = \frac{\sqrt{\ln{2}}}{4\pi\sqrt{\pi}} \frac{c^3}{\nu_{\rm{u,l}}^{3}\Delta \mathrm{v}} \frac{g_{\rm{u}}}{Q(T_{\rm{ex}})} A_{\rm{u,l}} N_{\rm{tot}} \rm{e}^{-E_{\rm{l}}/k_{\rm{B}}T_{\rm{ex}}} \Bigl( 1 - \rm{e}^{h\nu_{\rm{u,l}}/k_{\rm{B}}T_{\rm{ex}}}\Bigr)
\end{equation}
where $\Delta\mathrm{v}$ is the FWHM in velocity units, namely $\Delta\mathrm{v} = \Delta\nu \,c/\nu_{\rm{u,l}}$.

\newpage
\onecolumn
\section{Tables of identified transitions}
\setlength{\tabcolsep}{5pt}
\longtab[1]{
\label{MFused}
\renewcommand{\arraystretch}{1.5}
\begin{longtable}{lcccccrrrcrrrrrcc}
 \caption{Transitions of methyl formate (MF) identified in the spectrum.}
\\ \hline
  &  $\nu$  &$\mathrm{A_{E}}$ & $\mathrm{E_{U}}$ &   $g_{U}$ & &  & $\mathrm{Q_{U}}$ & &  & & & $\mathrm{Q_{L}}$ & &  &$\tau_{0} $\\ 
  &   \small{[MHz]} & \small{[$s^{-1}$]}  & \small{[K]} & & & \small{J} & \small{$\mathrm{K_{a}}$} & \small{$\mathrm{K_{c}}$} & & & \small{J}  & \small{$\mathrm{K_{a}}$} & \small{$\mathrm{K_{c}}$} & & \\
\hline

\endfirsthead
\caption{Continued.}
\\ \hline
  &  $\nu$  & $\mathrm{A_{E}}$  &   $\mathrm{E_{U}}$ &   $g_{U}$ & &  & $\mathrm{Q_{U}}$   & &  & & & $\mathrm{Q_{L}}$ & &  &$\tau_{0} $\\ 
  &   \small{[MHz]} & \small{[$s^{-1}$]}  & \small{[K]} & & & \small{J} & \small{$\mathrm{K_{a}}$} & \small{$\mathrm{K_{c}}$} & & & \small{J}  & \small{$\mathrm{K_{a}}$} & \small{$\mathrm{K_{c}}$} & & \\
\hline

\endhead
\hline
\endfoot
\hline
\endlastfoot
$\mathrm{CH_{3}OCHO \,\, v=0}$&          84233.3350       &  9.202e-07  &   49.80 &    46   & & 11 &    4 &    7 &    A &      & 11 &    3 &    8 &    A &   2.691e-02 \\
 $\mathrm{CH_{3}OCHO \,\, v=0}$&         84449.1690       &  7.957e-06  &   19.00 &    30   & &  7 &    2 &    6 &    E &      &  6 &    2 &    5 &    E &   1.734e-01 \\
 $\mathrm{CH_{3}OCHO \,\, v=0}$&         84454.7540       &  7.961e-06  &   18.98 &    30   & &  7 &    2 &    6 &    A &      &  6 &    2 &    5 &    A &   1.735e-01 \\
 $\mathrm{CH_{3}OCHO \,\, v_{18}=1}$&          85120.4370     &  2.359e-06  &  228.23 &    30   & &  7 &    6 &    1 &    A &      &  6 &    6 &    0 &    A &   1.972e-02 \\
 $\mathrm{CH_{3}OCHO \,\, v_{18}=1}$&          85120.4370     &  2.359e-06  &  228.23 &    30   & &  7 &    6 &    2 &    A &      &  6 &    6 &    1 &    A &   1.972e-02 \\
 $\mathrm{CH_{3}OCHO \,\, v_{18}=1}$&          85157.1350     &  2.372e-06  &  228.15 &    30   & &  7 &    6 &    1 &    E &      &  6 &    6 &    0 &    E &   1.982e-02 \\
 $\mathrm{CH_{3}OCHO \,\, v_{18}=1}$&          85185.4660     &  4.362e-06  &  220.90 &    30   & &  7 &    5 &    3 &    A &      &  6 &    5 &    2 &    A &   3.764e-02 \\
 $\mathrm{CH_{3}OCHO \,\, v_{18}=1}$&          85186.0630     &  4.363e-06  &  220.90 &    30   & &  7 &    5 &    2 &    A &      &  6 &    5 &    1 &    A &   3.764e-02 \\
 $\mathrm{CH_{3}OCHO \,\, v_{18}=1}$&          85286.2240     &  4.402e-06  &  220.68 &    30   & &  7 &    5 &    2 &    E &      &  6 &    5 &    1 &    E &   3.793e-02 \\
 $\mathrm{CH_{3}OCHO \,\, v_{18}=1}$&          85327.0290     &  6.026e-06  &  214.91 &    30   & &  7 &    4 &    4 &    A &      &  6 &    4 &    3 &    A &   5.323e-02 \\
 $\mathrm{CH_{3}OCHO \,\, v_{18}=1}$&          85360.7640     &  6.032e-06  &  214.91 &    30   & &  7 &    4 &    3 &    A &      &  6 &    4 &    2 &    A &   5.325e-02 \\
 $\mathrm{CH_{3}OCHO \,\, v_{18}=1}$&          85371.7300     &  7.308e-06  &  210.26 &    30   & &  7 &    3 &    5 &    A &      &  6 &    3 &    4 &    A &   6.586e-02 \\
 $\mathrm{CH_{3}OCHO \,\, v_{18}=1}$&          85506.2190     &  6.105e-06  &  214.56 &    30   & &  7 &    4 &    3 &    E &      &  6 &    4 &    2 &    E &   5.379e-02 \\
 $\mathrm{CH_{3}OCHO \,\, v_{18}=1}$&          85553.3800     &  4.439e-06  &  220.04 &    30   & &  7 &    5 &    3 &    E &      &  6 &    5 &    2 &    E &   3.812e-02 \\
 $\mathrm{CH_{3}OCHO \,\, v=0}$&         85638.3290       &  7.133e-07  &    8.54 &    18   & &  4 &    2 &    3 &    E &      &  3 &    1 &    2 &    E &   9.510e-03 \\
 $\mathrm{CH_{3}OCHO \,\, v=0}$&         85655.8030       &  7.142e-07  &    8.52 &    18   & &  4 &    2 &    3 &    A &      &  3 &    1 &    2 &    A &   9.519e-03 \\
 $\mathrm{CH_{3}OCHO \,\, v=0}$&         85773.3990       &  1.324e-06  &  155.82 &    86   & & 21 &    5 &    16&    A &      &  21&    4 &    17&    A &   4.333e-02 \\
 $\mathrm{CH_{3}OCHO \,\, v=0}$&         85785.3420       &  1.298e-06  &  142.79 &    82   & & 20 &    5 &    15&    A &      &  20&    4 &    16&    A &   4.294e-02 \\
 $\mathrm{CH_{3}OCHO \,\, v=0}$&         85919.2090       &  2.433e-06  &   40.44 &    30   & &  7 &    6 &    1 &    E &      &  6 &    6 &    0 &    E &   4.652e-02 \\
 $\mathrm{CH_{3}OCHO \,\, v=0}$&         86021.1240       &  4.507e-06  &   33.13 &    30   & &  7 &    5 &    2 &    E &      &  6 &    5 &    1 &    E &   8.884e-02 \\
 $\mathrm{CH_{3}OCHO \,\, v=0}$&         86027.7230       &  4.507e-06  &   33.12 &    30   & &  7 &    5 &    3 &    E &      &  6 &    5 &    2 &    E &   8.884e-02 \\
 $\mathrm{CH_{3}OCHO \,\, v=0}$&         86029.4420       &  4.509e-06  &   33.11 &    30   & &  7 &    5 &    3 &    A &      &  6 &    5 &    2 &    A &   8.888e-02 \\
 $\mathrm{CH_{3}OCHO \,\, v=0}$&         86030.1860       &  4.509e-06  &   33.11 &    30   & &  7 &    5 &    2 &    A &      &  6 &    5 &    1 &    A &   8.888e-02 \\
 $\mathrm{CH_{3}OCHO \,\, v_{18}=1}$&          86034.0130     &  7.532e-06  &  209.82 &    30   & &  7 &    3 &    4 &    E &      &  6 &    3 &    3 &    E &   6.697e-02 \\
 $\mathrm{CH_{3}OCHO \,\, v_{18}=1}$&          86155.0780     &  7.513e-06  &  210.32 &    30   & &  7 &    3 &    4 &    A &      &  6 &    3 &    3 &    A &   6.647e-02 \\
 $\mathrm{CH_{3}OCHO \,\, v_{18}=1}$&          86172.7060     &  7.513e-06  &  209.41 &    30   & &  7 &    3 &    5 &    E &      &  6 &    3 &    4 &    E &   6.671e-02 \\
 $\mathrm{CH_{3}OCHO \,\, v=0}$&         86210.0570       &  6.236e-06  &   27.15 &    30   & &  7 &    4 &    4 &    A &      &  6 &    4 &    3 &    A &   1.258e-01 \\
 $\mathrm{CH_{3}OCHO \,\, v=0}$&         86223.6550       &  6.202e-06  &   27.17 &    30   & &  7 &    4 &    3 &    E &      &  6 &    4 &    2 &    E &   1.250e-01 \\
 $\mathrm{CH_{3}OCHO \,\, v=0}$&         86224.1600       &  6.202e-06  &   27.16 &    30   & &  7 &    4 &    4 &    E &      &  6 &    4 &    3 &    E &   1.250e-01 \\
 $\mathrm{CH_{3}OCHO \,\, v=0}$&         86250.5520       &  6.245e-06  &   27.15 &    30   & &  7 &    4 &    3 &    A &      &  6 &    4 &    2 &    A &   1.258e-01 \\
 $\mathrm{CH_{3}OCHO \,\, v=0}$&         86265.7960       &  7.568e-06  &   22.51 &    30   & &  7 &    3 &    5 &    A &      &  6 &    3 &    4 &    A &   1.556e-01 \\
 $\mathrm{CH_{3}OCHO \,\, v=0}$&         86268.7390       &  7.501e-06  &   22.53 &    30   & &  7 &    3 &    5 &    E &      &  6 &    3 &    4 &    E &   1.542e-01 \\
 $\mathrm{CH_{3}OCHO \,\, v=0}$&         87143.2820       &  7.734e-06  &   22.60 &    30   & &  7 &    3 &    4 &    E &      &     6 &    3 &    3 &    E &   1.558e-01 \\
 $\mathrm{CH_{3}OCHO \,\, v=0}$&         87161.2850       &  7.808e-06  &   22.58 &    30   & &  7 &    3 &    4 &    A &      &     6 &    3 &    3 &    A &   1.573e-01 \\
 $\mathrm{CH_{3}OCHO \,\, v_{18}=1}$&        87419.3080     &  1.370e-06  &  329.27 &    82   & &  20&    5 &    15&    A &      &  20 &    4 &    16 &    A &   1.884e-02 \\
 $\mathrm{CH_{3}OCHO \,\, v=0}$&         87552.3790       &  4.968e-07  &   34.46 &    42   & &  10&    2 &    9 &    A &      &  10 &    1 &    10 &    A &   1.316e-02 \\
 $\mathrm{CH_{3}OCHO \,\, v=0}$&         87766.3820       &  1.331e-06  &   20.08 &    34   & &  8 &    0 &    8 &    E &   &    7 &    1 &    7 &    E &   3.031e-02 \\
 $\mathrm{CH_{3}OCHO \,\, v=0}$&         87769.0350       &  1.330e-06  &   20.06 &    34   & &   8 &    0 &    8 &    A &   &     7 &    1 &    7 &    A &   3.029e-02 \\
 $\mathrm{CH_{3}OCHO \,\, v_{18}=1}$&        88030.8000$^{\mathbf{\ast}}$     &  1.341e-06  &  329.14 &    82   & & 20&    5 &    15&    E &     &  20 &    4 &    16 &    E &   1.819e-02 \\
 $\mathrm{CH_{3}OCHO \,\, v=0}$&         88053.9730       &  1.344e-06  &  130.44 &    78   & & 19&    5 &    14&    E &     &  19 &    4 &    15 &    E &   4.244e-02 \\
 $\mathrm{CH_{3}OCHO \,\, v=0}$&         88054.4610       &  1.344e-06  &  130.44 &    78   & & 19&    5 &    14 &    A &      &  19 &    4 &    15 &    A &   4.242e-02 \\
 $\mathrm{CH_{3}OCHO \,\, v=0}$&         88116.0500       &  2.630e-07  &   34.46 &    42   & & 10&    2 &    9 &    A &     &  10 &    0 &    10 &    A &   6.879e-03 \\
 $\mathrm{CH_{3}OCHO \,\, v=0}$&         88175.5060       &  9.761e-07  &   43.22 &    42   & & 10&    4 &    6 &    E &     &  10 &    3 &    7 &    E &   2.451e-02 \\
 $\mathrm{CH_{3}OCHO \,\, v=0}$&         88180.4240       &  9.888e-07  &   43.21 &    42   & & 10&    4 &    6 &    A &     &  10 &    3 &    7 &    A &   2.482e-02 \\
 $\mathbf{CH_{3}OCHO \,\, v_{18}=1}$&           \textbf{88220.7530}     &  \textbf{9.621e-06}   &  \textbf{204.92} &     \textbf{30}    & &   \textbf{7} &    \textbf{1} &   \textbf{6} &    \textbf{E} &   & \textbf{6} &    \textbf{1} &    \textbf{5} &    \textbf{E} &   \textbf{8.320e-02} \\
 $\mathrm{CH_{3}OCHO \,\, v=0}$&         88358.4870       &  1.438e-06  &  169.52 &    90   & & 22 &    5 &    17 &    A &     &  22 &    4 &    18 &    A &   4.361e-02 \\
 $\mathrm{CH_{3}OCHO \,\, v=0}$&         88686.8870       &  8.591e-07  &   44.97 &    46   & & 11 &    3 &    9 &    E &      &  11 &    2 &    10 &    E &   2.317e-02 \\
 $\mathrm{CH_{3}OCHO \,\, v_{18}=1}$&          88720.3000     &  1.440e-06  &  355.62 &    90   & & 22 &    5 &    17 &    E &     &  22 &    4 &    18 &    E &   1.875e-02 \\
 $\mathrm{CH_{3}OCHO \,\, v_{18}=1}$&          88720.3000     &  1.440e-06  &  355.62 &    90   & & 22 &    5 &    17 &    E &     &  22 &    4 &    18 &    E &   1.875e-02 \\
 $\mathrm{CH_{3}OCHO \,\, v=0}$&         88723.2680       &  8.594e-07  &   44.96 &    46   & & 11 &    3 &    9 &    A &      &  11 &    2 &    10 &    A &   2.316e-02 \\
  $\mathbf{CH_{3}OCHO \,\, v_{18}=1}$&       \textbf{88770.8680}    &  \textbf{9.942e-06}   &  \textbf{207.91} &     \textbf{34}    & &   \textbf{8} &    \textbf{1} &   \textbf{8} &    \textbf{A} &   &    \textbf{7} &    \textbf{1} &    \textbf{7} &    \textbf{A} &   \textbf{9.495e-02} \\
 $\mathrm{CH_{3}OCHO \,\, v=0}$&         88843.1870       &  9.816e-06  &   17.96 &    30   & &   7 &    1 &    6 &    E &   &     6 &    1 &    5 &    E &   1.943e-01 \\
 $\mathbf{CH_{3}OCHO \,\, v=0}$&         \textbf{88851.6070}      &  \textbf{9.820e-06}   &   \textbf{17.94} &     \textbf{30}    & &   \textbf{7} &    \textbf{1} &    \textbf{6} &    \textbf{A} &   &     \textbf{6} &    \textbf{1} &    \textbf{5} &    \textbf{A} &   \textbf{1.944e-01} \\
 $\mathrm{CH_{3}OCHO \,\, v_{18}=1}$&          88862.4140     &  1.003e-05  &  207.12 &    34   & &   8 &    1 &    8 &    E &   &     7 &    1 &    7 &    E &   9.597e-02 \\
 $\mathrm{CH_{3}OCHO \,\, v_{18}=1}$&          89140.3750     &  9.382e-06  &  206.70 &    30   & &   7 &    2 &    5 &    E &   &     6 &    2 &    4 &    E &   7.883e-02 \\
 $\mathrm{CH_{3}OCHO \,\, v_{18}=1}$&          89731.6950     &  1.030e-05  &  207.83 &    34   & &   8 &    0 &    8 &    A &   &     7 &    0 &    7 &    A &   9.635e-02 \\
 $\mathrm{CH_{3}OCHO \,\, v=0}$&         89796.9180       &  5.046e-07  &   40.38 &    46   & & 11 &    1 &    10 &    A &     &  11 &    0 &    11 &    A &   1.355e-02 \\
 $\mathrm{CH_{3}OCHO \,\, v_{18}=1}$&          89829.7040     &  1.039e-05  &  207.04 &    34   & &   8 &    0 &    8 &    E &   &     7 &    0 &    7 &    E &   9.734e-02 \\
 $\mathrm{CH_{3}OCHO \,\, v=0}$&         90145.7230       &  9.741e-06  &   19.68 &    30   & &   7 &    2 &    5 &    E &   &     6 &    2 &    4 &    E &   1.859e-01 \\
 $\mathbf{CH_{3}OCHO \,\, v=0}$&         \textbf{90227.6590}      &  \textbf{1.050e-05}   &   \textbf{20.08} &     \textbf{34}    & &   \textbf{8} &    \textbf{0} &    \textbf{8} &    \textbf{E} &   &     \textbf{7} &    \textbf{0} &    \textbf{7} &   \textbf{E} &   \textbf{2.264e-01} \\
$\mathbf{CH_{3}OCHO \,\, v=0}$&          \textbf{90229.6240}      &  \textbf{1.051e-05}   &   \textbf{20.06} &     \textbf{34}    & &   \textbf{8} &    \textbf{0} &    \textbf{8} &    \textbf{A} &   &     \textbf{7} &    \textbf{0} &    \textbf{7} &    \textbf{A} &   \textbf{2.265e-01} \\
 $\mathrm{CH_{3}OCHO \,\, v=0}$&         91356.7660       &  1.034e-06  &   37.24 &    38   & &   9 &    4 &    5 &    A &   &     9 &    3 &    6 &    A &   2.248e-02 \\
 $\mathrm{CH_{3}OCHO \,\, v=0}$&         91366.4950       &  9.709e-07  &   37.26 &    38   & &   9 &    4 &    5 &    E &   &     9 &    3 &    6 &    E &   2.111e-02 \\
 $\mathrm{CH_{3}OCHO \,\, v=0}$&         91381.7380       &  4.084e-07  &   66.97 &    58   & & 14 &    2 &    12 &    E &     &  14 &    2 &    13 &    E &   1.185e-02 \\
 $\mathrm{CH_{3}OCHO \,\, v_{18}=1}$&        91531.2900       &  1.495e-06  &  207.12 &    34   & & 8 &    1 &    8 &    E &     &  7 &    0 &    7 &    E &   1.348e-02 \\
 $\mathrm{CH_{3}OCHO \,\, v=0}$&         91775.9350       &  1.540e-06  &   20.15 &    34   & & 8 &    1 &    8 &    E &     &  7 &    0 &    7 &    E &   3.207e-02 \\
 $\mathrm{CH_{3}OCHO \,\, v=0}$&         91776.8860       &  1.384e-06  &  138.67 &    82   & & 20 &    4 &    16 &    E &     &  20 &    3 &    17 &    E &   4.076e-02 \\
 $\mathrm{CH_{3}OCHO \,\, v=0}$&         91777.2300       &  1.539e-06  &   20.14 &    34   & & 8 &    1 &    8 &    A &     &  7 &    0 &    7 &    A &   3.205e-02 \\
 $\mathrm{CH_{3}OCHO \,\, v=0}$&         91825.2470       &  1.385e-06  &  138.67 &    82   & & 20 &    4 &    16 &    A &     &  20 &    3 &    17 &    A &   4.075e-02 \\
 $\mathrm{CH_{3}OCHO \,\, v=0}$&         92073.1010       &  1.448e-06  &  118.78 &    74   & & 18 &    5 &    13 &    A &     &  18 &    4 &    14 &    A &   4.181e-02 \\
  $\mathrm{CH_{3}OCHO \,\, v=0}$&          92074.0200       &  1.448e-06  &  118.79 &    74   & & 18 &    5 &    13 &    E &     &  18 &    4 &    14 &    E &   4.182e-02 \\
$\mathrm{CH_{3}OCHO \,\, v=0}$&          92884.2380       &  1.186e-06  &   99.73 &    70   & & 17 &    3 &    14 &    E &     &  17 &    2 &    15 &    E &   3.471e-02 \\
$\mathrm{CH_{3}OCHO \,\, v_{18}=1}$&       92936.1810     &  1.622e-06  &  369.85 &    94   & & 23 &    5 &    18 &    E &     &  23 &    4 &    19 &    E &   1.885e-02 \\
$\mathrm{CH_{3}OCHO \,\, v=0}$&          92940.3280       &  1.187e-06  &   99.72 &    70   & & 17 &    3 &    14 &    A &     &  17 &    2 &    15 &    A &   3.469e-02 \\
$\mathrm{CH_{3}OCHO \,\, v=0}$&          93205.7410       &  9.163e-07  &   66.97 &    58   & & 14 &    2 &    12 &    E &     &  14 &    1 &    13 &    E &   2.556e-02 \\
$\mathrm{CH_{3}OCHO \,\, v=0}$&          93261.7490       &  9.170e-07  &   66.96 &    58   & & 14 &    2 &    12 &    A &     &  14 &    1 &    13 &    A &   2.555e-02 \\
$\mathrm{CH_{3}OCHO \,\, v=0}$&          93660.0770       &  1.044e-06  &   31.89 &    34   & & 8 &    4 &    4 &    A &     &  8 &    3 &    5 &    A &   1.982e-02 \\
$\mathrm{CH_{3}OCHO \,\, v=0}$&          93701.1910       &  8.322e-07  &   31.91 &    34   & & 8 &    4 &    4 &    E &     &  8 &    3 &    5 &    E &   1.577e-02 \\
$\mathrm{CH_{3}OCHO \,\, v=0}$&          93702.6470       &  1.658e-06  &  183.88 &    94   & & 23 &    5 &    18 &    A &     &  23 &    4 &    19 &    A &   4.381e-02 \\
$\mathrm{CH_{3}OCHO \,\, v=0}$&          94387.5850       &  8.685e-07  &   28.12 &    38   & & 9 &    1 &    8 &    A &     &  8 &    2 &    7 &    A &   1.844e-02 \\
 $\mathrm{CH_{3}OCHO \,\, v=0}$&         94626.8600       &  1.000e-06  &   52.02 &    50   & & 12 &    3 &    10 &    E &     &  12 &    2 &    11 &    E &   2.498e-02 \\
 $\mathrm{CH_{3}OCHO \,\, v=0}$&         95175.9660       &  1.013e-06  &   27.15 &    30   & & 7 &    4 &    3 &    A &     &  7 &    3 &    4 &    A &   1.678e-02 \\
 $\mathrm{CH_{3}OCHO \,\, v_{18}=1}$&          95242.0050     &  1.169e-05  &  211.31 &    34   & & 8 &    2 &    7 &    A &     &  7 &    2 &    6 &    A &   9.565e-02 \\
 $\mathrm{CH_{3}OCHO \,\, v_{18}=1}$&          95902.1090     &  1.423e-06  &  305.40 &    74   & & 18 &    5 &    13 &    E &     &  18 &    4 &    14 &    E &   1.635e-02 \\
 $\mathrm{CH_{3}OCHO \,\, v=0}$&         96070.7250       &  1.204e-05  &   23.61 &    34   & & 8 &    2 &    7 &    E &     &  7 &    2 &    6 &    E &   2.255e-01 \\
 $\mathrm{CH_{3}OCHO \,\, v=0}$&         96076.8450       &  1.205e-05  &   23.59 &    34   & & 8 &    2 &    7 &    A &     &  7 &    2 &    6 &    A &   2.256e-01 \\
 $\mathrm{CH_{3}OCHO \,\, v=0}$&         96107.1720       &  6.072e-07  &   40.70 &    46   & & 11 &    2 &    10 &    A &     &  11 &    1 &    11 &    A &   1.423e-02 \\
 $\mathrm{CH_{3}OCHO \,\, v=0}$&         96507.8820       &  7.010e-07  &   27.16 &    30   & & 7 &    4 &    4 &    E &     &  7 &    3 &    5 &    E &   1.129e-02 \\
 $\mathrm{CH_{3}OCHO \,\, v=0}$&         96613.0200       &  9.009e-07  &   31.90 &    34   & & 8 &    4 &    5 &    E &     &  8 &    3 &    6 &    E &   1.606e-02 \\
 $\mathrm{CH_{3}OCHO \,\, v=0}$&         96637.8110       &  1.054e-06  &   27.15 &    30   & & 7 &    4 &    4 &    A &     &  7 &    3 &    5 &    A &   1.693e-02 \\
 $\mathrm{CH_{3}OCHO \,\, v=0}$&         96709.2590       &  1.134e-06  &   31.89 &    34   & & 8 &    4 &    5 &    A &     &  8 &    3 &    6 &    A &   2.018e-02 \\
 $\mathrm{CH_{3}OCHO \,\, v=0}$&         96956.7440       &  1.128e-06  &   37.23 &    38   & & 9 &    4 &    6 &    E &     &  9 &    3 &    7 &    E &   2.179e-02 \\
 $\mathrm{CH_{3}OCHO \,\, v=0}$&         97018.1250       &  1.202e-06  &   37.22 &    38   & & 9 &    4 &    6 &    A &     &  9 &    3 &    7 &    A &   2.321e-02 \\
 $\mathrm{CH_{3}OCHO \,\, v=0}$&         97199.1110       &  1.596e-06  &  107.82 &    70   & & 17 &    5 &    12 &    E &     &  17 &    4 &    13 &    E &   4.113e-02 \\
 $\mathrm{CH_{3}OCHO \,\, v=0}$&         97199.2220       &  1.595e-06  &  107.81 &    70   & & 17 &    5 &    12 &    A &     &  17 &    4 &    13 &    A &   4.112e-02 \\
 $\mathrm{CH_{3}OCHO \,\, v=0}$&         97318.5920       &  7.007e-07  &    8.64 &    18   & & 4 &    2 &    2 &    E &     &  3 &    1 &    3 &    E &   7.240e-03 \\
 $\mathrm{CH_{3}OCHO \,\, v_{18}=1}$&          97350.4170     &  5.868e-06  &  232.90 &    34   & & 8 &    6 &    2 &    A &     &  7 &    6 &    1 &    A &   4.168e-02 \\
 $\mathrm{CH_{3}OCHO \,\, v_{18}=1}$&          97350.4170     &  5.868e-06  &  232.90 &    34   & & 8 &    6 &    3 &    A &     &  7 &    6 &    2 &    A &   4.168e-02 \\
 $\mathrm{CH_{3}OCHO \,\, v_{18}=1}$&          97457.9670     &  8.196e-06  &  225.57 &    34   & & 8 &    5 &    4 &    A &     &  7 &    5 &    3 &    A &   6.003e-02 \\
 $\mathrm{CH_{3}OCHO \,\, v_{18}=1}$&          97597.1610     &  1.158e-05  &  214.94 &    34   & & 8 &    3 &    6 &    A &     &  7 &    3 &    5 &    A &   8.879e-02 \\
 $\mathrm{CH_{3}OCHO \,\, v_{18}=1}$&          97661.4010     &  1.014e-05  &  219.60 &    34   & & 8 &    4 &    5 &    A &     &  7 &    4 &    4 &    A &   7.602e-02 \\
 $\mathrm{CH_{3}OCHO \,\, v=0}$&         97694.2600       &  1.271e-06  &   43.16 &    42   & & 10 &    4 &    7 &    A &      &  10 &    3 &    8 &    A &   2.603e-02 \\
 $\mathrm{CH_{3}OCHO \,\, v=0}$&         97735.2450       &  1.207e-07  &  431.09 &   114   & & 28 &    17 &    11 &    E &   &    29 &    16 &    13 &    E &   1.168e-03 \\
 $\mathrm{CH_{3}OCHO \,\, v_{18}=1}$&          97738.7380     &  5.966e-06  &  232.10 &    34   & & 8 &    6 &    3 &    E &     &  7 &    6 &    2 &    E &   4.219e-02 \\
 $\mathrm{CH_{3}OCHO \,\, v_{18}=1}$&          97885.6630     &  8.342e-06  &  224.73 &    34   & & 8 &    5 &    4 &    E &     &  7 &    5 &    3 &    E &   6.080e-02 \\
$\mathrm{CH_{3}OCHO \,\, v=0}$&          98182.3360       &  3.234e-06  &   53.78 &    34   & & 8 &    7 &    1 &    E &     &  7 &    7 &    0 &    E &   5.061e-02 \\
$\mathrm{CH_{3}OCHO \,\, v=0}$&          98190.6580       &  3.236e-06  &   53.76 &    34   & & 8 &    7 &    2 &    A &     &  7 &    7 &    1 &    A &   5.063e-02 \\
$\mathrm{CH_{3}OCHO \,\, v=0}$&          98190.6580       &  3.236e-06  &   53.76 &    34   & & 8 &    7 &    1 &    A &     &  7 &    7 &    0 &    A &   5.063e-02 \\
$\mathrm{CH_{3}OCHO \,\, v=0}$&          98191.4600       &  3.235e-06  &   53.76 &    34   & & 8 &    7 &    2 &    E &     &  7 &    7 &    1 &    E &   5.062e-02 \\
$\mathrm{CH_{3}OCHO \,\, v=0}$&          98270.5010       &  6.053e-06  &   45.15 &    34   & & 8 &    6 &    2 &    E &     &  7 &    6 &    1 &    E &   9.830e-02 \\
$\mathrm{CH_{3}OCHO \,\, v=0}$&          98278.9210       &  6.055e-06  &   45.13 &    34   & & 8 &    6 &    3 &    E &     &  7 &    6 &    2 &    E &   9.831e-02 \\
$\mathrm{CH_{3}OCHO \,\, v=0}$&          98279.7620       &  6.056e-06  &   45.13 &    34   & & 8 &    6 &    3 &    A &     &  7 &    6 &    2 &    A &   9.833e-02 \\
$\mathrm{CH_{3}OCHO \,\, v=0}$&          98279.7620       &  6.056e-06  &   45.13 &    34   & & 8 &    6 &    2 &    A &     &  7 &    6 &    1 &    A &   9.833e-02 \\
$\mathrm{CH_{3}OCHO \,\, v_{18}=1}$&         98423.1650     &  1.990e-06  &  212.64 &    38   & & 9 &    0 &    9 &    A &     &  8 &    1 &    8 &    A &   1.693e-02 \\
$\mathrm{CH_{3}OCHO \,\, v=0}$&          98424.2070       &  8.469e-06  &   37.86 &    34   & & 8 &    5 &    3 &    E &     &  7 &    5 &    2 &    E &   1.417e-01 \\
$\mathrm{CH_{3}OCHO \,\, v=0}$&          98431.8030       &  8.471e-06  &   37.84 &    34   & & 8 &    5 &    4 &    E &     &  7 &    5 &    3 &    E &   1.417e-01 \\
$\mathrm{CH_{3}OCHO \,\, v=0}$&          98432.7600       &  8.473e-06  &   37.84 &    34   & & 8 &    5 &    4 &    A &     &  7 &    5 &    3 &    A &   1.417e-01 \\
$\mathrm{CH_{3}OCHO \,\, v=0}$&          98432.7600       &  8.473e-06  &   37.84 &    34   & & 8 &    5 &    4 &    A &     &  7 &    5 &    3 &    A &   1.417e-01 \\
$\mathrm{CH_{3}OCHO \,\, v=0}$&          98435.8020       &  8.473e-06  &   37.84 &    34   & & 8 &    5 &    3 &    A &     &  7 &    5 &    2 &    A &   1.417e-01 \\
$\mathrm{CH_{3}OCHO \,\, v_{18}=1}$&         98682.4210     &  1.189e-05  &  214.15 &    34   & & 8 &    3 &    6 &    E &     &  7 &    3 &    5 &    E &   8.949e-02 \\
 $\mathrm{CH_{3}OCHO \,\, v=0}$&         98682.6150       &  1.050e-05  &   31.89 &    34   & & 8 &    4 &    5 &    A &     &  7 &    4 &    4 &    A &   1.796e-01 \\
 $\mathrm{CH_{3}OCHO \,\, v=0}$&         98712.0010       &  1.016e-05  &   31.90 &    34   & & 8 &    4 &    5 &    E &     &  7 &    4 &    4 &    E &   1.737e-01 \\
 $\mathrm{CH_{3}OCHO \,\, v=0}$&         98747.9060       &  1.017e-05  &   31.91 &    34   & & 8 &    4 &    4 &    E &     &  7 &    4 &    3 &    E &   1.738e-01 \\
 $\mathrm{CH_{3}OCHO \,\, v=0}$&         98792.2890       &  1.053e-05  &   31.89 &    34   & & 8 &    4 &    4 &    A &     &  7 &    4 &    3 &    A &   1.798e-01 \\
  $\mathbf{CH_{3}OCHO \,\, v_{18}=1}$&          \textbf{98815.2900}     &  \textbf{1.204e-05}   &  \textbf{214.56} &     \textbf{34}    & & \textbf{8} &    \textbf{3} &    \textbf{5} &    \textbf{E} &     &  \textbf{7} &    \textbf{3} &    \textbf{4} &    \textbf{E} &   \textbf{9.017e-02} \\
 $\mathrm{CH_{3}OCHO \,\, v=0}$&         98839.5220       &  1.343e-06  &   49.71 &    46   & & 11 &    4 &    8 &    E &      &  11 &    3 &    9 &    E &   2.859e-02 \\
 $\mathrm{CH_{3}OCHO \,\, v_{18}=1}$&          99089.5180     &  1.213e-05  &  215.08 &    34   & & 8 &    3 &    5 &    A &     &  7 &    3 &    4 &    A &   9.016e-02 \\
 $\mathrm{CH_{3}OCHO \,\, v=0}$&         99133.2720       &  2.004e-06  &   24.91 &    38   & & 9 &    0 &    9 &    E &     &  8 &    1 &    8 &    E &   3.917e-02 \\
 $\mathrm{CH_{3}OCHO \,\, v=0}$&         99133.2720       &  2.004e-06  &   24.91 &    38   & & 9 &    0 &    9 &    E &     &  8 &    1 &    8 &    E &   3.917e-02 \\
 $\mathrm{CH_{3}OCHO \,\, v=0}$&         99135.7620       &  2.003e-06  &   24.89 &    38   & & 9 &    0 &    9 &    A &     &  8 &    1 &    8 &    A &   3.915e-02 \\
 $\mathrm{CH_{3}OCHO \,\, v=0}$&         99488.2149$^{\mathbf{\ast}}$       &  5.828e-08  &  257.08 &   114   & & 28 &    4 &    24 &    E &     &  27 &    6 &    21 &    E &   1.192e-03 \\
 $\mathrm{CH_{3}OCHO \,\, v_{18}=1}$&        99575.5480     &  1.391e-05  &  210.43 &    34   & & 8 &    1 &    7 &    A &     &  7 &    1 &    6 &    A &   1.045e-01 \\
 $\mathrm{CH_{3}OCHO \,\, v_{18}=1}$&        99577.4190     &  1.425e-05  &  211.90 &    38   & & 9 &    1 &    9 &    E &     &  8 &    1 &    8 &    E &   1.189e-01 \\
  $\mathbf{CH_{3}OCHO \,\, v_{18}=1}$&        \textbf{99869.1020}     &  \textbf{1.409e-05}&   \textbf{209.71} &     \textbf{34}    & & \textbf{8} &    \textbf{1} &    \textbf{7} &    \textbf{E} &     &  \textbf{7} &    \textbf{1} &    \textbf{6} &    \textbf{E} &   \textbf{1.056e-01} \\
 $\mathbf{CH_{3}OCHO \,\, v_{18}=1}$&        \textbf{100136.9130}     &  \textbf{1.443e-05}   &  \textbf{212.64} &     \textbf{38}    & & \textbf{9} &    \textbf{0} &    \textbf{9} &    \textbf{A} &     &  \textbf{8} &    \textbf{0} &    \textbf{8} &    \textbf{A} &   \textbf{1.186e-01} \\
  $\mathbf{CH_{3}OCHO \,\, v_{18}=1}$&       \textbf{100226.6840}     &  \textbf{1.455e-05}   &  \textbf{211.85} &     \textbf{38}    & & \textbf{9} &    \textbf{0} &    \textbf{9} &   \textbf{E} &      &  \textbf{8} &    \textbf{0} &    \textbf{8} &   \textbf{E} &   \textbf{1.199e-01} \\
 $\mathrm{CH_{3}OCHO \,\, v=0}$&        100294.6040       &  1.261e-05  &   27.41 &    34   & & 8 &    3 &    5 &    E &     &  7 &    3 &    4 &    E &   2.130e-01 \\
 $\mathrm{CH_{3}OCHO \,\, v=0}$&        100482.2410       &  1.433e-05  &   22.78 &    34   & & 8 &    1 &    7 &    E &     &  7 &    1 &    6 &    E &   2.463e-01 \\
 $\mathrm{CH_{3}OCHO \,\, v=0}$&        100681.5450       &  1.471e-05  &   24.91 &    38   & & 9 &    0 &    9 &    E &     &  8 &    0 &    8 &    E &   2.788e-01 \\
 $\mathrm{CH_{3}OCHO \,\, v=0}$&        100693.1270       &  1.439e-06  &   56.84 &    50   & & 12 &    4 &    9 &    A &      &  12 &    3 &    10 &    A &   3.108e-02 \\
$\mathrm{CH_{3}OCHO \,\, v=0}$&         101305.5060       &  2.152e-06  &  219.30 &   102   & & 25 &    6 &    19 &    E &     &  25 &    5 &    20 &    E &   4.504e-02 \\
$\mathrm{CH_{3}OCHO \,\, v_{18}=1}$&        101318.9660     &  1.164e-06  &  246.45 &    54   & & 13 &    3 &    11 &    E &     &  13 &    2 &    12 &    E &   1.141e-02 \\
$\mathrm{CH_{3}OCHO \,\, v=0}$&         101370.5050       &  1.172e-06  &   59.64 &    54   & & 13 &    3 &    11 &    E &     &  13 &    2 &    12 &    E &   2.662e-02 \\
$\mathrm{CH_{3}OCHO \,\, v=0}$&         101414.7460       &  1.172e-06  &   59.63 &    54   & & 13 &    3 &    11 &    A &     &  13 &    2 &    12 &    A &   2.662e-02 \\
$\mathrm{CH_{3}OCHO \,\, v=0}$&         101626.8840       &  2.171e-06  &   24.96 &    38   & & 9 &    1 &    9 &    E &     &  8 &    0 &    8 &    E &   4.037e-02 \\
$\mathrm{CH_{3}OCHO \,\, v=0}$&         101628.1490       &  2.169e-06  &   24.94 &    38   & & 9 &    1 &    9 &    A &     &  8 &    0 &    8 &    A &   4.034e-02 \\
$\mathrm{CH_{3}OCHO \,\, v=0}$&         101728.8600       &  2.009e-06  &  198.86 &    98   & & 24 &    5 &    19 &  E &     &  24 &    4 &    20 &    E &   4.394e-02 \\
$\mathrm{CH_{3}OCHO \,\, v=0}$&         101771.9320       &  2.011e-06  &  198.86 &    98   & & 24 &    5 &    19 &  A &     &  24&    4 &    20 &    A &   4.393e-02 \\
$\mathrm{CH_{3}OCHO \,\, v=0}$&         101839.8940       &  2.211e-06  &  235.46 &   106   & & 26 &    6 &    20 &  E &     &  26&    5 &    21 &    E &   4.425e-02 \\
$\mathrm{CH_{3}OCHO \,\, v=0}$&         102104.9920       &  3.166e-07  &  174.45 &    90   & & 22 &    6 &    17 &  E &     &  21&    7 &    14 &    E &   7.045e-03 \\
$\mathrm{CH_{3}OCHO \,\, v=0}$&         102117.7810       &  3.237e-07  &  174.45 &    90   & & 22 &    6 &    17 &  A &     &  21&    7 &    14 &    A &   7.200e-03 \\
$\mathbf{CH_{3}OCHO \,\, v_{18}=1}$&      \textbf{102179.5100}      &  \textbf{1.455e-05}   &  \textbf{212.27} &     \textbf{34}    & & \textbf{8} &     \textbf{2} &    \textbf{6} &   \textbf{A} &     &  \textbf{7} &    \textbf{2} &    \textbf{5} &    \textbf{A} &   \textbf{1.030e-01} \\
$\mathrm{CH_{3}OCHO \,\, v_{18}=1}$&      102483.2590$^{\mathbf{\ast}}$       &  2.267e-06  &  421.00 &   106   & & 26 &    6 &    20 &  A &     &  26 &    5 &    21 &    A &   1.942e-02 \\
$\mathbf{CH_{3}OCHO \,\, v_{18}=1}$&       \textbf{102503.1050}     &  \textbf{1.476e-05}   &  \textbf{211.62} &     \textbf{34}    & & \textbf{8} &     \textbf{2} &    \textbf{6} &   \textbf{E} &     &  \textbf{7} &    \textbf{2} &    \textbf{5} &    \textbf{E} &   \textbf{1.042e-01} \\
$\mathrm{CH_{3}OCHO \,\, v=0}$&         102736.8690       &  1.770e-06  &   97.51 &    66   & & 16 &    5 &    11 &  A &     &  16 &    4 &    12 &    A &   4.035e-02 \\
$\mathrm{CH_{3}OCHO \,\, v=0}$&         102897.0590       &  1.157e-06  &   14.95 &    26   & & 6 &    2 &    5 &    E &     &  5 &    1 &    4 &    E &   1.503e-02 \\
$\mathrm{CH_{3}OCHO \,\, v=0}$&         103114.8830       &  1.808e-06  &  151.70 &    86   & & 21&    4 &    17&    A &     &  21 &    3 &    18 &    A &   4.177e-02 \\
$\mathrm{CH_{3}OCHO \,\, v=0}$&         103228.3790       &  1.554e-06  &   64.59 &    54   & & 13&    4 &    10&    E &     &  13 &    3 &    11 &    E &   3.329e-02 \\
$\mathrm{CH_{3}OCHO \,\, v=0}$&         103262.0870       &  1.554e-06  &   64.58 &    54   & & 13&    4 &    10&    A &     &  13 &    3 &    11 &    A &   3.329e-02 \\
$\mathrm{CH_{3}OCHO \,\, v_{18}=1}$&      103367.4180       &  2.226e-06  &  405.14 &   102   & & 25&    6 &    19&    E &     &  25 &    5 &    20 &    E &   1.938e-02 \\
 $\mathrm{CH_{3}OCHO \,\, v=0}$&        103387.2000       &  2.211e-06  &  203.82 &    98   & & 24&    6 &    18&    E &     &  24 &    5 &    19 &    E &   4.579e-02 \\
 $\mathbf{CH_{3}OCHO \,\, v=0}$&        \textbf{103466.5720}      &  \textbf{1.517e-05}   &   \textbf{24.65} &     \textbf{34}    & & \textbf{8} &    \textbf{2} &    \textbf{6} &    \textbf{E} &     &  \textbf{7} &    \textbf{2} &    \textbf{5} &    \textbf{E}&   \textbf{2.439e-01} \\
 $\mathbf{CH_{3}OCHO \,\, v=0}$&        \textbf{103478.6630}      &  \textbf{1.517e-05}   &   \textbf{24.63} &     \textbf{34}    & & \textbf{8} &    \textbf{2} &    \textbf{6} &    \textbf{A} &     &  \textbf{7} &    \textbf{2} &    \textbf{5} &    \textbf{A} &   \textbf{2.440e-01} \\
 $\mathrm{CH_{3}OCHO \,\, v=0}$&        105281.1190$^{\mathbf{\ast}}$       &  2.411e-06  &  252.31 &   110   & & 27&    6 &    21&    E &     &  27&    5 &    22&    E &   4.344e-02 \\
 $\mathrm{CH_{3}OCHO \,\, v=0}$&        105299.7820       &  2.411e-06  &  252.31 &   110   & & 27&    6 &    21&    A &     &  27&    5 &    22&    A &   4.343e-02 \\
 $\mathrm{CH_{3}OCHO \,\, v=0}$&        105363.7220       &  1.220e-06  &   75.92 &    62   & & 15&    2 &    13&    E &     &  15&    1 &    14&    E &   2.739e-02 \\
 $\mathrm{CH_{3}OCHO \,\, v=0}$&        105424.6230       &  1.221e-06  &   75.90 &    62   & & 15&    2 &    13&    A &     &  15&    1 &    14&    A &   2.738e-02 \\
 $\mathrm{CH_{3}OCHO \,\, v=0}$&        105521.7890       &  1.595e-06  &  110.73 &    74   & & 18&    3 &    15&    A &     &  18&    2 &    16&    A &   3.641e-02 \\
 $\mathrm{CH_{3}OCHO \,\, v=0}$&        106018.8790       &  1.917e-06  &    9.53 &    14   & & 3 &    3 &    1 &    E &     &  2 &    2 &    1 &    E &   1.294e-02 \\
 $\mathrm{CH_{3}OCHO \,\, v=0}$&        106031.7050       &  1.914e-06  &    9.54 &    14   & & 3 &    3 &    0 &    E &     &  2 &    2 &    0 &    E &   1.292e-02 \\
 $\mathrm{CH_{3}OCHO \,\, v=0}$&        106125.3440       &  2.256e-06  &    9.52 &    14   & & 3 &    3 &    0 &    A &     &  2 &    2 &    1 &    A &   1.520e-02 \\
 $\mathrm{CH_{3}OCHO \,\, v=0}$&        106668.1340       &  1.698e-06  &   72.92 &    58   & & 14&    4 &    11&    A &     &  14&    3 &    12 &    A &   3.528e-02 \\
 $\mathrm{CH_{3}OCHO \,\, v_{18}=1}$&         107022.1590     &  1.700e-05  &  215.71 &    38   & & 9 &    2 &    8 &    E &     &  8 &    2 &    7 &    E &   1.208e-01 \\
 $\mathrm{CH_{3}OCHO \,\, v=0}$&        108045.9590       &  1.946e-06  &   87.87 &    62   & & 15&    5 &    10&    E &     &  15&    4 &    11&    E &   3.938e-02 \\
 $\mathrm{CH_{3}OCHO \,\, v_{18}=1}$&         108163.3340$^{\mathbf{\ast}}$     &  1.505e-06  &  284.40 &    66   & & 16&    5 &    11&    E &     &  16&    4 &    12&    E &   1.335e-02 \\
$\mathrm{CH_{3}OCHO \,\, v=0}$&         108539.6410       &  7.720e-07  &  165.28 &    90   & & 22&    4 &    18&    E &     &  22&    4 &    19&    E &   1.585e-02 \\
$\mathrm{CH_{3}OCHO \,\, v=0}$&         108883.5670       &  1.377e-06  &   67.80 &    58   & & 14&    3 &    12&    A &     &  14&    2 &    13&    A &   2.810e-02 \\
$\mathrm{CH_{3}OCHO \,\, v_{18}=1}$&      109390.4360$^{\mathbf{\ast}}$       &  4.033e-06  &  257.04 &    38   & & 9 &    8 &    1 &    E &     &  8 &    8 &    0 &    E &   2.277e-02 \\
$\mathrm{CH_{3}OCHO \,\, v_{18}=1}$&      109531.4130     &  7.602e-06  &  246.82 &    38   & & 9 &    7 &    3 &    A &     &  8 &    7 &    2 &    A &   4.483e-02 \\
$\mathrm{CH_{3}OCHO \,\, v_{18}=1}$&      109531.4130     &  7.602e-06  &  246.82 &    38   & & 9 &    7 &    2 &    A &     &  8 &    7 &    1 &    A &   4.483e-02 \\
$\mathrm{CH_{3}OCHO \,\, v_{18}=1}$&      109608.5400     &  1.070e-05  &  238.16 &    38   & & 9 &    6 &    4 &    A &     &  8 &    6 &    3 &    A &   6.557e-02 \\
$\mathrm{CH_{3}OCHO \,\, v_{18}=1}$&      109608.5400     &  1.070e-05  &  238.16 &    38   & & 9 &    6 &    3 &    A &     &  8 &    6 &    2 &    A &   6.557e-02 \\
$\mathrm{CH_{3}OCHO \,\, v_{18}=1}$&      109662.9440     &  1.076e-05  &  238.08 &    38   & & 9 &    6 &    3 &    E &     &  8 &    6 &    2 &    E &   6.589e-02 \\
$\mathrm{CH_{3}OCHO \,\, v_{18}=1}$&      109770.9950     &  1.337e-05  &  230.84 &    38   & & 9 &    5 &    5 &    A &     &  8 &    5 &    4 &    A &   8.441e-02 \\
$\mathrm{CH_{3}OCHO \,\, v_{18}=1}$&      109924.8170     &  7.719e-06  &  246.10 &    38   & & 9 &    7 &    3 &    E &     &  8 &    7 &    2 &    E &   4.534e-02 \\
$\mathrm{CH_{3}OCHO \,\, v_{18}=1}$&      110050.3320     &  1.088e-05  &  237.38 &    38   & & 9 &    6 &    4 &    E &     &  8 &    6 &    3 &    E &   6.636e-02 \\
$\mathrm{CH_{3}OCHO \,\, v_{18}=1}$&      110238.7130     &  1.948e-05  &  217.19 &    42   & & 10&    1 &    10&    E &     &  9 &    1 &    9 &    E &   1.433e-01 \\
$\mathrm{CH_{3}OCHO \,\, v_{18}=1}$&      110262.6420     &  1.361e-05  &  230.02 &    38   & & 9 &    5 &    5 &    E &     &  8 &    5 &    4 &    E &   8.546e-02 \\
$\mathrm{CH_{3}OCHO \,\, v=0}$&         110447.1800       &  4.151e-06  &   69.04 &    38   & & 9 &    8 &    1 &    E &     &  8 &    8 &    0 &    E &   5.363e-02 \\
$\mathrm{CH_{3}OCHO \,\, v=0}$&         110525.7410       &  7.828e-06  &   59.09 &    38   & & 9 &    7 &    2 &    E &     &  8 &    7 &    1 &    E &   1.056e-01 \\
$\mathrm{CH_{3}OCHO \,\, v_{18}=1}$&        110526.1900     &  1.483e-06  &  206.74 &    30   & & 7 &    2 &    6 &    A &     &  6 &    1 &    5 &    A &   8.128e-03 \\
$\mathrm{CH_{3}OCHO \,\, v=0}$&         110535.1860       &  7.832e-06  &   59.07 &    38   & & 9 &    7 &    2 &    A &     &  8 &    7 &    1 &    A &   1.057e-01 \\
$\mathrm{CH_{3}OCHO \,\, v=0}$&         110535.1860       &  7.832e-06  &   59.07 &    38   & & 9 &    7 &    3 &    A &     &  8 &    7 &    2 &    A &   1.057e-01 \\
$\mathrm{CH_{3}OCHO \,\, v=0}$&         110536.0030       &  7.832e-06  &   59.07 &    38   & & 9 &    7 &    3 &    E &     &  8 &    7 &    2 &    E &   1.057e-01 \\
$\mathrm{CH_{3}OCHO \,\, v=0}$&         110550.2030       &  1.465e-06  &   19.00 &    30   & & 7 &    2 &    6 &    E &     &  6 &    1 &    5 &    E &   1.869e-02 \\
$\mathrm{CH_{3}OCHO \,\, v_{18}=1}$&        110571.6320     &  1.956e-05  &  217.94 &    42   & & 10&    0 &    10&    A &     &  9 &    0 &    9 &    A &   1.425e-01 \\
$\mathbf{CH_{3}OCHO \,\, v=0}$&          \textbf{110652.8130}       &  \textbf{1.104e-05}   &   \textbf{50.46} &     \textbf{38}    & & \textbf{9} &    \textbf{6} &    \textbf{3} &    \textbf{E} &     &  \textbf{8} &    \textbf{6} &    \textbf{2} &    \textbf{E} &   \textbf{1.546e-01} \\
$\mathbf{CH_{3}OCHO \,\, v_{18}=1}$&         \textbf{110655.3100}     &  \textbf{1.971e-05}   &  \textbf{217.16} &     \textbf{42}    & & \textbf{10}&    0 &    \textbf{10}&    \textbf{E} &      &  \textbf{9} &    \textbf{0} &    \textbf{9} &    \textbf{E} &   \textbf{1.439e-01} \\
$\mathrm{CH_{3}OCHO \,\, v=0}$&         110662.3150       &  1.104e-05  &   50.45 &    38   & & 9 &    6 &    4 &    E &     &  8 &    6 &    3 &    E &   1.546e-01 \\
$\mathrm{CH_{3}OCHO \,\, v=0}$&         110663.2730       &  1.105e-05  &   50.44 &    38   & & 9 &    6 &    4 &    A &     &  8 &    6 &    3 &    A &   1.546e-01 \\
 $\mathrm{CH_{3}OCHO \,\, v=0}$&        110663.4290       &  1.105e-05  &   50.44 &    38   & & 9 &    6 &    3 &    A &     &  8 &    6 &    2 &    A &   1.546e-01 \\
 $\mathbf{CH_{3}OCHO \,\, v_{18}=1}$&        \textbf{110776.4990}     &  \textbf{1.927e-05}   &  \textbf{215.75} &     \textbf{38}    & & \textbf{9} &    \textbf{1} &    \textbf{8} &    \textbf{A} &     &  \textbf{8} &    \textbf{1} &    \textbf{7} &    \textbf{A} &   \textbf{1.278e-01} \\
 $\mathrm{CH_{3}OCHO \,\, v=0}$&        110788.6640       &  1.972e-05  &   30.27 &    42   & & 10&    1 &    10&    E &     &  9 &    1 &    9 &    E &   3.333e-01 \\
 $\mathrm{CH_{3}OCHO \,\, v=0}$&        110790.5260       &  1.973e-05  &   30.26 &    42   & & 10&    1 &    10&    A &     &  9 &    1 &    9 &    A &   3.334e-01 \\
  $\mathbf{CH_{3}OCHO \,\, v=0}$&          \textbf{110873.9550}       &  \textbf{1.382e-05}   &   \textbf{43.18}
  &    \textbf{38}    & & \textbf{9} &    \textbf{5} &    \textbf{4} &    \textbf{E} &     &  \textbf{8} &    \textbf{5} &    \textbf{3} &    \textbf{E} &   \textbf{1.991e-01} \\
  $\mathbf{CH_{3}OCHO \,\, v=0}$&          \textbf{110879.7660}       &  \textbf{1.774e-05}   &   \textbf{32.58} &     \textbf{38}    & & \textbf{9} &    \textbf{3} &    \textbf{7} &    \textbf{E} &     &  \textbf{8} &    \textbf{3} &    \textbf{6} &    \textbf{E} &   \textbf{2.680e-01} \\
  $\mathbf{CH_{3}OCHO \,\, v=0}$&          \textbf{110880.4470}       &  \textbf{1.383e-05}   &   \textbf{43.16} &     \textbf{38}    & & \textbf{9} &    \textbf{5} &    \textbf{5} &    \textbf{A} &     &  \textbf{8} &    \textbf{5} &    \textbf{4} &    \textbf{A} &   \textbf{1.992e-01} \\
  $\mathbf{CH_{3}OCHO \,\, v=0}$&          \textbf{110887.0920}       &  \textbf{1.775e-05}   &   \textbf{32.57} &     \textbf{38}    & & \textbf{9} &    \textbf{3} &    \textbf{7} &    \textbf{A} &     &  \textbf{8} &    \textbf{3} &    \textbf{6} &    \textbf{A} &  \textbf{2.682e-01} \\
  $\mathbf{CH_{3}OCHO \,\, v=0}$&          \textbf{110890.2560}       &  \textbf{1.383e-05}   &   \textbf{43.16} &     \textbf{38}    & & \textbf{9} &    \textbf{5} &    \textbf{4} &    \textbf{A} &     &  \textbf{8} &    \textbf{5} &    \textbf{3} &    \textbf{A} &   \textbf{1.992e-01} \\
 $\mathrm{CH_{3}OCHO \,\, v=0}$&        110924.2620       &  1.878e-06  &   81.85 &    62   & & 15&    4 &    12&    E &     &  15&    3 &    13&    E &   3.706e-02 \\
 $\mathrm{CH_{3}OCHO \,\, v_{18}=1}$&       111094.1240       &  1.952e-05  &  215.04 &    38   & & 9 &    1 &    8 &    E &     &  8 &    1 &    7 &    E &   1.292e-01 \\
 $\mathrm{CH_{3}OCHO \,\, v=0}$&        111169.9030       &  1.994e-05  &   30.25 &    42   & & 10&    0 &    10&    E &     &  9 &    0 &    9 &    E &   3.348e-01 \\
 $\mathrm{CH_{3}OCHO \,\, v=0}$&        111171.6340       &  1.994e-05  &   30.23 &    42   & & 10&    0 &    10&    A &     &  9 &    0 &    9 &    A &   3.348e-01 \\
 $\mathrm{CH_{3}OCHO \,\, v=0}$&        111195.9620       &  1.618e-05  &   37.22 &    38   & & 9 &    4 &    6 &    A &     &  8 &    4 &    5 &    A &   2.381e-01 \\
 $\mathrm{CH_{3}OCHO \,\, v=0}$&        111223.4910       &  1.532e-05  &   37.23 &    38   & & 9 &    4 &    6 &    E &     &  8 &    4 &    5 &    E &   2.253e-01 \\
  $\mathbf{CH_{3}OCHO \,\, v=0}$&          \textbf{111408.4120}       &  \textbf{1.540e-05}   &   \textbf{37.26} &     \textbf{38}    & & \textbf{9} &    \textbf{4} &    \textbf{5} &    \textbf{E} &     &  \textbf{8} &    \textbf{4} &    \textbf{4} &    \textbf{E} &   \textbf{2.258e-01} \\
  $\mathbf{CH_{3}OCHO \,\, v=0}$&          \textbf{111453.3000}       &  \textbf{1.629e-05}   &   \textbf{37.24} &     \textbf{38}    & & \textbf{9} &    \textbf{4} &    \textbf{5} &    \textbf{A} &     &  \textbf{8} &    \textbf{4} &    \textbf{4} &    \textbf{A} &   \textbf{2.387e-01} \\
 $\mathrm{CH_{3}OCHO \,\, v=0}$&        111492.3860       &  8.206e-07  &   54.62 &    54   & & 13&    1 &    12&    A &     &  13&    0 &    13&    A &   1.578e-02 \\
 $\mathbf{CH_{3}OCHO \,\, v=0}$&        \textbf{111682.1890}      &  \textbf{1.979e-05}   &   \textbf{28.12} &     \textbf{38}    & & \textbf{9} &    \textbf{1} &    \textbf{8} &    \textbf{A} &     &  \textbf{8} &    \textbf{1} &    \textbf{7} &    \textbf{A} &   \textbf{3.008e-01} \\
 $\mathrm{CH_{3}OCHO \,\, v=0}$&        111718.1400$^{\mathbf{\ast}}$       &  2.778e-06  &  269.80 &   114   & & 28&    6 &    22&    E &     &  28&    5 &    23&    E &   4.261e-02 \\
 $\mathrm{CH_{3}OCHO \,\, v=0}$&        111734.0020       &  2.972e-06  &   30.27 &    42   & & 10&    1 &    10&    E &     &  9 &    0 &    9 &    E &   4.939e-02 \\
 $\mathrm{CH_{3}OCHO \,\, v=0}$&        111735.3070       &  2.970e-06  &   30.26 &    42   & & 10&    1 &    10&    A &     &  9 &    0 &    9 &    A &   4.937e-02 \\
 $\mathbf{CH_{3}OCHO \,\, v_{18}=1}$&        \textbf{112011.9660}     &  \textbf{1.808e-05}   &  \textbf{219.94} &     \textbf{38}    & & \textbf{9} &    \textbf{3} &    \textbf{6} &    \textbf{E} &     &  \textbf{8} &    \textbf{3} &    \textbf{5} &    \textbf{E} &   \textbf{1.152e-01} \\
 $\mathrm{CH_{3}OCHO \,\, v_{18}=1}$&       112306.9410     &  1.839e-05  &  220.47 &    38   & & 9 &    3 &    6 &    A &     &  8 &    3 &    5 &    A &   1.162e-01 \\
 $\mathrm{CH_{3}OCHO \,\, v=0}$&        112672.7590       &  2.091e-06  &   78.87 &    58   & & 14&    5 &    9 &    E &     &  14&    4 &    10&    E &   3.792e-02 \\
 $\mathrm{CH_{3}OCHO \,\, v=0}$&        112676.8560       &  2.110e-06  &   78.86 &    58   & & 14&    5 &    9 &    A &     &  14&    4 &    10&    A &   3.826e-02 \\
 $\mathrm{CH_{3}OCHO \,\, v=0}$&        113743.1070       &  1.918e-05  &   32.87 &    38   & & 9 &    3 &    6 &    E &     &  8 &    3 &    5 &    E &   2.751e-01 \\
 $\mathbf{CH_{3}OCHO \,\, v=0}$&        \textbf{113756.6100}      &  \textbf{1.919e-05}   &   \textbf{32.86} &     \textbf{38}    & & \textbf{9} &    \textbf{3} &    \textbf{6} &    \textbf{A} &     &  \textbf{8} &    \textbf{3} &    \textbf{5} &    \textbf{A} &   \textbf{2.753e-01} \\
 $\mathrm{CH_{3}OCHO \,\, v=0}$&        115822.8920       &  2.370e-06  &  165.28 &    90   & & 22&    4 &    18&    E &     &  22&    3 &    19&    E &   4.278e-02 \\
\hline\hline
\multicolumn{16}{c}{\textit{Possible minor contributions}}\\
$\mathrm{CH_{3}OCHO \,\, v_{18}=1}$&       84450.4708       &  1.400e-07  &  1294.17&     214   & & 53 &    20 &    34 &  A &        & 52&     21&    31&    A &   6.972e-05 \\
$\mathrm{CH_{3}OCHO \,\, v=0}$&          85158.3954       &  9.588e-08  &  1169.87&     198   & & 49 &    26&    23&    A &      &  50&    25&    26&    A &   7.603e-05 \\
$\mathrm{CH_{3}OCHO \,\, v=0}$&          85158.3955       &  9.588e-08  &  1169.87&     198   & & 49 &    26&    24&    A &      &  50&    25&    25&    A &   7.603e-05 \\
$\mathrm{CH_{3}OCHO \,\, v=0}$&          87141.3675       &  1.457e-07  &  1395.28&     238   & & 59 &    23&    36&    A &      &  58&    24&    35&    A &   4.806e-05 \\
$\mathrm{CH_{3}OCHO \,\, v=0}$&          87141.3681       &  1.457e-07  &  1395.28&     238   & & 59 &    23&    37&    A &      &  58&    24&    34&    A &   4.806e-05 \\
$\mathrm{CH_{3}OCHO \,\, v_{18}=1}$&       87142.7650         &  1.686e-07  &  348.04 &    86   & & 21 &    6 &    15&    E &      &  20 &    7 &    13 &    E &   2.249e-03 \\
$\mathrm{CH_{3}OCHO \,\, v_{18}=1}$&       87143.6480         &  1.354e-06  &  342.04 &    86   & & 21 &    5 &    16&    E &      &  21 &    4 &    17 &    E &   1.855e-02 \\
$\mathrm{CH_{3}OCHO \,\, v_{18}=1}$&       87160.8350       &  1.274e-06  &  207.04 &    34   & &  8 &    0 &    8 &    E &      &     7 &    1 &    7 &    E &   1.267e-02 \\
$\mathrm{CH_{3}OCHO \,\, v_{18}=1}$&       87162.1070     &  6.876e-08  &  348.58 &    66   & &  16&    11&    5 &    E &      &  17 &    10 &    7 &    E &   7.015e-04 \\
$\mathrm{CH_{3}OCHO \,\, v_{18}=1}$&       87766.6099       &  1.663e-07  &  939.91 &   178   & & 44 &    16 &    28 &    E &   &    43 &    17 &    26 &    E &   3.145e-04 \\
$\mathrm{CH_{3}OCHO \,\, v_{18}=1}$&         92885.9950     &  1.120e-07  &  635.64 &   118   & & 29 &    17 &    13 &    A &   &    30 &    16 &    14 &    A &   4.943e-04 \\
$\mathrm{CH_{3}OCHO \,\, v_{18}=1}$&         92885.9950     &  1.120e-07  &  635.64 &   118   & & 29 &    17 &    12 &    A &   &    30 &    16 &    15 &    A &   4.943e-04 \\
$\mathrm{CH_{3}OCHO \,\, v_{18}=1}$&         96508.0050     &  1.077e-07  &  466.85 &    90   & & 22 &    14 &    9 &    E &     &  23 &    13 &    11 &    E &   7.186e-04 \\
$\mathrm{CH_{3}OCHO \,\, v=0}$&          97198.6390       &  2.451e-07  &  468.66 &   142   & & 35 &    12 &    23 &    A &   &    34 &    13 &    22 &    A &   2.522e-03 \\
$\mathrm{CH_{3}OCHO \,\, v_{18}=1}$&       97738.4514         &  2.464e-07  &  735.89 &   154   & & 38 &    13 &    26 &    A &   &    37 &    14 &    23 &    A &   8.161e-04 \\
$\mathrm{CH_{3}OCHO \,\, v=0}$&          99136.5850       &  1.400e-07  &  780.68 &   158   & & 39 &    22 &    18 &    A &   &    40 &    21 &    19 &    A &   3.780e-04 \\
$\mathrm{CH_{3}OCHO \,\, v=0}$&          99136.5850       &  1.400e-07  &  780.68 &   158   & & 39 &    22 &    17 &    A &   &    40 &    21 &    20 &    A &   3.780e-04 \\
$\mathrm{CH_{3}OCHO \,\, v_{18}=1}$&       99577.0130     &  4.151e-07  &  238.87 &    50   & & 12 &    3 &    10 &    E &     &  12 &    1 &    11 &    E &   4.035e-03 \\
$\mathrm{CH_{3}OCHO \,\, v_{18}=1}$&      100224.6730     &  1.240e-07  &  514.14 &    98   & & 24 &    15 &    9 &    E &     &  25 &    14 &    11 &    E &   6.749e-04 \\
$\mathrm{CH_{3}OCHO \,\, v=0}$&         100694.6657       &  4.615e-08  &   14.84 &    22   & & 5 &    3 &    3 &    E &     &  5 &    1 &    4 &    E &   5.295e-04 \\
$\mathrm{CH_{3}OCHO \,\, v=0}$&         103367.7950       &  1.466e-07  &  550.71 &   130   & & 32&    19&    14&    A &     &  33 &    18&    15 &    A &   8.445e-04 \\
$\mathrm{CH_{3}OCHO \,\, v=0}$&         103367.7950       &  1.466e-07  &  550.71 &   130   & & 32&    19&    13&    A &     &  33 &    18&    16 &    A &   8.445e-04 \\
$\mathrm{CH_{3}OCHO \,\, v_{18}=1}$&        106668.0137     &  2.968e-07  &  1152.25&     202   & & 50&    18&    32&    E &     & 49 &    19&    30 &    E &   1.661e-04 \\
$\mathrm{CH_{3}OCHO \,\, v_{18}=1}$&      110775.2970     &  6.274e-07  &  254.57 &    58   & & 14&    3 &    12&    E &     &  14&    1 &    13&    E &   5.333e-03 \\
$\mathrm{CH_{3}OCHO \,\, v_{18}=1}$&      111224.9860       &  4.800e-07  &  256.70 &    58   & & 14&    3 &    11&    E &     &  13&    4 &    10&    E &   4.009e-03 \\
$\mathrm{CH_{3}OCHO \,\, v_{18}=1}$&      112011.6610     &  4.052e-07  &  565.28 &   130   & & 32&    10&    22&    E &     & 31 &    11&    20&    E &   1.863e-03 \\

\end{longtable}
\tablefoot{$\nu$: rest frequency of the transitions in units MHz; $\mathrm{A_{E}}$: Einstein coefficient for spontaneous emission; $\mathrm{E_{U}}$: energy of the upper state of the transition in units K; $g_{U}$: degeneracy of the upper state of the transition; $\mathrm{Q_{U}}$: quantum numbers of the upper level; $\mathrm{Q_{L}}$: quantum numbers of the lower level; $\tau_{0} $: optical depth at the center of the line, calculated with the parameters $\mathrm{N_{tot}}$, $\mathrm{T_{ex}}$ and FWHM for AA given from best fit of XCLASS in Table \ref{fitresults}. In the upper part of the table are listed the  transitions of methyl formate identified in the spectrum. In the lowest part of the table (below the horizontal line) are listed the transitions of MF that lie inside the  width of the detected transitions, that hence can give a contribution to the line intensity.} 
}
\newpage
\setlength{\tabcolsep}{5pt}
\longtab[2]{
\label{AAused}
\renewcommand{\arraystretch}{1.5}
\begin{longtable}{lcccccrrrcrrrrrcc}
\caption{Transitions of acetic acid (AA) identified in the spectrum.}\\
\hline
  &  $\nu$  &$\mathrm{A_{E}}$ & $\mathrm{E_{U}}$ &   $g_{U}$ & &  & $\mathrm{Q_{U}}$ & &  & & & $\mathrm{Q_{L}}$ & &  &$\tau_{0} $\\ 
  &   \small{[MHz]} & \small{[$s^{-1}$]}  & \small{[K]} & & & \small{J} & \small{$\mathrm{K_{a}}$} & \small{$\mathrm{K_{c}}$} & & & \small{J}  & \small{$\mathrm{K_{a}}$} & \small{$\mathrm{K_{c}}$} & & \\
\hline

\endfirsthead
\caption{Continued.}
 \\ \hline
  &  $\nu$  & $\mathrm{A_{E}}$  &   $\mathrm{E_{U}}$ &   $g_{U}$ & &  & $\mathrm{Q_{U}}$   & &  & & & $\mathrm{Q_{L}}$ & &  &$\tau_{0} $\\ 
  &   \small{[MHz]} & \small{[$s^{-1}$]}  & \small{[K]} & & & \small{J} & \small{$\mathrm{K_{a}}$} & \small{$\mathrm{K_{c}}$} & & & \small{J}  & \small{$\mathrm{K_{a}}$} & \small{$\mathrm{K_{c}}$} & & \\
\hline

\endhead
\hline
\endfoot
\hline
\endlastfoot
$\mathrm{CH_{3}COOH \,\, v_{18}=1}$     &    85311.7558$^{\mathbf{\ast}}$     &    4.279e-06  &  406.30  &    51 & &   25&  17&  9&  E&   &  25&  16&  10&  E&         6.306e-03 \\ 
$\mathbf{CH_{3}COOH \,\, v_{18}=1}$     &    \textbf{85319.1640}    &    \textbf{2.864e-06}   &  \textbf{181.72}  &   \textbf{27} & &    \textbf{13}&  \textbf{4}&  \textbf{9}&  \textbf{A}&    &  \textbf{13}&  \textbf{3}&  \textbf{10}&  \textbf{A1}&         \textbf{4.736e-03} \\  
$\mathbf{CH_{3}COOH \,\, v_{18}=1}$     &   \textbf{85319.2754}     &    \textbf{2.864e-06}   &  \textbf{181.72}  &   \textbf{27} & &    \textbf{13}&  \textbf{5}&  \textbf{9}&  \textbf{A}&    &  \textbf{13}&  \textbf{4}&  \textbf{10}&  \textbf{A1}&         \textbf{4.736e-03} \\  
$\mathbf{CH_{3}COOH \,\, v=0}$        &    \textbf{85322.0066}    &    \textbf{4.489e-06}   &  \textbf{207.81}  &   \textbf{43} & &    \textbf{21}&  \textbf{13}&  \textbf{9}&  \textbf{E}&   &  \textbf{21}&  \textbf{12}&  \textbf{10}&  \textbf{E}&         \textbf{1.083e-02} \\  
$\mathbf{CH_{3}COOH \,\, v=0}$        &    \textbf{85633.0990}    &    \textbf{4.629e-06}   &  \textbf{229.22}  &   \textbf{45} & &    \textbf{22} &  \textbf{14}&  \textbf{9}&  \textbf{A}&    &  \textbf{22}&  \textbf{13}&  \textbf{10}&  \textbf{A2}&        \textbf{1.080e-02} \\  
$\mathrm{CH_{3}COOH \,\, v_{18}=1}$     &    86366.1725     &    2.029e-08  &  295.37  &    41 & &   20&  11&  9&  A&   &  20&  11&  10&  A1&        3.401e-05 \\ 
$\mathrm{CH_{3}COOH \,\, v_{18}=2}$     &    86366.2526     &    3.825e-06  &  359.03  &    39 & &   19&  10&  10&  E&    &  19&  9&  11&  E&        4.928e-03 \\ 
$\mathbf{CH_{3}COOH \,\, v=0}$        &    \textbf{90246.2358}    &    \textbf{8.102e-06}   &   \textbf{20.33}  &   \textbf{17} & &    \textbf{8}&  \textbf{0}&  \textbf{8}&  \textbf{A}&   &  \textbf{7}&  \textbf{1}&  \textbf{7}&  \textbf{A1}&  \textbf{1.294e-02} \\ 
$\mathbf{CH_{3}COOH \,\, v=0}$        &    \textbf{90246.2394}    &    \textbf{2.777e-06}   &   \textbf{20.33}  &   \textbf{17} & &    \textbf{8}&  \textbf{1}&  \textbf{8}&  \textbf{A}&   &  \textbf{7}&  \textbf{1}&  \textbf{7}&  \textbf{A1}&    \textbf{4.435e-03} \\ 
$\mathbf{CH_{3}COOH \,\, v=0}$        &    \textbf{90246.2662}    &    \textbf{2.777e-06} &   \textbf{20.33}&   \textbf{17} & &    \textbf{8}&  \textbf{0}&  \textbf{8}&  \textbf{A}&   &  \textbf{7}&  \textbf{0}&  \textbf{7}&  \textbf{A1}&  \textbf{4.435e-03} \\ 
$\mathbf{CH_{3}COOH \,\, v=0}$        &    \textbf{90246.2697}    &    \textbf{8.102e-06}   &   \textbf{20.33}  &   \textbf{17} & &    \textbf{8}&  \textbf{1}&  \textbf{8}&  \textbf{A}&   &  \textbf{7}&  \textbf{0}&  \textbf{7}&  \textbf{A1}&    \textbf{1.294e-02} \\ 
$\mathrm{CH_{3}COOH \,\, v_{18}=1}$     &    92565.0961$^{\mathbf{\ast}}$     &    6.664e-06  &  512.21  &    59 & &   29&  21&  9&  A&  &    29&  20&  10&  A1&       6.776e-03 \\   
$\mathrm{CH_{3}COOH \,\, v=0}$        &    93525.2612     &    6.144e-06  &  315.37  &    53 & &   26&  15&  11&  E&  &   26&  14&  12&  E&      1.062e-02 \\   
$\mathrm{CH_{3}COOH \,\, v=0}$        &    93760.4223     &    5.487e-07  &   58.24  &    25 & &   12&  3&  9&  E&    &  12&  3&  10&  E&        1.052e-03 \\ 
$\mathrm{CH_{3}COOH \,\, v=0}$        &    93760.4269     &    2.774e-06  &   58.24  &    25 & &   12&  3&  9&  E&    &  12&  2&  10&  E&        5.318e-03 \\ 
$\mathrm{CH_{3}COOH \,\, v=0}$        &    93760.6848     &    2.774e-06  &   58.24  &    25 & &   12&  4&  9&  E&    &  12&  3&  10&  E&        5.318e-03 \\ 
$\mathrm{CH_{3}COOH \,\, v=0}$        &    93760.6895     &    5.487e-07  &   58.24  &    25 & &   12&  4&  9&  E&    &  12&  2&  10&  E&        1.052e-03 \\ 
$\mathbf{CH_{3}COOH \,\, v=0}$        &    \textbf{94499.3172}    &    \textbf{6.050e-07}   &   \textbf{57.88}  &   \textbf{25} & &    \textbf{12}&  \textbf{3}&  \textbf{9}&  \textbf{A}&    &  \textbf{12}&  \textbf{3}&  \textbf{10}&  \textbf{A2}&         \textbf{1.143e-03} \\  
$\mathbf{CH_{3}COOH \,\, v=0}$        &    \textbf{94499.3276}    &    \textbf{2.817e-06}   &   \textbf{57.88}&   \textbf{25} & &    \textbf{12}&  \textbf{3}&  \textbf{9}&  \textbf{A}&    &  \textbf{12}&  \textbf{2}&  \textbf{10}&  \textbf{A2}&         \textbf{5.323e-03} \\  
$\mathbf{CH_{3}COOH \,\, v=0}$        &    \textbf{94499.8209}    &    \textbf{2.817e-06}   &   \textbf{57.88}  &   \textbf{25} & &    \textbf{12}&  \textbf{4}&  \textbf{9}&  \textbf{A}&    &  \textbf{12}&  \textbf{3}&  \textbf{10}&  \textbf{A2}&         \textbf{5.323e-03} \\  
$\mathbf{CH_{3}COOH \,\, v=0}$        &    \textbf{94499.8312}    &    \textbf{6.050e-07}   &   \textbf{57.88}  &   \textbf{25} & &    \textbf{12}&  \textbf{4}&  \textbf{9}&  \textbf{A}&    &  \textbf{12}&  \textbf{2}&  \textbf{10}&  \textbf{A2} &        \textbf{1.143e-03} \\  
$\mathrm{CH_{3}COOH \,\, v=0}$        &    94788.8006$^{\mathbf{\ast}}$     &    5.483e-06  &  260.36  &    47 & &   23&  17&  7&  A&  &    23&  16&  8&  A2&      9.836e-03 \\   
$\mathbf{CH_{3}COOH \,\, v=0}$        &   \textbf{100168.7044}    &    \textbf{9.428e-06}&    \textbf{24.4}0  &   \textbf{17} & &    \textbf{8}&  \textbf{1}&  \textbf{7}&  \textbf{E}&   &  \textbf{7}&  \textbf{2}&  \textbf{6}&  \textbf{E}&   \textbf{1.206e-02} \\
$\mathbf{CH_{3}COOH \,\, v=0}$        &   \textbf{100168.9638}    &    \textbf{3.303e-06}   &   \textbf{24.40}  &   \textbf{17} & &    \textbf{8}&  \textbf{2}&  \textbf{7}&  \textbf{E}&   &  \textbf{7}&  \textbf{2}&  \textbf{6}&  \textbf{E}&   \textbf{4.227e-03} \\
$\mathbf{CH_{3}COOH \,\, v=0}$        &   \textbf{100170.6980}    &    \textbf{3.303e-06}   &   \textbf{24.40}  &   \textbf{17} & &    \textbf{8}&  \textbf{1}&  \textbf{7}&  \textbf{E}&   &  \textbf{7}&  \textbf{1}&  \textbf{6}&  \textbf{E}&   \textbf{4.227e-03} \\
$\mathbf{CH_{3}COOH \,\, v=0}$        &   \textbf{100170.9574}    &    \textbf{9.429e-06}   &   \textbf{24.40}  &   \textbf{17} & &    \textbf{8}&  \textbf{2}&  \textbf{7}&  \textbf{E}&   &  \textbf{7}&  \textbf{1}&  \textbf{6}&  \textbf{E}&   \textbf{1.206e-02} \\
$\mathrm{CH_{3}COOH \,\, v_{18}=1}$     &   100286.6070     &    1.476e-05  &  138.77  &    19 & &   9&  0&  9&  A&   &  8&  1&  8&  A1&         1.437e-02 \\ 
$\mathrm{CH_{3}COOH \,\, v_{18}=1}$     &   100286.6070     &    3.923e-07  &  138.77  &    19 & &   9&  1&  9&  A&   &  8&  1&  8&  A1&         3.819e-04 \\ 
$\mathrm{CH_{3}COOH \,\, v_{18}=1}$     &   100286.6071     &    3.923e-07  &  138.77  &    19 & &   9&  0&  9&  A&   &  8&  0&  8&  A1&         3.819e-04 \\ 
$\mathrm{CH_{3}COOH \,\, v_{18}=1}$     &   100286.6072     &    1.476e-05  &  138.77  &    19 & &   9&  1&  9&  A&   &  8&  0&  8&  A1&         1.437e-02 \\ 
$\mathbf{CH_{3}COOH \,\, v=0}$        &   \textbf{100855.4272}    &    \textbf{1.162e-05}   &   \textbf{25.67}  &   \textbf{19} & &    \textbf{9}&  \textbf{0}&  \textbf{9}&  \textbf{E}&   &  \textbf{8}&  \textbf{1}&  \textbf{8}&  \textbf{E}&   \textbf{1.633e-02} \\
$\mathbf{CH_{3}COOH \,\, v=0}$        &   \textbf{100855.4274}    &    \textbf{3.732e-06}   &   \textbf{25.67}  &   \textbf{19} & &    \textbf{9}&  \textbf{1}&  \textbf{9}&  \textbf{E}&   &  \textbf{8}&  \textbf{1}&  \textbf{8}&  \textbf{E}&   \textbf{5.243e-03} \\
$\mathbf{CH_{3}COOH \,\, v=0}$        &   \textbf{100855.4291}    &    \textbf{3.732e-06}   &   \textbf{25.67}  &   \textbf{19} & &    \textbf{9}&  \textbf{0}&  \textbf{9}&  \textbf{E}&   &  \textbf{8}&  \textbf{0}&  \textbf{8}&  \textbf{E}&   \textbf{5.243e-03} \\
$\mathbf{CH_{3}COOH \,\, v=0}$        &   \textbf{100855.4293}    &    \textbf{1.162e-05}   &   \textbf{25.67}  &   \textbf{19} & &   \textbf{9}&  \textbf{1}&  \textbf{9}&  \textbf{E}&    &  \textbf{8}&  \textbf{0}&  \textbf{8}&  \textbf{E}&         \textbf{ 1.633e-02} \\
$\mathbf{CH_{3}COOH \,\, v=0}$        &   \textbf{100897.4541}    &    \textbf{1.145e-05}   &   \textbf{25.17}  &   \textbf{19} & &    \textbf{9}&  \textbf{0}&  \textbf{9}&  \textbf{A}&   &  \textbf{8}&  \textbf{1}&  \textbf{8} &  \textbf{A2}&  \textbf{1.610e-02} \\  
$\mathbf{CH_{3}COOH \,\, v=0}$        &   \textbf{100897.4545}    &    \textbf{3.932e-06}   &   \textbf{25.17}  &   \textbf{19} & &    \textbf{9}&  \textbf{1}&  \textbf{9}&  \textbf{A}&   &  \textbf{8}&  \textbf{1}&  \textbf{8}&  \textbf{A2}&    \textbf{5.529e-03} \\ 
$\mathbf{CH_{3}COOH \,\, v=0}$        &   \textbf{100897.4577}    &    \textbf{3.932e-06}   &   \textbf{25.17}  &   \textbf{19} & &    \textbf{9}&  \textbf{0}&  \textbf{9}&  \textbf{A}&   &  \textbf{8}&  \textbf{0}&  \textbf{8}&  \textbf{A2}&   \textbf{5.529e-03} \\  
$\mathbf{CH_{3}COOH \,\, v=0}$        &   \textbf{100897.4581}    &    \textbf{1.145e-05}   &   \textbf{25.17}  &   \textbf{19} & &    \textbf{9}&  \textbf{1}&  \textbf{9}&  \textbf{A}&   &  \textbf{8}&  \textbf{0}&  \textbf{8}&  \textbf{A2}&  \textbf{1.610e-02} \\ 
$\mathrm{CH_{3}COOH \,\, v=0}$        &   101310.2851     &    5.235e-06  &  224.85  &    43 & &   21&  17&  5&  A&  &    21&  16&  6&  A2&      8.474e-03 \\   
$\mathrm{CH_{3}COOH \,\, v=0}$        &   102231.5843     &    5.200e-06  &  110.83  &    33 & &   16&  7&  10&  E&   &  16&  6&  11&  E&        9.290e-03 \\ 
$\mathrm{CH_{3}COOH \,\, v=0}$        &   102231.7521     &    9.609e-07  &  110.83  &    33 & &   16&  7&  10&  E&   &  16&  5&  11&  E&        1.714e-03 \\ 
$\mathrm{CH_{3}COOH \,\, v_{18}=1}$     &   102355.3056$^{\mathbf{\ast}}$     &    8.276e-06  &  471.65  &    57 & &   28&  17&  12&  E&  &   28&  16&  13&  E&      7.621e-03 \\   
$\mathrm{CH_{3}COOH \,\, v=0}$        &   102853.7378     &    8.988e-07  &   95.11  &    31 & &   15&  5&  10&  E&   &  15&  5&  11&  E&        1.571e-03 \\ 
$\mathrm{CH_{3}COOH \,\, v=0}$        &   102853.7688     &    4.779e-06  &   95.11  &    31 & &   15&  5&  10&  E&   &  15&  4&  11&  E&        8.351e-03 \\ 
$\mathrm{CH_{3}COOH \,\, v=0}$        &   102854.8133     &    4.779e-06  &   95.11  &    31 & &   15&  6&  10&  E&   &  15&  5&  11&  E&        8.351e-03 \\ 
$\mathrm{CH_{3}COOH \,\, v=0}$        &   102854.8444     &    8.988e-07  &   95.11  &    31 & &   15&  6&  10&  E&   &  15&  4&  11&  E&        1.571e-03 \\ 
$\mathrm{CH_{3}COOH \,\, v=0}$        &   103356.4540     &    8.109e-07  &   80.35  &    29 & &   14&  4&  10&  E&   &  14&  4&  11&  E&        1.379e-03 \\ 
$\mathrm{CH_{3}COOH \,\, v=0}$        &   103356.4587     &    4.223e-06  &   80.35  &    29 & &   14&  4&  10&  E&   &  14&  3&  11&  E&        7.183e-03 \\ 
$\mathrm{CH_{3}COOH \,\, v=0}$        &   103356.6709     &    4.223e-06  &   80.35  &    29 & &   14&  5&  10&  E&   &  14&  4&  11&  E&        7.183e-03 \\ 
$\mathrm{CH_{3}COOH \,\, v=0}$        &   103356.6757     &    8.109e-07  &   80.35  &    29 & &   14&  5&  10&  E&   &  14&  3&  11&  E&        1.379e-03 \\ 
$\mathbf{CH_{3}COOH \,\, v=0}$        &   \textbf{104069.5729}    &    \textbf{2.624e-06}   &   \textbf{53.74}  &   \textbf{25} & &    \textbf{12}&  \textbf{2}&  \textbf{10}&  \textbf{E}&   &  \textbf{12}&  \textbf{1}&  \textbf{11}&  \textbf{E}&  \textbf{4.148e-03} \\  
$\mathbf{CH_{3}COOH \,\, v=0}$        &   \textbf{104069.5729}    &    \textbf{5.195e-07}   &   \textbf{53.74}  &   \textbf{25} & &    \textbf{12}&  \textbf{2}&  \textbf{10}&  \textbf{E}&   &  \textbf{12}&  \textbf{2}&  \textbf{11}&  \textbf{E}&  \textbf{8.212e-04} \\  
$\mathbf{CH_{3}COOH \,\, v=0}$        &   \textbf{104069.5776}    &    \textbf{5.195e-07}   &   \textbf{53.74}  &   \textbf{25} & &    \textbf{12}&  \textbf{3}&  \textbf{10}&  \textbf{E}&   &  \textbf{12}&  \textbf{1}&  \textbf{11}&  \textbf{E}&  \textbf{8.212e-04} \\  
$\mathbf{CH_{3}COOH \,\, v=0}$        &   \textbf{104069.5776}    &    \textbf{2.624e-06}   &   \textbf{53.74}  &   \textbf{25} & &    \textbf{12}&  \textbf{3}&  \textbf{10}&  \textbf{E}&   &  \textbf{12}&  \textbf{2}&  \textbf{11}&  \textbf{E}&  \textbf{4.148e-03} \\  
$\mathbf{CH_{3}COOH \,\, v=0}$        &   \textbf{104073.5696}    &    \textbf{4.768e-06}   &  \textbf{208.43}  &  \textbf{ 41} & &    \textbf{20}&  \textbf{17}&  \textbf{4}&  \textbf{A}&  &    \textbf{20}&  \textbf{16}&  \textbf{5}&  \textbf{A1}&   \textbf{7.369e-03} \\   
$\mathbf{CH_{3}COOH \,\, v=0}$        &   \textbf{104077.7341}    &    \textbf{9.004e-07}   &   \textbf{80.05}  &   \textbf{29} & &    \textbf{14}&  \textbf{4}&  \textbf{10}&  \textbf{A}&  &    \textbf{14}&  \textbf{4}&  \textbf{11}&  \textbf{A1}&  \textbf{1.512e-03} \\    
$\mathbf{CH_{3}COOH \,\, v=0}$        &   \textbf{104077.7454}    &    \textbf{4.262e-06}   &   \textbf{80.05}  &   \textbf{29} & &    \textbf{14}&  \textbf{4}&  \textbf{10}&  \textbf{A}&  &    \textbf{14}&  \textbf{3}& \textbf{ 11}&  \textbf{A1}&   \textbf{7.156e-03} \\   
$\mathbf{CH_{3}COOH \,\, v=0}$        &   \textbf{104078.1731}    &    \textbf{4.262e-06}   &  \textbf{ 80.05}  &   \textbf{29} & &    \textbf{14}&  \textbf{5}&  \textbf{10}&  \textbf{A}&  &    \textbf{14}&  \textbf{4}&  \textbf{11}&  \textbf{A1}&  \textbf{7.156e-03} \\    
$\mathbf{CH_{3}COOH \,\, v=0}$        &   \textbf{104078.1843}    &    \textbf{9.004e-07}   &   \textbf{80.05}  &   \textbf{29} & &    \textbf{14}&  \textbf{5}&  \textbf{10}&  \textbf{A}&  &    \textbf{14}&  \textbf{3}&  \textbf{11}&  \textbf{A1}&   \textbf{1.512e-03} \\   
$\mathbf{CH_{3}COOH \,\, v=0}$        &   \textbf{104078.5877}    &   \textbf{ 7.766e-06}   & \textbf{ 266.33}  &  \textbf{ 49} & &    \textbf{24}&  \textbf{14}&  \textbf{11}&  \textbf{A}& &   \textbf{24}&  \textbf{13}&  \textbf{12}&  \textbf{A2}&      \textbf{1.182e-02} \\    
$\mathbf{CH_{3}COOH \,\, v=0}$        &   \textbf{106876.9005}    &    \textbf{7.735e-06}   &  \textbf{220.51}  &   \textbf{45} & &    \textbf{22}&  \textbf{11}&  \textbf{11}&  \textbf{A}& &    \textbf{22}&  \textbf{10}&  \textbf{12}&  \textbf{A2}&       \textbf{1.195e-02} \\    
$\mathbf{CH_{3}COOH \,\, v=0}$        &   \textbf{106877.6308}    &    \textbf{5.343e-06}   &   \textbf{24.23}  &   \textbf{15} & &    \textbf{7}&  \textbf{3}&  \textbf{4}&  \textbf{E}&   &  \textbf{6}&  \textbf{4}&  \textbf{3}&  \textbf{E}&          \textbf{5.305e-03} \\
$\mathbf{CH_{3}COOH \,\, v=0}$        &   \textbf{110817.2439}    &    \textbf{1.325e-05}   &   \textbf{29.72}  &   \textbf{19} & &    \textbf{9}&  \textbf{1}&  \textbf{8}&  \textbf{E}&   &  \textbf{8}&  \textbf{2}&  \textbf{7}&  \textbf{E}&          \textbf{1.523e-02} \\
$\mathbf{CH_{3}COOH \,\, v=0}$        &   \textbf{110817.2758}    &    \textbf{4.559e-06}   &   \textbf{29.72}  &   \textbf{19} & &    \textbf{9}&  \textbf{2}&  \textbf{8}&  \textbf{E}&   &  \textbf{8}&  \textbf{2}&  \textbf{7}&  \textbf{E}&          \textbf{5.238e-03} \\
$\mathbf{CH_{3}COOH \,\, v=0}$        &   \textbf{110817.5033}    &    \textbf{4.559e-06}   &   \textbf{29.72}  &   \textbf{19} & &    \textbf{9}&  \textbf{1}&  \textbf{8}&  \textbf{E}&   &  \textbf{8}&  \textbf{1}&  \textbf{7}&  \textbf{E}&          \textbf{5.238e-03} \\
$\mathbf{CH_{3}COOH \,\, v=0}$        &   \textbf{110817.5351}    &    \textbf{1.325e-05}   &   \textbf{29.72}  &   \textbf{19} & &    \textbf{9}&  \textbf{2}&  \textbf{8}&  \textbf{E}&   &  \textbf{8}&  \textbf{1}&  \textbf{7}&  \textbf{E}&          \textbf{1.523e-02} \\
$\mathrm{CH_{3}COOH \,\, v=0}$        &   111548.5353     &    5.370e-06  &   30.53  &    21 & &   10&  1&  10&  A&  &    9&  1&  9&  A1&        6.712e-03 \\ 
$\mathrm{CH_{3}COOH \,\, v=0}$        &   111548.5353     &    1.569e-05  &   30.53  &    21 & &   10&  0&  10&  A&  &    9&  1&  9&  A1&        1.962e-02 \\ 
$\mathrm{CH_{3}COOH \,\, v=0}$        &   111548.5357     &    5.370e-06  &   30.53  &    21 & &   10&  0&  10&  A&  &    9&  0&  9&  A1&        6.712e-03 \\ 
$\mathrm{CH_{3}COOH \,\, v=0}$        &   111548.5357     &    1.569e-05  &   30.53  &    21 & &   10&  1&  10&  A&  &    9&  0&  9&  A1&        1.962e-02 \\ 
$\mathbf{CH_{3}COOH \,\, v=0}$        &   \textbf{114638.0194}    &    \textbf{9.614e-07}&    \textbf{75.06}  &   \textbf{29} & &    \textbf{14}&  \textbf{3}&  \textbf{11}&  \textbf{A}&  &    \textbf{14}&  \textbf{3}&  \textbf{12}&  \textbf{A2}&      \textbf{1.354e-03} \\    
$\mathbf{CH_{3}COOH \,\, v=0}$        &   \textbf{114638.0196}    &    \textbf{4.383e-06}   &   \textbf{75.06}  &   \textbf{29} & &    \textbf{14}&  \textbf{3}&  \textbf{11}&  \textbf{A}&  &    \textbf{14}&  \textbf{2}&  \textbf{12}&  \textbf{A2}&      \textbf{6.174e-03} \\    
$\mathbf{CH_{3}COOH \,\, v=0}$        &   \textbf{114638.0306}    &    \textbf{4.383e-06} &   \textbf{75.06}  &   \textbf{29} & &    \textbf{14}&  \textbf{4}&  \textbf{11}&  \textbf{A}&  &    \textbf{14}&  \textbf{3}&  \textbf{12}&  \textbf{A2}&      \textbf{6.174e-03} \\    
$\mathbf{CH_{3}COOH \,\, v=0}$        &   \textbf{114638.0308}    &    \textbf{9.614e-07}   &   \textbf{75.06}  &   \textbf{29} & &    \textbf{14}&  \textbf{4}&  \textbf{11}&  \textbf{A}&  &    \textbf{14}&  \textbf{2}&  \textbf{12}&  \textbf{A2}&        \textbf{1.354e-03} \\
\hline\hline
\multicolumn{16}{c}{\textit{Possible minor contributions}}\\
$\mathrm{CH_{3}COOH \,\, v=0}$        &    85310.1480     &    1.056e-08  &  343.85  &    59 & &   29&  10&  19&  A&  &   28&  13&  16&  A2&       2.221e-05 \\   
$\mathrm{CH_{3}COOH \,\, v=0}$        &    85311.0468     &    1.057e-08  &  343.85  &    59 & &   29&  11&  19&  A&  &   28&  12&  16&  A2&       2.221e-05 \\ 
$\mathrm{CH_{3}COOH \,\, v_{18}=1}$     &    85319.1596     &    7.229e-09  &  181.72  &    27 & &   13&  4&  9&  A&    &  13&  4&  10&  A1&         1.195e-05 \\
$\mathrm{CH_{3}COOH \,\, v_{18}=1}$     &    85319.2798     &    7.229e-09  &  181.72  &    27 & &   13&  5&  9&  A&    &  13&  3&  10&  A1&         1.195e-05 \\
$\mathrm{CH_{3}COOH \,\, v=0}$        &    93758.5173     &    3.794e-07  &  233.68  &    45 & &   22&  15&  7&  E&   &  22&  14&  9&  E&        7.281e-04 \\
$\mathrm{CH_{3}COOH \,\, v_{18}=1}$     &   102230.2426     &    1.674e-06  &  303.68  &    41 & &   20&  14&  6&  E&   &  20&  13&  8&  E&        1.949e-03 \\
$\mathrm{CH_{3}COOH \,\, v=0}$        &   103356.3522     &    1.146e-06  &  290.63  &    51 & &   25&  15&  11&  A& &    25&  13&  12&  A2&       1.697e-03 \\
$\mathrm{CH_{3}COOH \,\, v=0}$        &   103357.1153     &    6.178e-06  &  264.88  &    47 & &   23&  18&  6&  E&   &  23&  17&  7&  E&        9.118e-03 \\
$\mathrm{CH_{3}COOH \,\, v_{18}=1}$     &   104068.5286     &    1.816e-07  &  132.37  &    13 & &   6&  4&  2&  A&   &  5&  4&  1&  A2&             1.148e-04 \\
$\mathrm{CH_{3}COOH \,\, v_{18}=2}$     &   104070.3084     &    7.319e-07  &  234.46  &    19 & &   9&  4&  5&  E&   &  9&  3&  6&  E&          4.805e-04 \\
\end{longtable}
\tablefoot{$\nu$: rest frequency of the transitions in units MHz; $\mathrm{A_{E}}$: Einstein coefficient for spontaneous emission; $\mathrm{E_{U}}$: energy of the upper state of the transition in units K; $g_{U}$: degeneracy of the upper state of the transition; $\mathrm{Q_{U}}$: quantum numbers of the upper level; $\mathrm{Q_{L}}$: quantum numbers of the lower level; $\tau_{0} $: optical depth at the center of the line, calculated with the parameters $\mathrm{N_{tot}}$, $\mathrm{T_{ex}}$ and FWHM for AA given from best fit of XCLASS in Table \ref{fitresults}. In the upper part of the table are listed the  transitions of methyl formate identified in the spectrum. In the lowest part of the table (below the horizontal line) are listed the transitions of AA that lie inside the  width of the detected transitions, that hence can give a contribution to the line intensity.} 
}
\newpage
\setlength{\tabcolsep}{5pt}

\longtab[3]{
\label{GLused}
\renewcommand{\arraystretch}{1.5}
\begin{longtable}{lcccccrrrcrrrc}
\caption{ Transitions of glycolaldehyde (GA) identified in the spectrum.}\\
\hline
 &   $\nu$  & $\mathrm{A_{E}}$  &   $\mathrm{E_{U}}$ &   $g_{U}$ & &\multicolumn{3}{c}{ $\mathrm{Q_{U}}$} & &\multicolumn{3}{c}{$\mathrm{Q_{L}}$} & $\tau_{0}$\\ 
  &   \small{[MHz]} & \small{[$s^{-1}$]}  & \small{[K]} & & & \small{J} & \small{$\mathrm{K_{a}}$} & \small{$\mathrm{K_{c}}$} & & \small{J} & \small{$\mathrm{K_{a}}$} & \small{$\mathrm{K_{c}}$} & \\
\hline
\endfirsthead
\caption{Continued.} \\
\hline
&  $\nu$  & $\mathrm{A_{E}}$  &   $\mathrm{E_{U}}$ &   $g_{U}$ & &\multicolumn{3}{c}{ $\mathrm{Q_{U}}$} & &\multicolumn{3}{c}{$\mathrm{Q_{L}}$} & $\tau_{0}$\\ 
  &   \small{[MHz]} & \small{[$s^{-1}$]}  & \small{[K]} & & & \small{J} & \small{$\mathrm{K_{a}}$} & \small{$\mathrm{K_{c}}$} & & \small{J} & \small{$\mathrm{K_{a}}$} & \small{$\mathrm{K_{c}}$} & \\
\hline
\endhead
\endfoot
\endlastfoot
$\mathrm{CH_{2}OHCHO\,\,v=0}$   &    86862.3880   & 1.25263e-05   &  25.28  &   15  & &  7  & 4   &  3  & &  7 &   3 &  4  &  3.650e-02  \\
$\mathbf{CH_{2}OHCHO\,\,v=0}$   &    \textbf{88530.4087}  & \textbf{1.41325e-05}   &  \textbf{29.73}  &   \textbf{17}  & &  \textbf{8}  & \textbf{4}    &  \textbf{5} & &  \textbf{8} &  \textbf{3} & \textbf{6}  &  \textbf{4.342e-02}  \\
$\mathbf{CH_{2}OHCHO\,\,v=0}$   &    \textbf{88691.2622}  & \textbf{1.31001e-05}   &  \textbf{48.77}  &   \textbf{25}  & &  \textbf{12} & \textbf{3}    & \textbf{10} & & \textbf{12} &  \textbf{2} & \textbf{11} &  \textbf{5.088e-02}  \\
$\mathrm{CH_{2}OHCHO\,\,v=0}$   &    90187.0934   & 2.32285e-05   & 172.97  &   47  & &  23 & 5   & 18  & & 23 &   4 &  19 &  6.265e-02  \\
$\mathrm{CH_{2}OHCHO\,\,v=0}$   &    90507.2338   & 1.70268e-05   &  93.78  &   35  & &  17 & 3   & 14  & & 17 &   2 &  15 &  6.275e-02  \\
$\mathbf{CH_{2}OHCHO\,\,v=0}$   &    \textbf{93048.4534}  & \textbf{2.68455e-05}   & \textbf{191.55}  &   \textbf{49}  & &  \textbf{24} & \textbf{6}    & \textbf{18} & & \textbf{24} &  \textbf{5} & \textbf{19} &  \textbf{6.144e-02}  \\
$\mathbf{CH_{2}OHCHO\,\,v=0}$   &    \textbf{93052.6650}  & \textbf{2.70046e-05}   &  \textbf{23.34}  &   \textbf{19}  & &  \textbf{9}  & \textbf{0}    &  \textbf{9} & &  \textbf{8} &  \textbf{1} & \textbf{8}  &  \textbf{8.827e-02}  \\
$\mathrm{CH_{2}OHCHO\,\,v=0}$   &    95070.0722   & 2.89224e-05   &  23.37  &   19  & &  9  & 1   &  9  & &  8 &   0 &  8  &  9.058e-02  \\
$\mathbf{CH_{2}OHCHO\,\,v=0}$   &    \textbf{95756.2274}  & \textbf{2.80545e-05}   & \textbf{177.57}  &   \textbf{47}  & &  \textbf{23} & \textbf{6}    & \textbf{17} & & \textbf{23} &  \textbf{5} & \textbf{18} &  \textbf{6.484e-02}  \\
$\mathrm{CH_{2}OHCHO\,\,v=0}$   &   101116.3718   & 2.64001e-05   & 142.69  &   43  & &  21 & 4   & 17  & & 21 &   3 &  18 &  6.567e-02  \\
$\mathbf{CH_{2}OHCHO\,\,v=0}$   &   \textbf{101232.1850}  & \textbf{1.69814e-05}   &  \textbf{71.31}  &   \textbf{31}  & &  \textbf{15} & \textbf{2}    & \textbf{13} & & \textbf{15} &  \textbf{1} & \textbf{14} &  \textbf{5.284e-02}  \\
$\mathbf{CH_{2}OHCHO\,\,v=0}$   &   \textbf{101527.8545}  & \textbf{2.53591e-05}   &  \textbf{73.82}  &   \textbf{29}  & &  \textbf{14} & \textbf{5}    &  \textbf{9} & & \textbf{14} &  \textbf{4} & \textbf{10} &  \textbf{7.197e-02}  \\
$\mathbf{CH_{2}OHCHO\,\,v=0}$   &   \textbf{102572.9322}  & \textbf{1.82454e-05}   &  \textbf{63.60}  &   \textbf{29}  & &  \textbf{14} & \textbf{3}    & \textbf{12} & & \textbf{14} &  \textbf{2} & \textbf{13} &  \textbf{5.493e-02}  \\
$\mathbf{CH_{2}OHCHO\,\,v=0}$   &   \textbf{102614.3647}  & \textbf{2.28639e-05}   & \textbf{104.10}  &   \textbf{37}  & &  \textbf{18} & \textbf{3}    & \textbf{15} & & \textbf{18} &  \textbf{2} & \textbf{16} &  \textbf{6.410e-02}  \\
$\mathrm{CH_{2}OHCHO\,\,v=0}$   &   103391.3370   & 3.82298e-05   &  28.34  &   21  & &  10 & 0   & 10  & &  9 &   1 &  9  &  1.078e-01  \\
$\mathrm{CH_{2}OHCHO\,\,v=0}$   &   103461.3577   & 2.44027e-05   &  76.75  &   31  & &  15 & 4   & 12  & & 15 &   3 &  13 &  6.972e-02  \\
$\mathbf{CH_{2}OHCHO\,\,v=0}$   &   \textbf{103667.9627}  & \textbf{2.22157e-05}   &  \textbf{31.93}  &   \textbf{21}  & &  \textbf{10} & \textbf{1}    &  \textbf{9} & &  \textbf{9} & \textbf{2} &  \textbf{8}  &  \textbf{6.062e-02}  \\
$\mathrm{CH_{2}OHCHO\,\,v=0}$   &   104587.7362   & 3.96541e-05   &  28.36  &   21  & &  10 & 1   & 10  & &  9 &   0 &  9  &  1.093e-01  \\
$\mathbf{CH_{2}OHCHO\,\,v=0}$   &   \textbf{112210.4945}  & \textbf{3.09425e-05}   &  \textbf{58.62}  &   \textbf{25}  & &  \textbf{12} & \textbf{5}    &  \textbf{8} & & \textbf{12} &  \textbf{4} & \textbf{9}  &  \textbf{6.987e-02}  \\
$\mathbf{CH_{2}OHCHO\,\,v=0}$   &   \textbf{114594.1032}  & \textbf{3.58410e-05}   &  \textbf{91.19}  &   \textbf{33}  & &  \textbf{16} & \textbf{5}    & \textbf{12} & & \textbf{16} &  \textbf{4} & \textbf{13} &  \textbf{7.961e-02}  \\
\hline\hline
\multicolumn{14}{c}{\textit{Possible minor contributions}}\\
$\mathrm{CH_{2}OHCHO\,\,v_{18}=1}$  &    88691.4466   & 2.02859e-06   & 407.00  &   41  & &  20 & 4   & 17  & & 19 &   5 &  14 &  8.041e-04  \\
$\mathrm{CH_{2}OHCHO\,\,v_{18}=1}$  &    90505.4741   & 3.36633e-06   & 599.99  &   61  & &  30 & 10    & 20  & & 29 &    11 &  19 &  4.272e-04  \\
$\mathrm{CH_{2}OHCHO\,\,v=0}$   &   114595.5269   & 2.61361e-07   & 290.40  &   63  & &  31 & 4   & 27  & & 32 &   3 &  30 &  2.366e-04  \\
\end{longtable}
\tablefoot{$\nu$: rest frequency of the transitions in units MHz; $\mathrm{A_{E}}$: Einstein coefficient for spontaneous emission; $\mathrm{E_{U}}$: energy of the upper state of the transition in units K; $g_{U}$: degeneracy of the upper state of the transition; $\mathrm{Q_{U}}$: quantum numbers of the upper level; $\mathrm{Q_{L}}$: quantum numbers of the lower level; $\tau_{0}$: optical depth at the center of the line, calculated with the parameters $\mathrm{N_{tot}}$, $\mathrm{T_{ex}}$ and FWHM for GA given in Table \ref{fitresults}. In the upper part of the table are listed the transitions of GA identified in the spectrum. In the lowest part of the table (below the horizontal line) are listed the transitions of GA that lie inside the  width of the detected transitions, that hence can give a contribution to the line intensity. }
}
\vfill
\clearpage
\newpage
\section{Full spectrum}
\begin{figure*}[h!]
\centering
\includegraphics[width=\hsize]{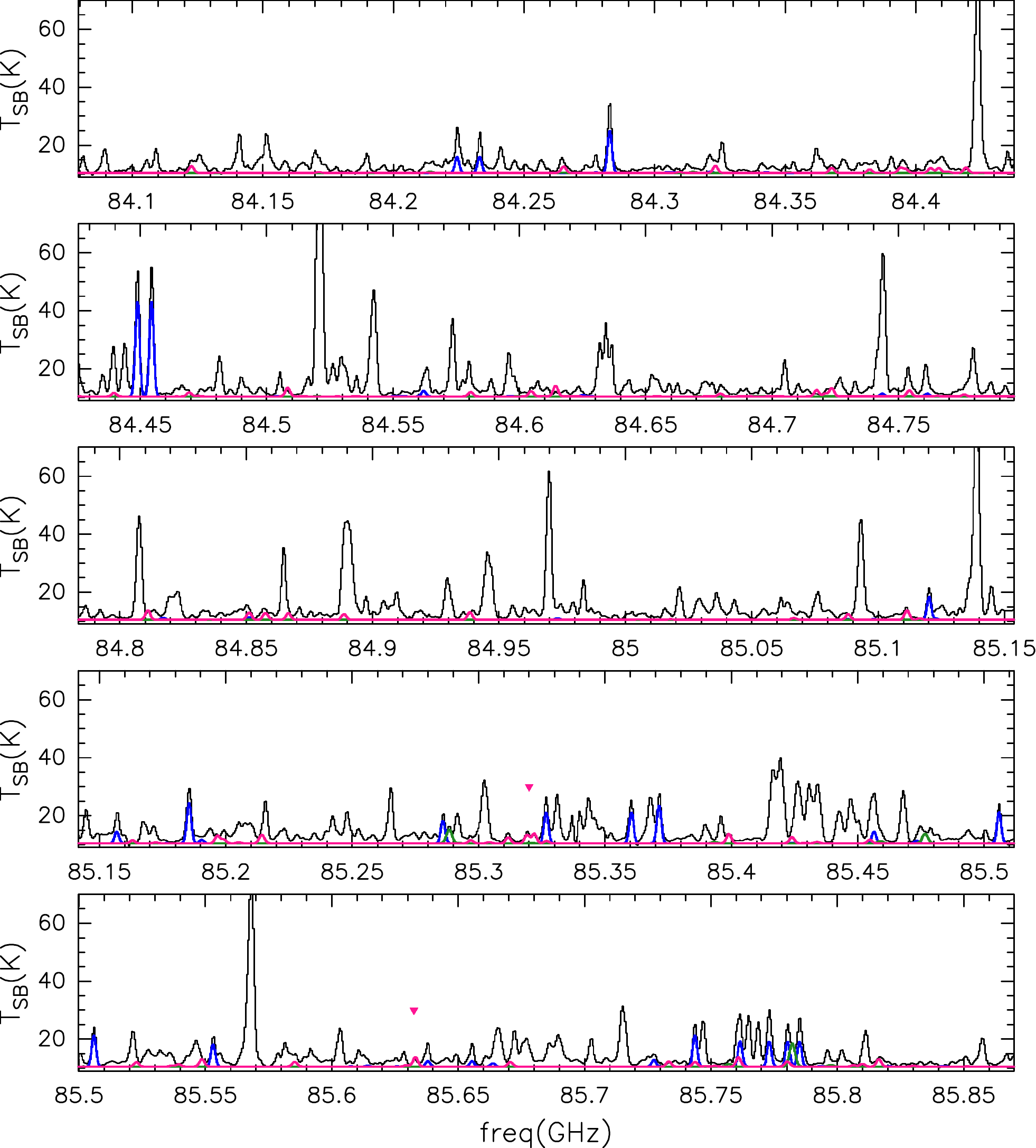}\\
\caption{Total spectra of the GUAPOS project in black. In blue the synthetic spectrum of the best fit of methyl formate (MF); in green the the synthetic spectrum of the best fit of glycolaldehyde (GA); in pink the synthetic spectrum of the best fit of acetic acid (AA). The colored triangles indicate the transitions used to constrain the fitting procedure of the 3 different molecules. Closer views of those transitions are given in Fig. \ref{figsingleMF}, 5 and 6.}
\label{fullfigure}
\end{figure*}
\clearpage
\newpage
\begin{figure*}
\includegraphics[width=\hsize]{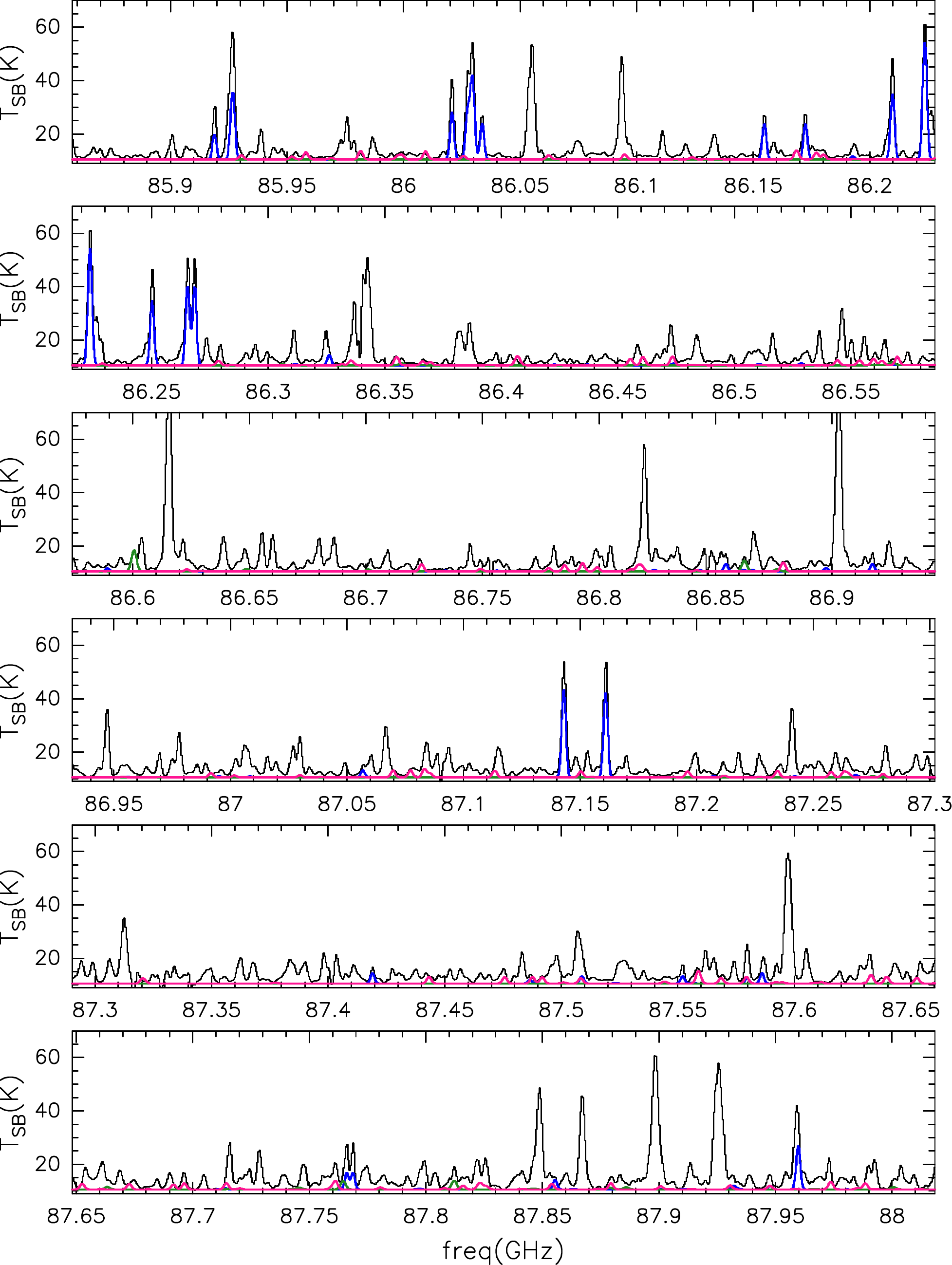}
\caption{ Continue}
\end{figure*}

\newpage
\begin{figure*}
\centering
\includegraphics[width=\hsize]{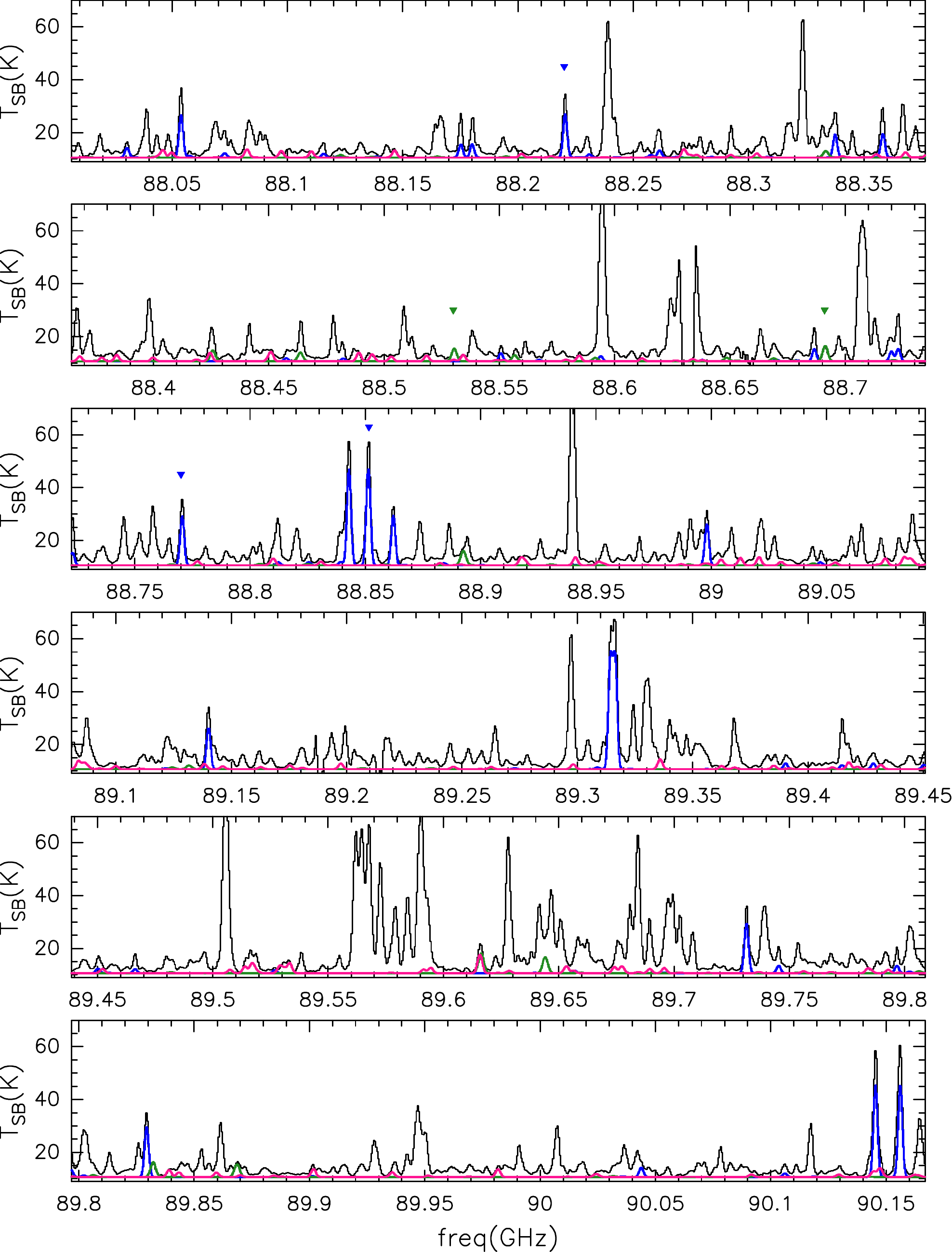}
\caption{ Continue}
\end{figure*}

\newpage
\begin{figure*}
\includegraphics[width=\hsize]{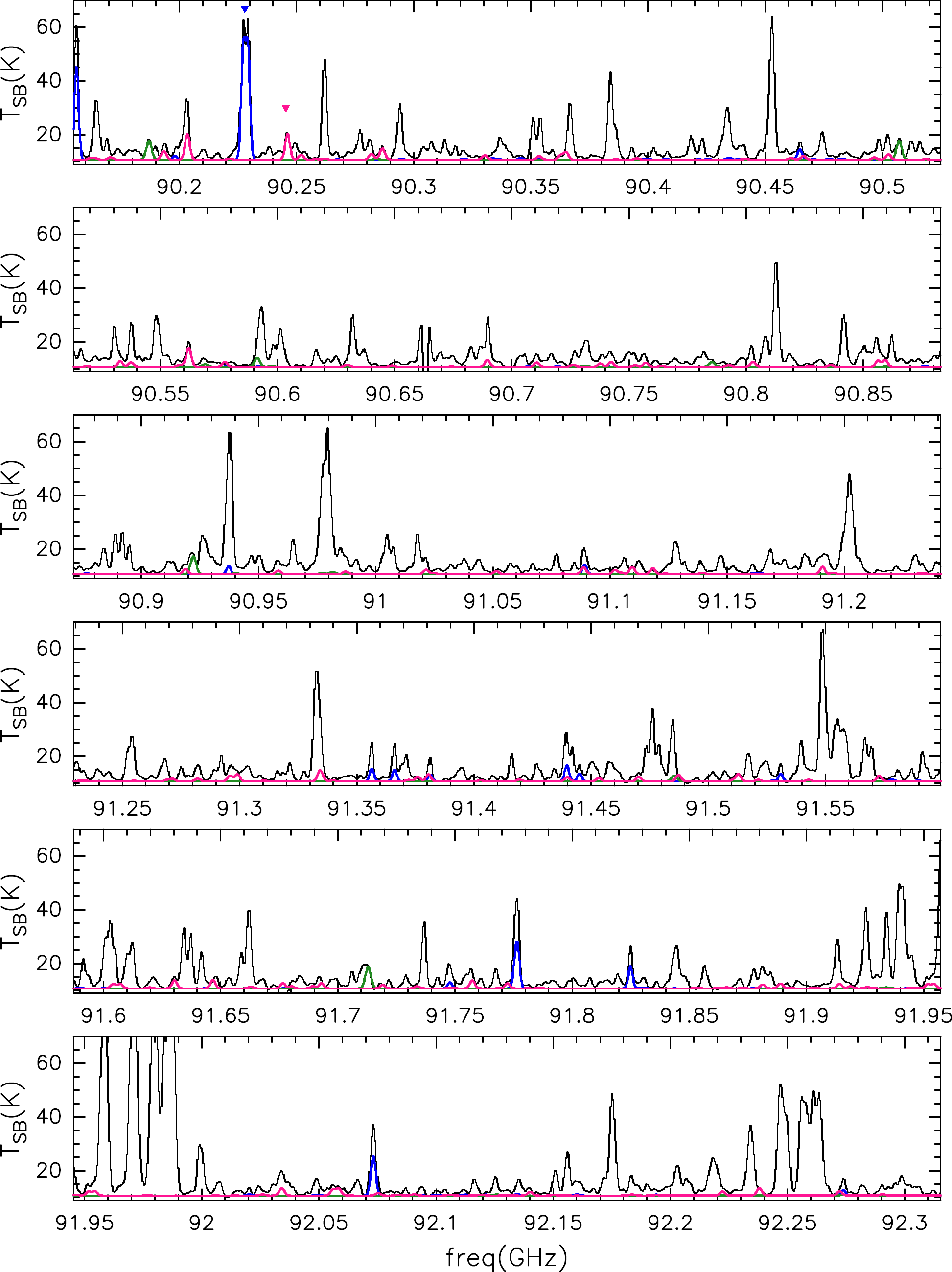}
\caption{ Continue}
\end{figure*}

\newpage
\begin{figure*}
\centering
\includegraphics[width=\hsize]{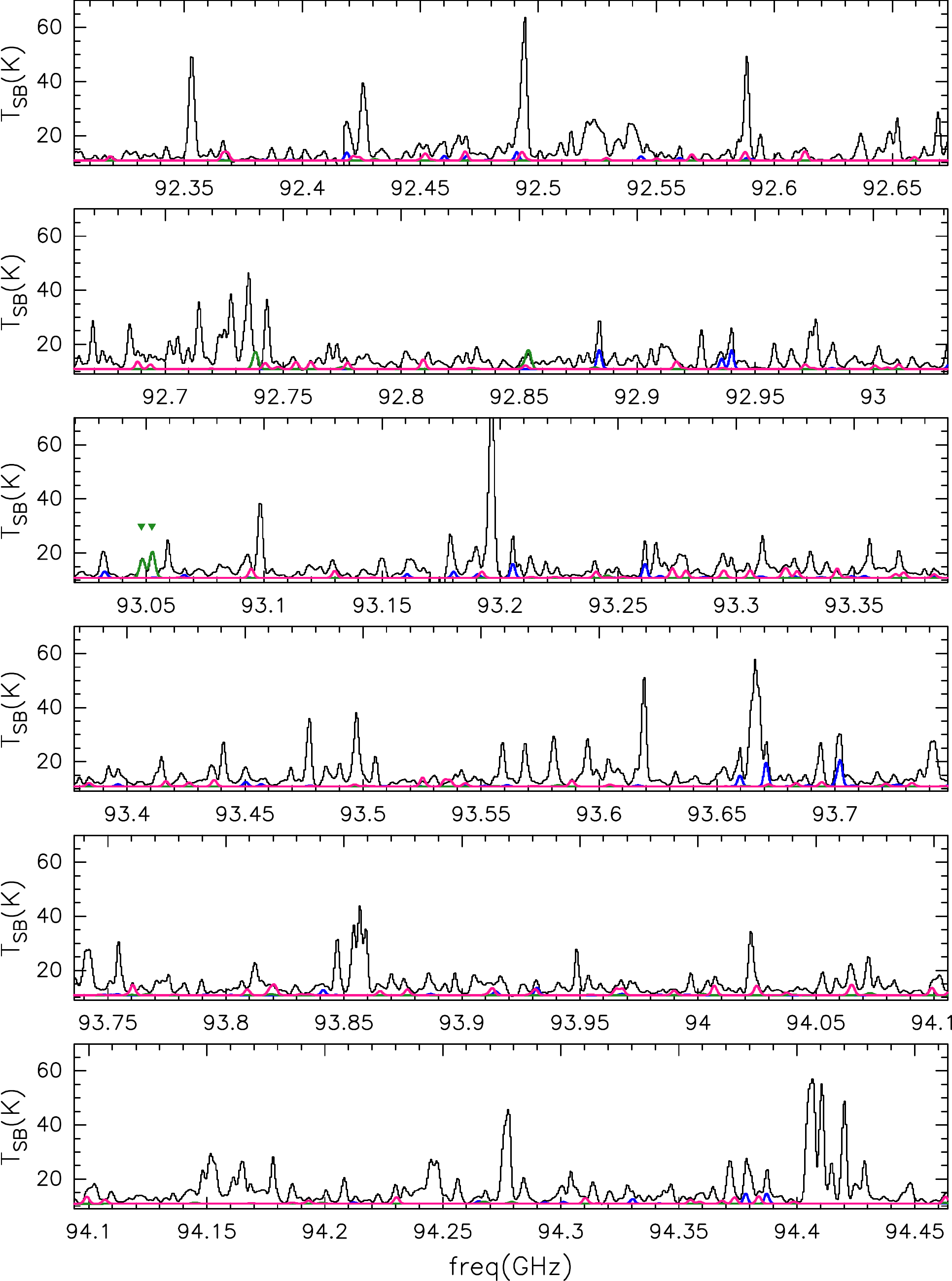}
\caption{ Continue}
\end{figure*}
\newpage
\begin{figure*}
\centering
\includegraphics[width=\hsize]{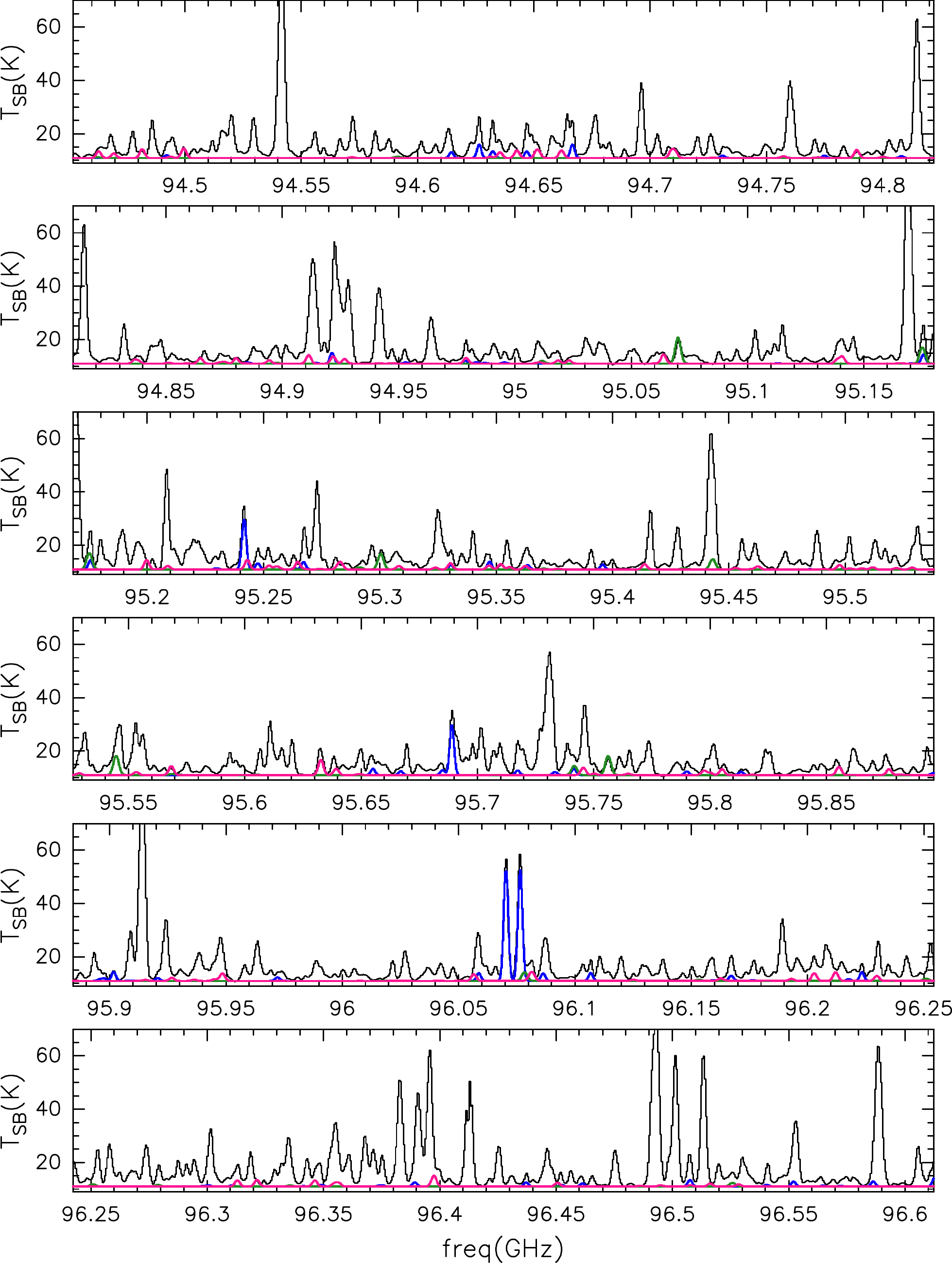}
\caption{ Continue}
\end{figure*}
\newpage
\begin{figure*}
\centering
\includegraphics[width=\hsize]{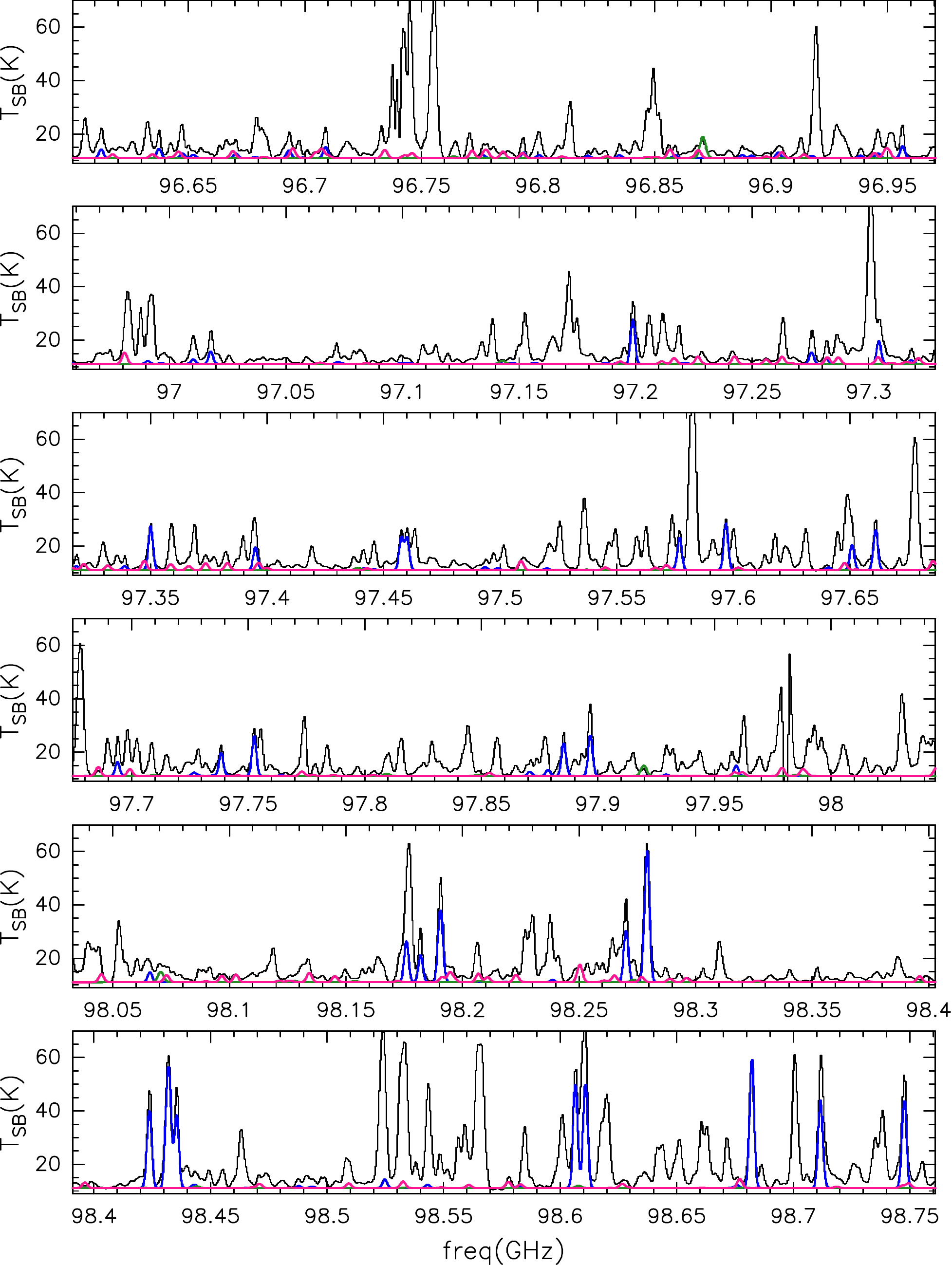}
\caption{ Continue}
\end{figure*}
\newpage
\begin{figure*}
\centering
\includegraphics[width=\hsize]{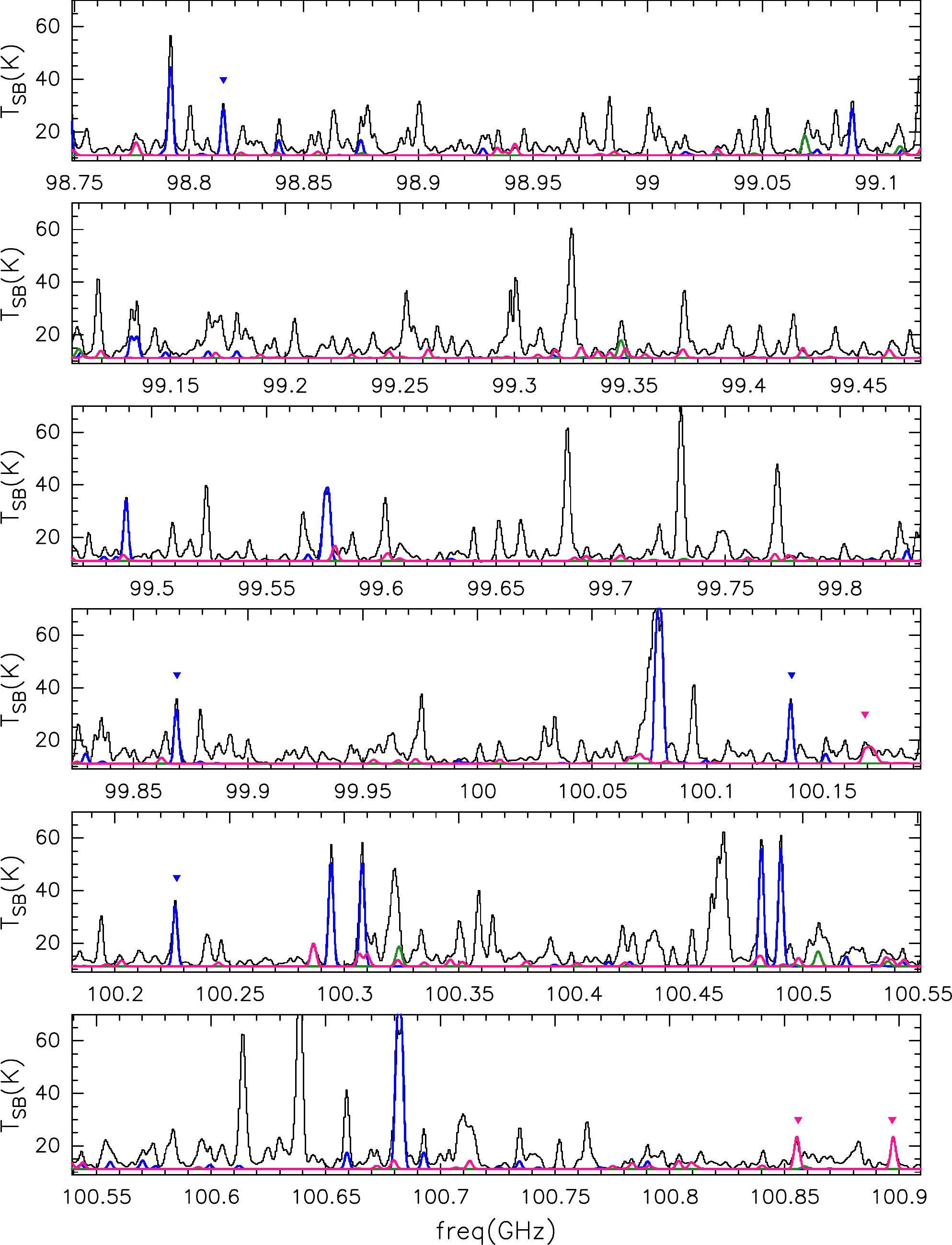}
\caption{ Continue}
\end{figure*}
\newpage
\begin{figure*}
\centering
\includegraphics[width=\hsize]{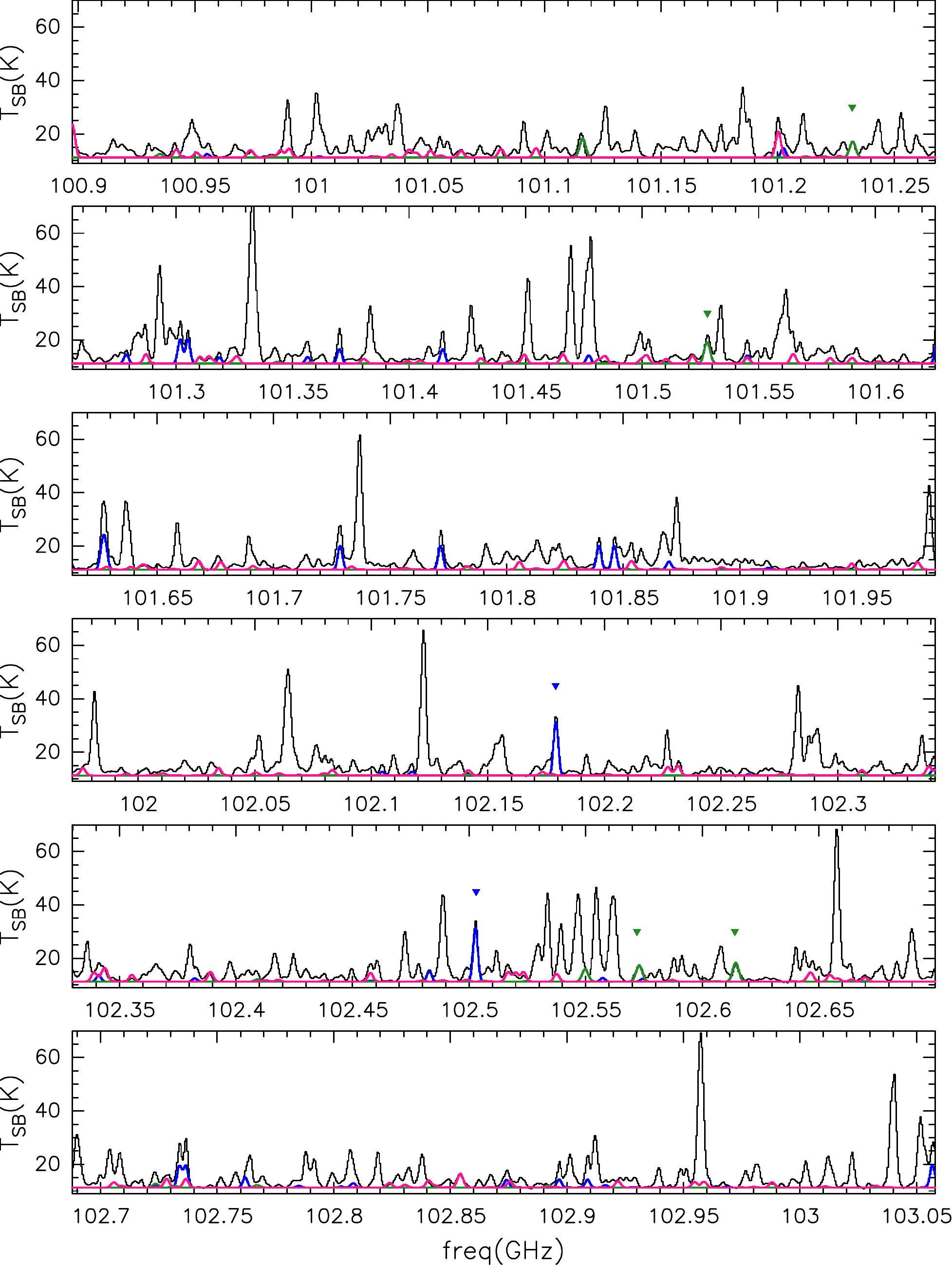}
\caption{ Continue}
\end{figure*}
\newpage
\begin{figure*}
\centering
\includegraphics[width=\hsize+0.2cm]{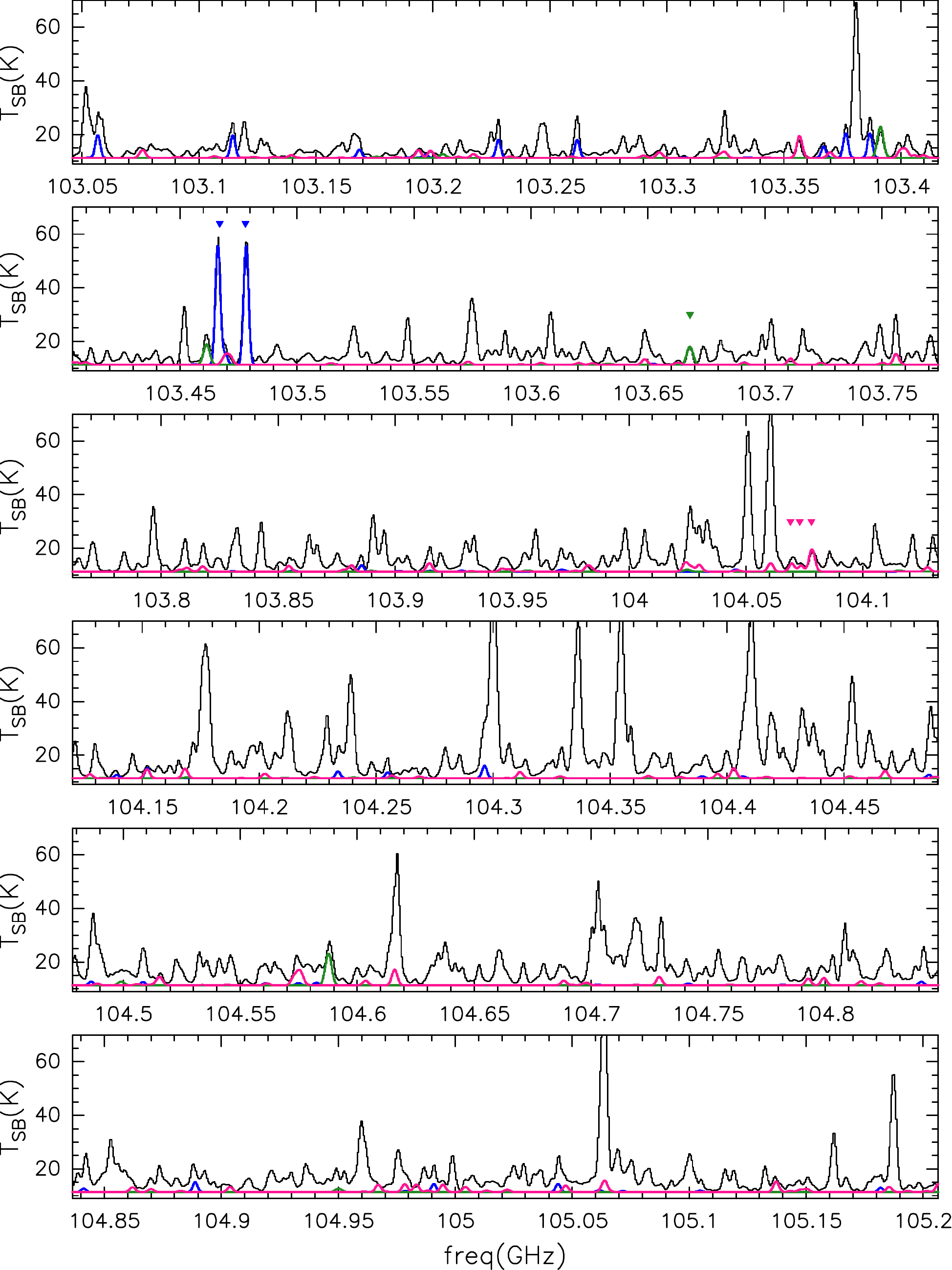}
\caption{ Continue}
\end{figure*}
\newpage
\begin{figure*}
\centering
\includegraphics[width=\hsize+0.2cm]{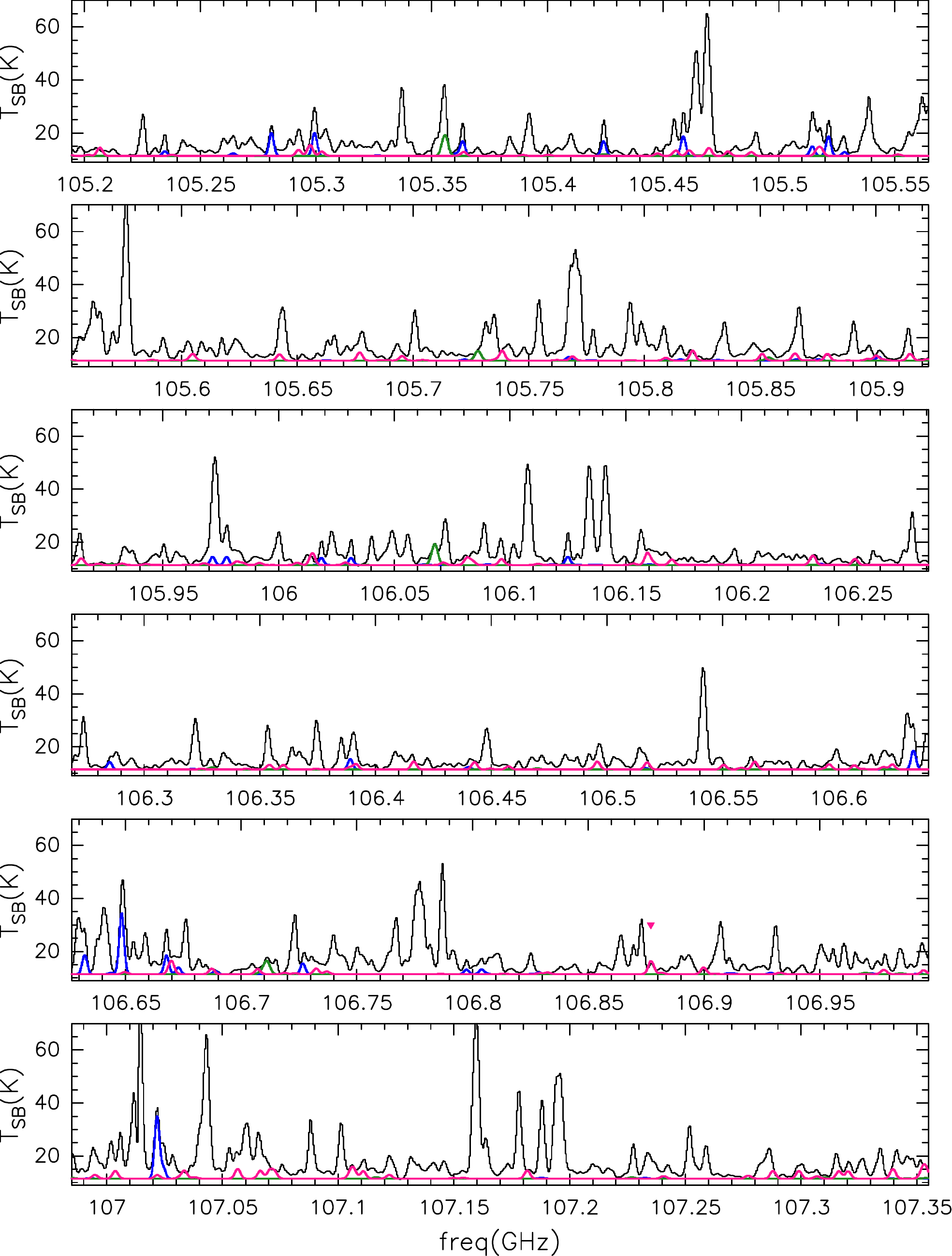}
\caption{ Continue}
\end{figure*}
\newpage
\begin{figure*}
\centering
\includegraphics[width=\hsize+0.2cm]{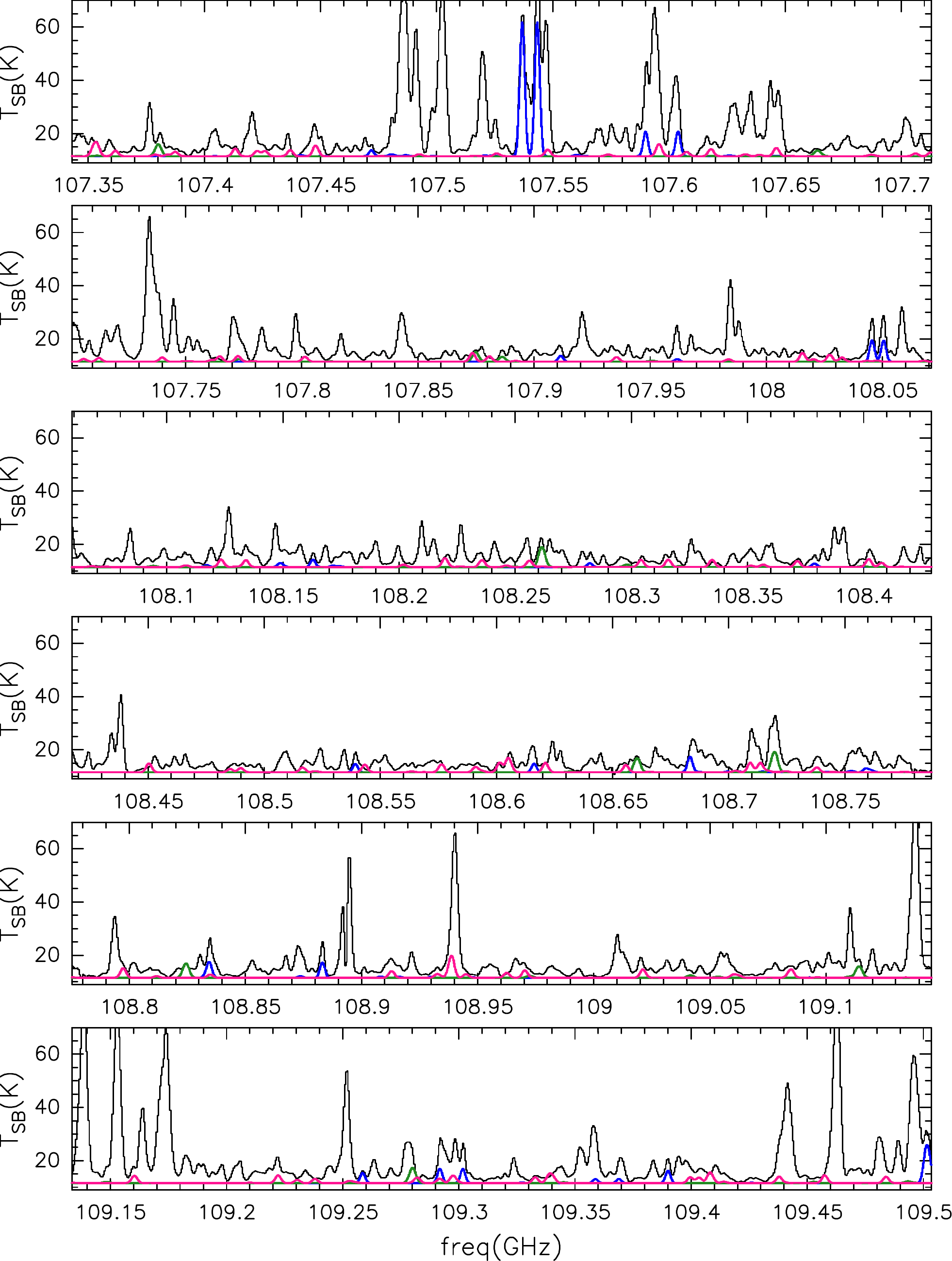}
\caption{ Continue}
\end{figure*}
\newpage
\begin{figure*}
\centering
\includegraphics[width=\hsize+0.2cm]{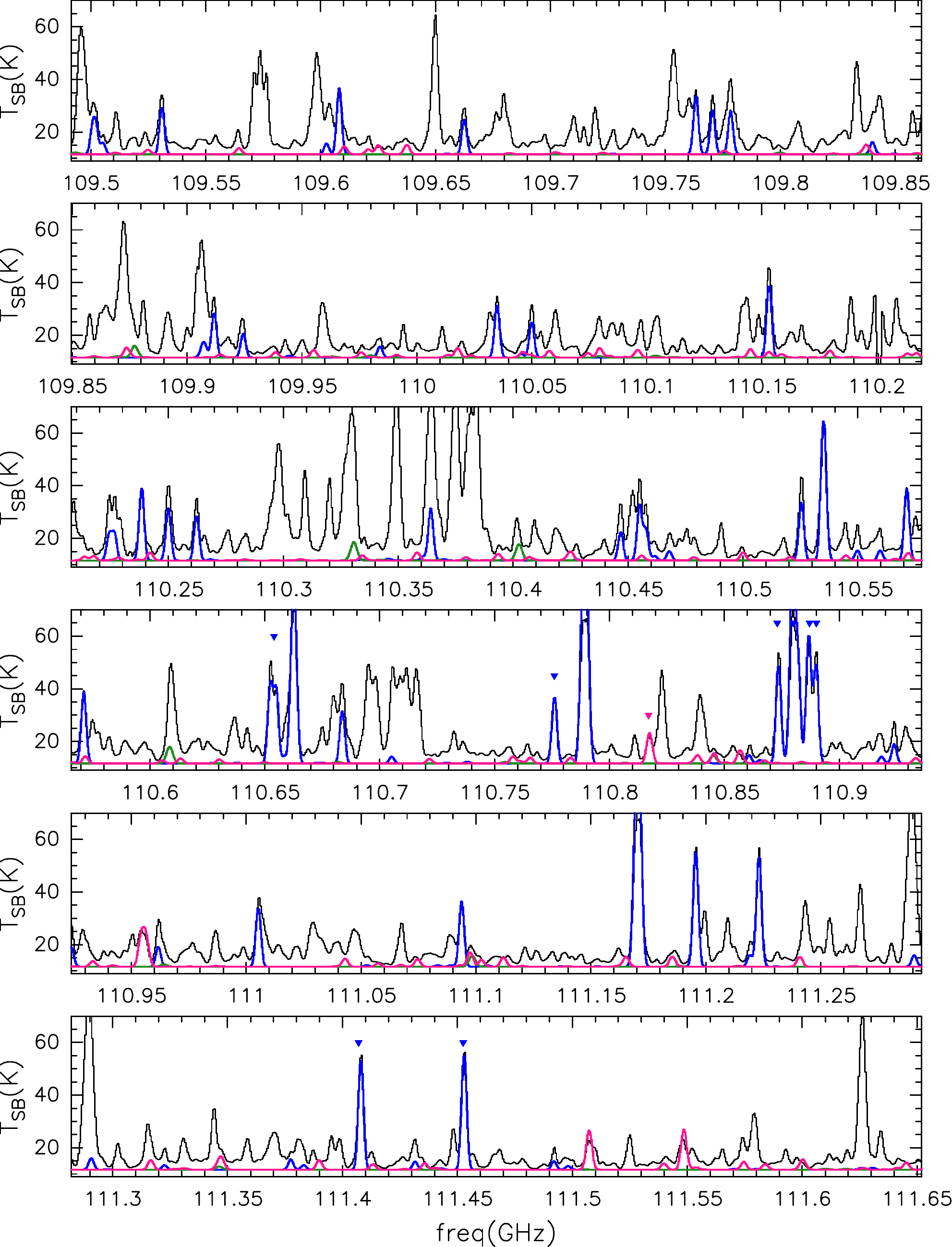}
\caption{ Continue}
\end{figure*}
\newpage
\begin{figure*}
\centering
\includegraphics[width=\hsize+0.2cm]{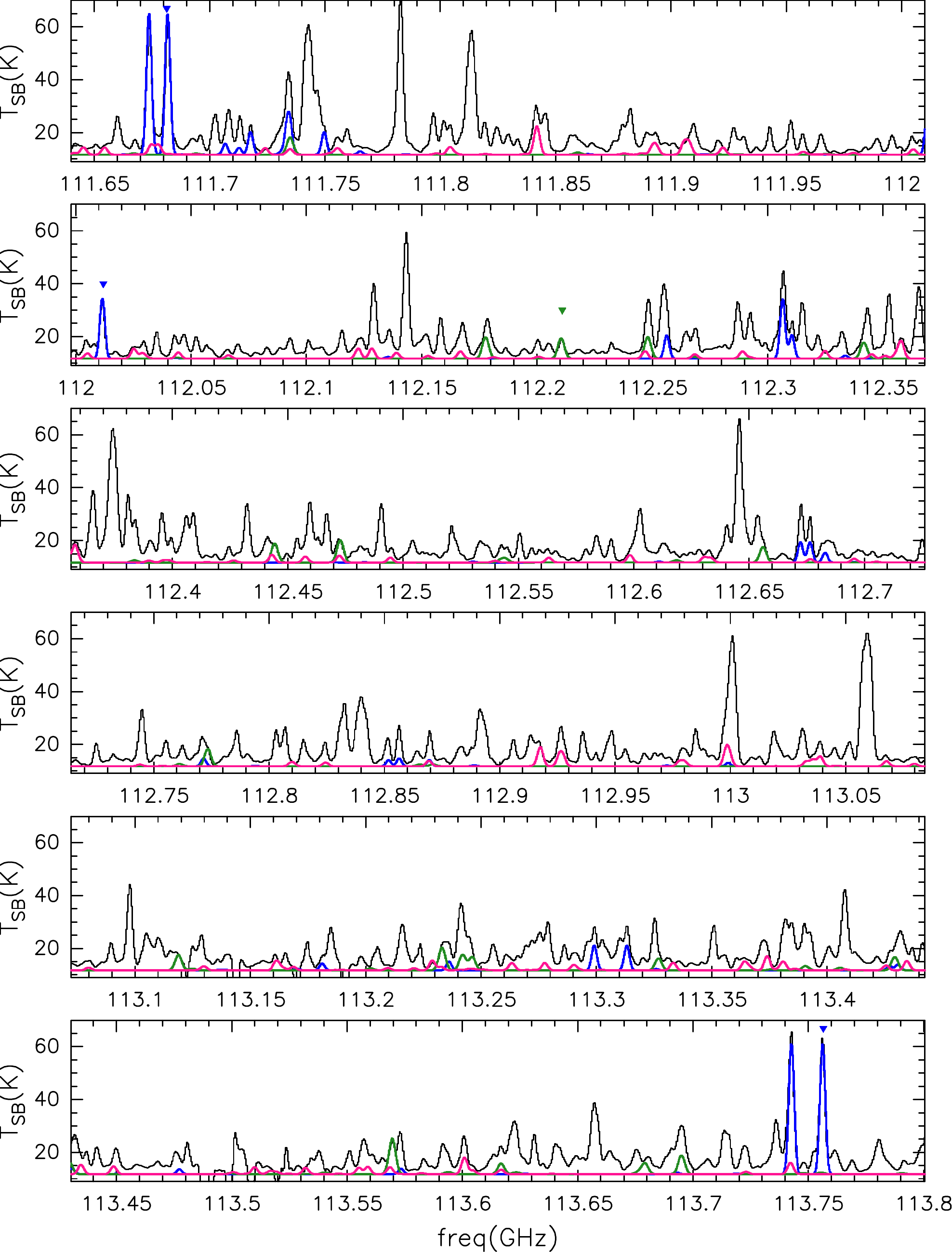}
\caption{ Continue}
\end{figure*}
\newpage
\begin{figure*}
\centering
\includegraphics[width=\hsize]{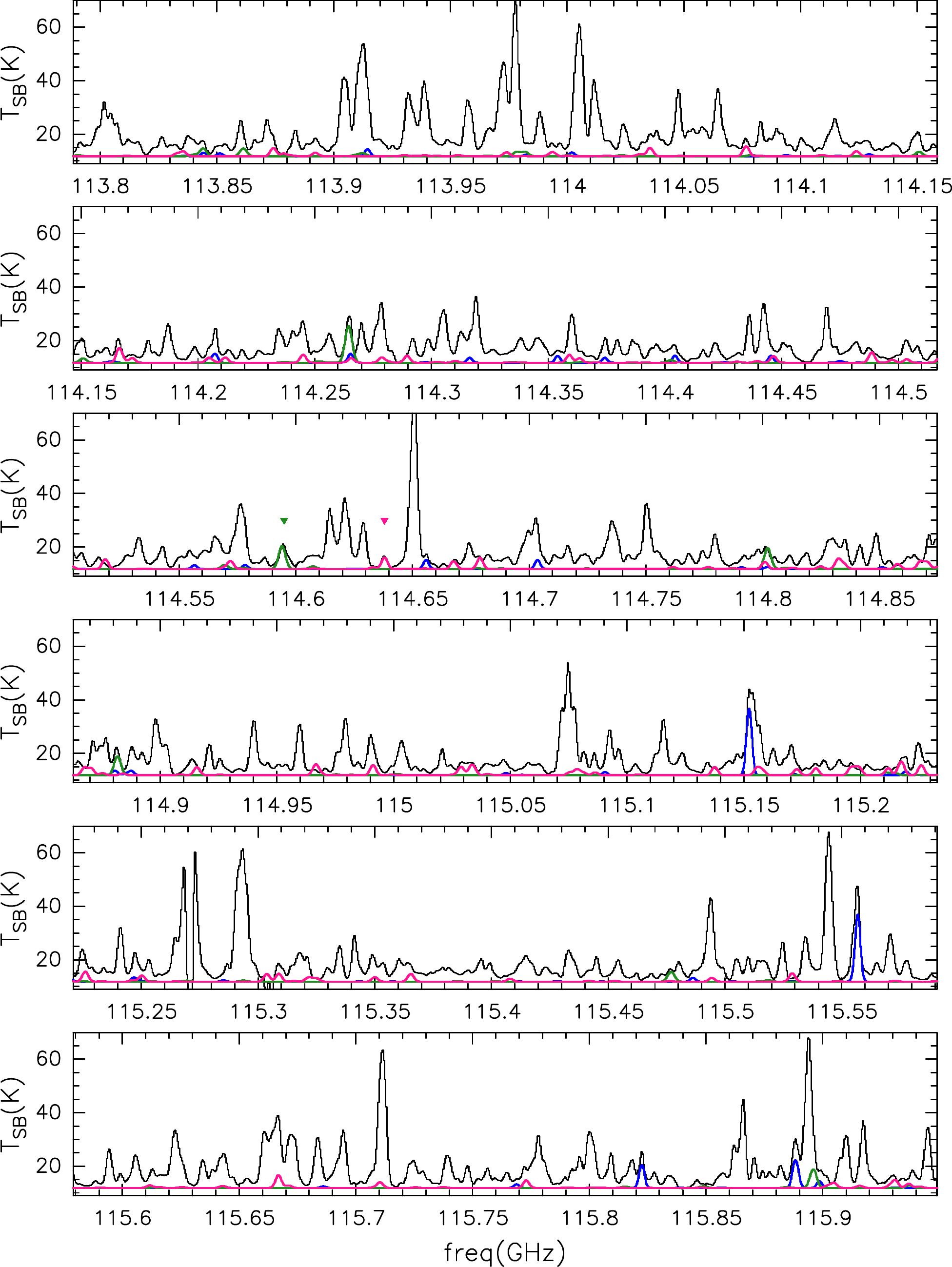}
\caption{ Continue}
\end{figure*}
\clearpage
\end{appendix}
\end{document}